\documentclass[fleqn,usenatbib]{mnras}

\usepackage{graphicx}	
\usepackage{amsmath}	
\usepackage{amssymb}	
\usepackage{bm}	        
\usepackage{mathtools}
\usepackage[utf8]{inputenc}
\usepackage[english]{babel}
\usepackage[section] {placeins}
\usepackage{textcomp,gensymb}
\usepackage{color}
\usepackage{xcolor}
\usepackage{multirow}
\usepackage{makeidx}
\usepackage{xspace}
\usepackage{float}
\usepackage{verbatim}
\usepackage{subcaption}
\usepackage{etoolbox}
\usepackage[normalem]{ulem} 

\usepackage{hyperref} 

\usepackage{etoolbox}
\makeatletter
\newcommand\sendemail[3]{
\edef\@tempa{mailto:#1?subject=#2 }%
\edef\@tempb{\expandafter\html@spaces\@tempa\@empty}%
\href{\@tempb}{#3}}

\catcode\%=11
\def\html@spaces#1 #2{#1
\catcode\%=14
\makeatother

\newcommand{\reffig}[1]{Fig.~\ref{#1}}
\newcommand{\refpan}[1]{Panel~\subref{#1}}
\newcommand{\reftab}[1]{Table~\ref{#1}}
\newcommand{\refapp}[1]{Appendix~\ref{#1}}

\newcommand{\orcid}[2]{\href{http://orcid.org/#2}{#1}}

\newcommand{\re}{$R_\mathrm{e}$\xspace}
\newcommand{\sige}{$\sigma_\mathrm{e}$\xspace}
\newcommand{\Lumsynth}{$L_\mathrm{synth}$\xspace}
\newcommand{\logl}{$\log L$\xspace}
\newcommand{\logre}{$\log R_\mathrm{e}$\xspace}
\newcommand{\logsige}{$\log \sigma_\mathrm{e}$\xspace}
\newcommand{\deltalogl}{$\Delta \, \log L$\xspace}

\newcommand{\Sigvir}{$\Sigma_\mathrm{vir}$\xspace}

\newcommand{\logage}{$\log age$\xspace}
\newcommand{\ZH}{$\rm{[Z/H]}$\xspace}
\newcommand{\aFe}{$\rm{[\alpha/Fe]}$\xspace}
\newcommand{\logups}{$\log \Upsilon_\star$\xspace}
\newcommand{\ups}{$\Upsilon_\star$\xspace}

\defcitealias{cappellari+2006}{C06}
\defcitealias{magoulas+2012}{M12}
\defcitealias{cappellari+2013a}{C13}
\defcitealias{barone+2018}{B18}

\title[The SAMI Fundamental Plane]{The SAMI Galaxy Survey: stellar population and structural trends across the Fundamental Plane}

\author[\orcid{F. D'Eugenio}{0000-0003-2388-8172}~et al.]{\parbox{\textwidth}{
\orcid{Francesco D'Eugenio}{0000-0003-2388-8172}$^{1,2}$\thanks{E-mail: francesco.deugenio@gmail.com}
\orcid{Matthew Colless}{0000-0001-9552-8075}$^{2,3}$,
\orcid{Nicholas Scott}{0000-0001-8495-8547}$^{4,3}$,
\orcid{Arjen van der Wel}{0000-0002-5027-0135}$^{1}$,\\
\orcid{Roger L. Davies}{0000-0001-7897-3812}$^{5}$,
\orcid{Jesse van de Sande}{0000-0003-2552-0021}$^{4,3}$,
\orcid{Sarah M. Sweet}{0000-0002-1576-2505}$^{6}$,
\orcid{Sree Oh}{0000-0002-4731-9604}$^{2,3}$,
\orcid{Brent Groves}{0000-0002-9768-0246}$^{7}$\\
\orcid{Rob Sharp}{0000-0003-4877-7866}$^{2}$,
\orcid{Matt S. Owers}{0000-0002-2879-1663}$^{8,9}$,
\orcid{Joss Bland-Hawthorn}{0000-0001-7516-4016}$^{4,3}$,
\orcid{Scott M. Croom}{0000-0003-2880-9197}$^{4,3}$,
\orcid{Sarah Brough}{0000-0002-9796-1363}$^{3,10}$,
\orcid{Julia J. Bryant}{0000-0003-1627-9301}$^{4,3,11}$,
Michael Goodwin$^{11}$,
\orcid{Jon S. Lawrence}{0000-0002-6998-6993}$^{11}$,
\orcid{Nuria P. F. Lorente}{0000-0003-0450-4807}$^{11}$ and
\orcid{Samuel N. Richards}{0000-0002-5368-0068}$^{12}$
}
\vspace{0.4cm}
\\
\parbox{\textwidth}{
$^{1}$Sterrenkundig Observatorium, Universiteit Gent, Krijgslaan 281 S9, B-9000 Gent, Belgium\\
$^{2}$Research School of Astronomy and Astrophysics, Australian National University, Canberra, ACT 2611, Australia\\
$^{3}$ARC Centre of Excellence for All Sky Astrophysics in 3 Dimensions (ASTRO 3D), Australia\\
$^{4}$Sydney Institute for Astronomy, School of Physics, The University of Sydney, NSW, 2006, Australia\\
$^{5}$Sub-department of Astrophysics, Department of Physics, University of Oxford, Denys Wilkinson Building, Keble Road, Oxford OX1 3RH, UK\\
$^{6}$School of Mathematics and Physics, University of Queensland, Brisbane, QLD 4072, Australia\\
$^{7}$International Centre for Radio Astronomy Research (ICRAR), University of Western Australia, Crawley, WA 6009, Australia
$^{8}$Department of Physics and Astronomy, Macquarie University, NSW 2109, Australia\\
$^{9}$Astronomy, Astrophysics and Astrophotonics Research Centre, Macquarie University, Sydney, NSW 2109, Australia\\
$^{10}$School of Physics, University of New South Wales, NSW 2052, Australia\\
$^{11}$Australian Astronomical Optics, Macquarie University, Sydney, NSW 2109, Australia\\
$^{12}$SOFIA Science Center, USRA, NASA Ames Research Center, Building N232, M/S 232-12, P.O. Box 1, Moffett Field, CA 94035-0001, USA\\
}
}

\date{Accepted 2021 April 20. Received 2021 April 14; in original form 2021 March 4}

\pubyear{2021}

\begin{document}
\label{firstpage}
\pagerange{\pageref{firstpage}--\pageref{lastpage}}
\maketitle

\begin{abstract}
  We study the Fundamental Plane (FP) for a volume- and luminosity-limited
  sample of 560 early-type galaxies from the SAMI survey.
  Using $r-$band sizes and luminosities from new Multi-Gaussian Expansion (MGE)
  photometric measurements, and treating luminosity as the dependent variable,
  the FP has coefficients
  $a=1.294\pm0.039$, $b=0.912\pm0.025$, and zero-point $c=7.067\pm0.078$.
  We leverage the high signal-to-noise of SAMI integral field spectroscopy,
  to determine how structural and stellar-population observables affect the 
  scatter about the FP. The FP residuals correlate most strongly (8$\sigma$
  significance) with luminosity-weighted simple-stellar-population (SSP) age.
  In contrast, the 
  structural observables surface mass density, rotation-to-dispersion ratio,
  S{\'e}rsic index and projected shape all show little or no significant
  correlation. We connect the FP residuals to the empirical relation between age 
  (or stellar mass-to-light ratio \ups) and surface mass density, the best
  predictor of SSP age amongst parameters based on FP observables. We show that
  the FP residuals (anti-)correlate with the residuals of the relation between
  surface density and \ups. This correlation implies that part of the FP scatter
  is due to the broad age and \ups distribution at any given surface mass
  density.
  Using virial mass and \ups we construct a simulated FP and compare it
  to the observed FP. We find that, while the empirical relations between observed
  stellar population relations and FP observables are responsible for most (75\%) of
  the FP scatter, on their own they do not explain the observed tilt of the FP
  away from the virial plane.
\end{abstract}




\section{Introduction}\label{s.i}

The Fundamental Plane (FP) is a 2-dimensional empirical relation between three
galaxy observables: physical size ($R$), root mean square velocity along the line
of sight ($\sigma$) and surface brightness \citep[$I$;][]{djorgovski+davis1987,
dressler+1987}. By connecting distance-independent $\sigma$ and $I$ to
distance-dependent $R$, the FP can be used as a distance indicator in cosmology
\citep[e.g.][]{hudson+1999, colless+2001, beutler+2011, johnson+2014}.
Tension between different determinations of the cosmological parameters
\citep{planck+2016, riess+2016}, as well as its use to map peculiar velocities,
mean that the FP remains a critical tool for cosmology \citep[e.g.][]{springob+2014, scrimgeour+2016, said+2020}.
In addition, there are several reasons why the FP remains a critical benchmark
for galaxy evolution studies. The tightness of the FP \citetext{scatter of
20--25\%, \citealp{jorgensen+1996}, \citealp{hyde+bernardi2009},
\citealp{magoulas+2012}, hereafter: \citetalias{magoulas+2012}} provides a strong
constraint to theory, limiting the
rate and strength of physical processes that drive galaxies away from the plane
\citep[e.g.][]{kobayashi2005}. Moreover, the FP enables us to probe - albeit
not directly - the scaling relations of dark matter and initial mass function
\citep[IMF; e.g.][]{prugniel+simien1997, graves+faber2010}, thus constraining
the present-day structure of local galaxies. Finally, the advent of
large-aperture ($8$-$10 \, \mathrm{m}$) telescopes opened a window to study how
the FP changes over cosmic time \citep{vandokkum+stanford2003, vanderwel+2004,
wuyts+2004}, a subject that continues to this day, with studies reporting either
evolution \citep{saracco+2020}, weak evolution \citep{saglia+2010, saglia+2016,
oldham+2017, dallabonta+2018} or no evolution \citep{holden+2010,
prichard+2017, degraaff+2020}. 
For all these reasons, a better understanding of the FP will improve our
understanding of galaxies and potentially increase its precision and
accuracy as a tool for cosmology.

The FP is physically rooted in the virialised nature of galaxies: the scalar
virial theorem links dynamical mass, size and specific kinetic energy, thus
constraining these observables on the virial (or mass) plane \citetext{
\citealp{djorgovski+davis1987}, \citealp{dressler+1987},
\citealp{cappellari+2006}, hereafter: \citetalias{cappellari+2006},
\citealp{cappellari+2013a}, hereafter: \citetalias{cappellari+2013a}}.
If we use projected half-light radius (\re) to measure size and 
root mean square velocity along the line of sight inside an aperture of radius
\re (\sige) to measure kinetic energy, the virial mass can be expressed by
\begin{equation}\label{eq.i.mvir}
    \log M_\mathrm{vir} = \log \kappa + 2 \log \sigma_\mathrm{e} + \log R_\mathrm{e} - \log G
\end{equation}
where $G$ is the gravitational constant and $\kappa$ is a parameter which
encodes the possibility of `non-homology', that is of systematic differences in
galaxy structure along and orthogonal to the FP \citep[cf.][]{bender+1992,
graham+colless1997, prugniel+simien1997}. Equation~(\ref{eq.i.mvir}) defines a geometric plane in
the logarithmic space of $(\sigma_\mathrm{e}, R_\mathrm{e}, M_\mathrm{vir})$. To obtain the FP,
we further introduce the stellar mass-to-light ratio \citep[$\Upsilon_\star$,
assuming a fiducial Chabrier initial mass function, IMF;][]{chabrier2003}, the stellar
mass-to-light ratio assuming a non-standard IMF ($\Upsilon_{\mathrm{IMF}}$) and
the stellar-to-total mass fraction within one effective radius ($f_\star$). With
these definitions, the FP can be expressed as
\begin{equation}\label{eq.i.logl}
\begin{split}
    \log L = \log \kappa + & 2 \log \sigma_\mathrm{e} + \log R_\mathrm{e} - \log G \\
        - & \log \Upsilon_{\star} + \log f_\star
        - \log (\Upsilon_{\mathrm{IMF}}/\Upsilon_\star)
\end{split}
\end{equation}
where $\kappa$, $f_\star$, $\Upsilon_\star$ and
$\Upsilon_{\mathrm{IMF}}/\Upsilon_\star$ may be (possibly indirect) functions of
both \sige and \re. Empirically, the FP is commonly expressed as
\begin{equation}\label{eq.i.fp.cosmo}
    \log R_\mathrm{e} = a \log \sigma_\mathrm{e} + b \log I_\mathrm{e} + c
\end{equation}
where $I_\mathrm{e}$ is the mean surface brightness inside one \re (in this
formulation, the virial prediction is $a=2$ and $b=-1$). In this work however,
following \citetalias{cappellari+2013a}, we use the alternative expression
\begin{equation}\label{eq.i.fp}
    \log L = a \log \sigma_\mathrm{e} + b \log R_\mathrm{e} + c
\end{equation}
because it reduces correlated noise between \re and $L$ and is easier to
interpret. While the observed mass plane is consistent with the virial plane
\citetext{$a=2$ and $b=1$, \citetalias{cappellari+2013a}}, the FP has
systematically different coefficients \citep{djorgovski+davis1987,
dressler+1987}. Geometrically, this
difference means that the FP is tilted (or rotated) with respect to the virial
plane. By comparing equations (\ref{eq.i.mvir}) and (\ref{eq.i.logl}), the FP
tilt must arise from systematic variations of $\kappa$, \ups, $f_\star$ and/or
$\Upsilon_{\mathrm{IMF}}/\Upsilon_\star$ with \sige and \re.

In addition to its tilt, the observed FP also differs from the mass plane in
that the latter is consistent with no intrinsic scatter \citepalias{
cappellari+2013a}, whereas the FP has finite scatter \citep{jorgensen+1996}.
The FP intrinsic scatter is critical to cosmology, because it represents a hard
limit to its precision as a distance estimator \citepalias{magoulas+2012}.
Therefore, understanding
the origin of this scatter is important to understanding if it can be reduced,
thereby improving the FP as a tool for cosmology.
From equation~(\ref{eq.i.logl}), the FP scatter must originate from
galaxy-to-galaxy variations in $\kappa$, \ups, $f_\star$ and/or
$\Upsilon_{\mathrm{IMF}}/\Upsilon_\star$ at fixed \sige and \re.
Therefore, the fact that the FP is `tight' requires either that these galaxy
observables have narrow distributions at fixed \sige and \re, or, alternatively,
that these distributions are correlated in such a way as to decrease the FP
scatter \citetext{FP `fine-tuning', \citealp{ciotti+1996} - but see
\citealp{chiu+2017} for a different view}. This fine-tuning requirement provides
an additional constraint to galaxy evolution theory.

A satisfactory understanding of the FP requires (i) determining $\kappa$, \ups,
$f_\star$ and $\Upsilon_{\mathrm{IMF}}/\Upsilon_\star$ as functions of \sige
and \re and (ii) using these functions in equation~(\ref{eq.i.logl}) to
reproduce the observed FP. The FP intrinsic scatter must also be consistent
with the combined (and possibly correlated) scatter of these input galaxy
properties at fixed \sige and \re. Unfortunately, measuring $f_\star$ and
$\Upsilon_{\mathrm{IMF}}/\Upsilon_\star$ is challenging, because
dark matter (in the definition of $f_\star$) cannot be observed directly and
because low-mass stars (for $\Upsilon_{\mathrm{IMF}}/\Upsilon_\star$) require
high quality observations \citep[cf.][]{conroy+vandokkum2012}.
A more practicable approach is to combine measurements of $\kappa$ and
\ups with the observed FP and to infer, by subtraction, the missing contribution
due to dark matter and IMF trends \citep[e.g.][]{prugniel+simien1997,
graves+faber2010}; these inferred trends are a useful benchmark for theory.

Single-fibre spectroscopy surveys enabled us to measure the relation
between the FP observable $\sigma$ and stellar population age and
metallicity (which, together, determine \ups). The strong observed trends
\citep{nelan+2005, gallazzi+2006, thomas+2010} must be reflected in the FP, as
confirmed by \citet{graves+2009} and \citet{springob+2012}.
The key role of stellar population properties on the FP is highlighted by
the fact that both early-type as well as late-type galaxies lie on the same
stellar-mass plane \citep[e.g.][]{bezanson+2015}, and that, for
non star-forming galaxies, the evolution of the FP is consistent with the
passive evolution of their stellar populations \citep[e.g.][]{
vandokkum+vandermarel2007, vandesande+2014}.

As for the origin of the FP scatter, a direct view is provided by the study of
the FP residuals, defined as the difference between the left- and right-hand
side of equation~(\ref{eq.i.fp}). FP residuals correlate with both
stellar-population light-weighted age \citep[a proxy for \ups, ][]{forbes+1998,
graves+2009, graves+faber2010, springob+2012} as well as with S{\'e}rsic index
$n$ \citep[a measure of non-homology $\kappa$,][]{prugniel+simien1997}. The
strength and significance of these correlations can be used to compare the
relative importance of stellar-population and structural differences to the FP
scatter. These residual trends are also important because if age and $n$ contain
information about the FP scatter, it should be possible, in principle, to factor
this information into the FP and to improve its precision as a distance
indicator \citepalias{magoulas+2012}. However, a consistent assessment of the
relative importance of stellar populations and non-homology requires an unbiased
determination of both these observables, as well as the FP residuals.

Non-homology is conveniently captured by S{\'e}rsic index $n$, which is
measured from photometry \citep{bertin+2002}. Until recently, however, studies
of the FP residuals with stellar-population properties had to rely on
single-fibre spectra with relatively low signal-to-noise \citep{springob+2012}
or, alternatively, on stacking observations of different galaxies, which hides
the intrinsic galaxy-to-galaxy variability \citep{graves+2009, graves+faber2010}.
Moreover, fibres of fixed apparent size, coupled with age and metallicity
gradients within galaxies \citep[e.g.][]{carollo+1993, mehlert+2003,
sanchez-blazquez+2007, zibetti+2020}, introduce a size-dependent bias. This
`aperture bias' is particularly problematic, because size \re appears directly
in the FP equation~(\ref{eq.i.fp}). Integral-field spectroscopy (IFS)
enabled us to measure precise and accurate stellar population properties for
individual galaxies. By adding the light inside an aperture that matches the
size of each galaxy, we can derive a `synthetic' spectrum with both high
signal-to-noise ratio and negligible aperture bias
\citep[e.g.][]{mcdermid+2015}. However, the first generation of IFS surveys
\citetext{SAURON, \citealp{dezeeuw+2002} and ATLAS$^{\rm 3D}$, \citealp{
cappellari+2011a}} were limited to $\approx 250$ galaxies.

The advent of large IFS surveys has delivered high signal-to-noise spectra for
thousands of galaxies, without aperture bias \citetext{CALIFA,
\citealp{sanchez+2012}, SAMI, \citealp{croom+2012} and MaNGA,
\citealp{bundy+2015}}. These surveys helped clarify the link between
galaxy structure and stellar population properties \citep{zibetti+2020}.
Previous studies have focused on the link between stellar-population
properties with either $\sigma$ \citep{gallazzi+2006, ganda+2007, mcdermid+2015}
or stellar mass \citep{gallazzi+2005}. However, more recently, it became clear
that at fixed $\sigma$, galaxy size also affects stellar population age.
\citet[][hereafter: \citetalias{barone+2018}]{barone+2018} have shown that while
the best predictor of light-weighted stellar-population metallicity is
gravitational potential (proportional to $\sigma^2$), age is driven by surface
mass density (proportional to $\sigma^2/R_\mathrm{e}$). Given that age and metallicity
jointly determine \ups, the systematic variations of age (and \ups) at fixed \re
must also be reflected in the FP tilt, or, alternatively, we need to explain the
absence of such effect.

As for the FP residuals, their determination is also prone to bias, because the
FP parameters depend on a number of assumptions: sample selection, photometric
band, measurement uncertainties and optimisation method \citetext{
\citealp{jorgensen+1996}, \citealp{colless+2001}, \citealp{hyde+bernardi2009},
\citetalias{magoulas+2012}}. For these reasons, understanding the FP requires a
careful consideration of the impact of these assumptions on the results.

In this work we leverage the consistent apertures and high signal-to-noise
of the integral field SAMI Galaxy Survey and state-of-the-art probabilistic
models (i)~to conduct a comparative analysis of the FP residuals and (ii)~to
investigate how stellar population trends affect the tilt and scatter of the FP.
In \S~\ref{s.ds} we present the photometric measurements that have recently been released as part of the SAMI third public data release \citep{croom+2021} and describe the observables and sample selection criteria used specifically in this work. In \S~\ref{s.da} we explain and justify the methods used in the
analysis. \S~\ref{s.r} illustrates the results: the residuals of the FP correlate
most strongly (8$\sigma$ significance) with stellar population age, whereas
structural variables show little or no significant correlation. We connect this
trend to the known relation between stellar population age and surface mass
density, and show that stellar population relations, on their own, explain most
($\approx$75\%) of the FP intrinsic scatter.
After  discussing the implications (\S~\ref{s.d}), we conclude with a summary of our
findings (\S~\ref{s.c}).

Throughout this paper, we use a flat $\Lambda$CDM cosmology with $H_0 = 70 \;
\mathrm{km \; s^{-1} \; Mpc^{-1}}$ and $\Omega_m = 0.3$. This is the
standard cosmology adopted by the SAMI Galaxy Survey \citep{bryant+2015}.
Unless otherwise
specified, all magnitudes are in the AB system \citep{oke+gunn1983} and stellar
masses and mass-to-light ratios assume a Chabrier IMF \citep{chabrier2003}.

\section{Data and Sample}\label{s.ds}

The aim of this section is twofold: (i)~to describe in detail the photometric
measurements of size, shape and flux made available in the third public data
release of the
SAMI Galaxy Survey and (ii)~to describe the additional measurements and the
sample selection criteria that are specific to the science goals of this paper.
To avoid confusion, we stress that for goal (i) we use the SAMI \textit{parent}
sample of $\sim$10,000 galaxies, whereas for goal (ii) we use the
sample from the SAMI Galaxy Survey, i.e.\ the subset of $\sim$3000
galaxies with available integral-field spectroscopy data.

This section is organised as follows. We provide a brief introduction to the
SAMI Galaxy Survey
(\S~\ref{s.ds.ss.sami}), then proceed to describe the photometric measurements
performed on the SAMI parent sample (\S~\ref{s.ds.ss.phot}). Afterwards, we
delve into the specifics of this paper: we present spectroscopic measurements
based on SAMI observations (\S~\ref{s.ds.ss.spec}) and we estimate the
correlation between the measurement uncertainties (\S~\ref{s.ds.ss.corr}). In
\S~\ref{s.ds.ss.ancillary} we briefly introduce additional observables that are
necessary to our analysis, but that have already been presented in previous works.
The final section explains the selection criteria for the SAMI Fundamental Plane
sample (\S~\ref{s.ds.ss.samp}).

\subsection{The SAMI Galaxy Survey}\label{s.ds.ss.sami}

Our sample is drawn from the SAMI Galaxy Survey, the integral field spectroscopy
survey based on the Sydney-AAO Multi-object Integral field spectroscopy
instrument \citep[hereafter, the SAMI instrument;][]{croom+2012}. The SAMI
Galaxy Survey (hereafter, simply SAMI) observed a mass-selected sample of
$\sim$3000 galaxies drawn from a larger parent sample of
$\sim$10,000 galaxies, spanning a wide range in both stellar mass and
environment. The parent sample includes galaxies between $10^7 \, \mathrm{M_\odot}$ and
$10^{12} \, \mathrm{M_\odot}$ \citep[the lower mass limit increases with
redshift; see][]{bryant+2015}. As for environment, SAMI consists of both field
and group galaxies \citep{bryant+2015} as well as cluster galaxies
\citep{owers+2017}. Because of their heterogeneous selection, SAMI galaxies
have different photometry. Field and group galaxies are from the
Galaxy and Mass Assembly Survey \citep[GAMA;][]{driver+2011} and use Sloan
Digital Sky Survey (SDSS) Data Release 7 optical imaging \citep{abazajian+2009},
reprocessed as described in \citet{hill+2011}.
Cluster galaxies have been selected from eight clusters; of these, the photometry for Abell~85, Abell~119, Abell~168 and Abell~2399 is from SDSS DR9 \citep{ahn+2012},
whereas photometry for APMCC0917, EDCC442, Abell~3880 and Abell~4038 is from the
VLT Survey Telescope's ATLAS Survey \citep[VST;][]{shanks+2013, shanks+2015}.

In this work, we use data from SAMI internal data release 0.12, consisting of
2153 field and group galaxies and 918 cluster members, for a total of 3071
galaxies. In addition to these galaxies, there are 311 repeat observations; in
these cases, we always select the galaxy observed under the best atmospheric
seeing. This data has been released to the community as part of SAMI's public
data release 3 \citep[DR3,][]{croom+2021}.

\subsection{SAMI photometry}\label{s.ds.ss.phot}

In this section we describe in detail the procedure adopted to measure projected
\emph{circularised} half-light radii (\re) and luminosities, using the Multi Gaussian
Expansion algorithm \citep[MGE; ][]{emsellem+1994}. We then describe the
calibration between different photometric surveys
(\S~\ref{s.ds.ss.phot.sss.calib}), the estimation of the measurement
uncertainties (\S~\ref{s.ds.ss.phot.sss.remag}) and the measurement of
galaxy surface density (\S~\ref{s.ds.ss.phot.sss.sig}).
Notice that all the \textit{photometric} measurements are performed on the SAMI
\textit{parent} sample, because this larger dataset helps to characterise the
quality of our photometric measurements with higher precision.

To facilitate comparison with the literature \citetext{e.g.\ \citetalias{
cappellari+2013a}, \citealp{scott+2015}}, our Fundamental Plane analysis uses
$r-$band photometry only. Nevertheless,  we measure $g-$, $r-$ and $i-$band \re
and total magnitudes ($m$) for each of the 9332 galaxies in the SAMI parent
sample. Notice that 436 galaxies in cluster Abell~85 have both SDSS and VST photometry,
so the parent sample contains up to 9768 images for each band
(some bands have incomplete coverage). For brevity, we discuss in depth only
$r-$band measurements, because the results are analogous and independent between
the three bands.

\subsubsection{Masking and PSF characterisation}\label{s.ds.ss.phot.sss.maskpsf}

For each galaxy, we retrieve a square cutout $400\,\mathrm{arcsec}$ on a side,
centred on the galaxy.
The size of these cutouts is chosen to guarantee the presence of a number of field stars
sufficient to characterise the point-spread function (PSF).

We show two randomly selected FP galaxies in \reffig{f.ds.succfit}; these are
galaxy 383585 from GAMA (using SDSS photometry, panel~\subref{f.ds.succfit.a})
and galaxy 9239900277 from cluster Abell~2399 (again using SDSS photometry,
panel~\subref{f.ds.succfit.b}).

First, we use {\sc \href{https://www.astromatic.net/software/sextractor}{SExtractor}}
\citep{bertin+arnouts1996} to retrieve all the
sources in the image.
{\sc SExtractor} also provides a map of all detected sources; after
removing the target galaxy, we use this map to create a mask of contaminating
sources (masked sources are shaded in yellow in
\reffig{f.ds.succfit}). Notice that interloping sources can be particularly large
for cluster galaxies, as shown for galaxy 9403800025 from cluster Abell~4038
(\reffig{f.ds.probfit.a}).
Secondly, to characterise the PSF, we retrieve a square cutout around all the
stars in the image, defined as sources having a {\sc SExtractor} keyword
{\sc class\_star}$>0.9$. We model each star as the superposition of 2--5
circular Gaussians, using {\sc \href{https:pypi.org/project/mgefit/}{mgefit}}
\citep[the MGE algorithm implemented by][]{cappellari2002} and we find the
reduced $\chi^2$ for each fit. From this set of stars, we select
the best compromise between magnitude, reduced $\chi^2$ and distance from the
target galaxy; the model PSF is then the best-fit MGE to the
selected star. In Figs~\ref{f.ds.succfit} and \ref{f.ds.probfit}, the
full-width half-maximum (FWHM) of the PSF is represented by the diameter of the
grey circle in the bottom left corner of each panel. Over the whole SAMI parent
sample, we find a median PSF FWHM of $1.18 \, \mathrm{arcsec}$ with a standard
deviation of $0.20 \, \mathrm{arcsec}$ (for SDSS photometry) and of $0.97 \,
\mathrm{arcsec}$ with a standard deviation of $0.19 \, \mathrm{arcsec}$
(for VST photometry); both these results are in qualitative agreement with the
relevant literature \citep{kelvin+2012, shanks+2015}.

\subsubsection{Multi Gaussian Expansion photometry}\label{s.ds.ss.phot.sss.mge}

Once the image mask and model PSF have been determined, we fit the galaxy flux, again using {\sc mgefit}. In order to minimise systematic errors due to
substructures (e.g.\ bars, spiral arms) we use the regularisation feature of
{\sc mgefit} \citep[described in][]{scott+2009}; this is the only difference
between the photometry we use for the FP analysis and the photometry released as part
of the SAMI DR3. The reason why SAMI DR3 uses unregularised fits is that these
yield the lowest $\chi^2$ and more realistic galaxy shapes. Regularised fits,
on the other hand, while being biased to more circular shapes, can be used to
build more robust dynamical models \citep[][their figures~2 and~3]{scott+2009}.
Given the possibility to expand this work to include dynamical models, here we
opted for regularised photometry.

\begin{figure}
  \centering
  \includegraphics[type=pdf,ext=.pdf,read=.pdf,width=0.5\textwidth]{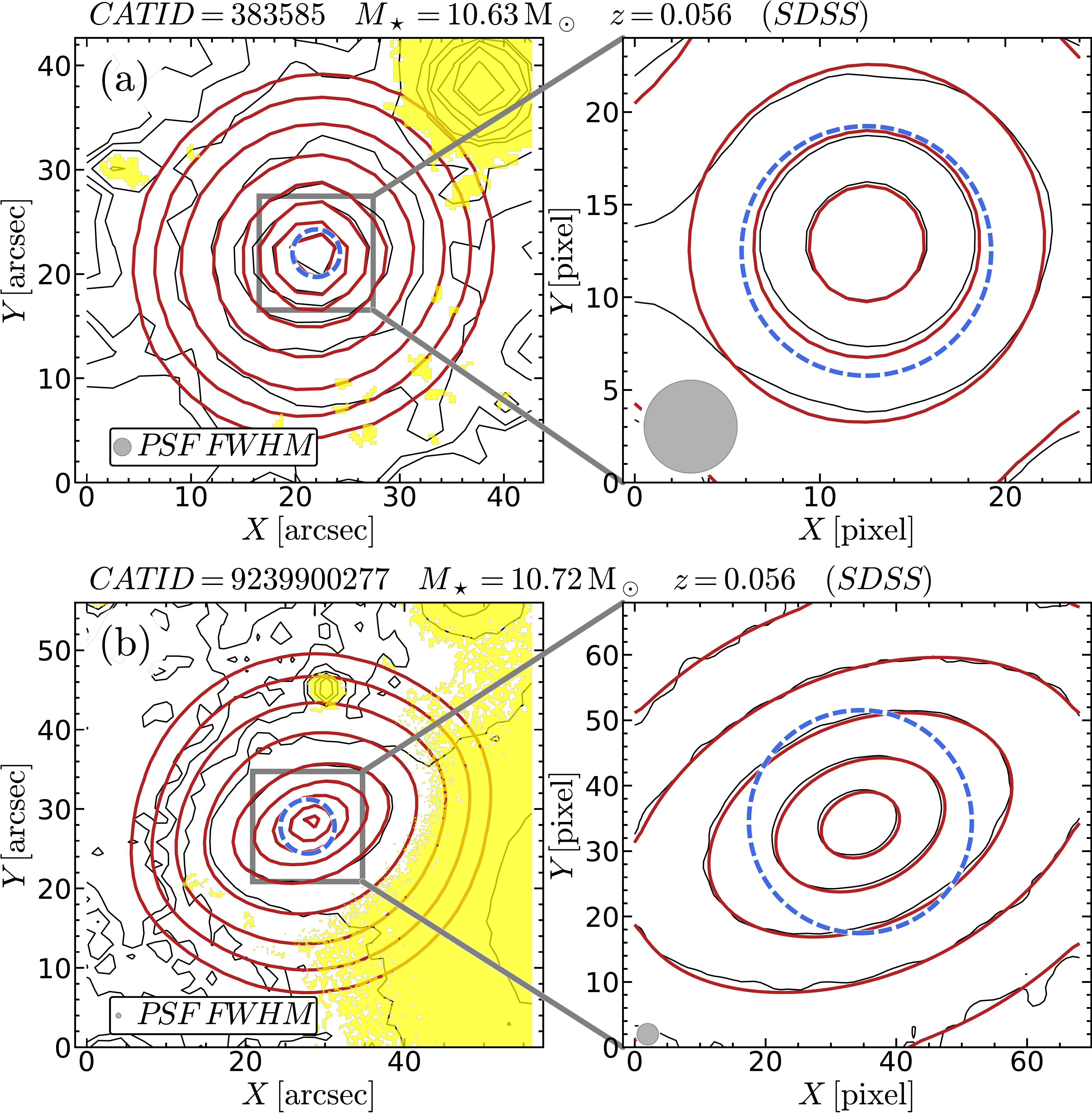}
  {\phantomsubcaption\label{f.ds.succfit.a}
   \phantomsubcaption\label{f.ds.succfit.b}}
  \caption{Example of two successful $r-$band MGE fits: galaxy 383585 from GAMA
  (with SDSS photometry, panel~\subref{f.ds.succfit.a}) and galaxy 9239900277
  from cluster Abell~2399 (again with SDSS photometry, panel~\subref{f.ds.succfit.b}).
  The left column shows the full extent of the
  galaxy, the right column shows an enlarged version of the grey square ($4\,\mathrm{R_e}$ on a side). The black/red contours are observed/model
  isophotes spaced by $1 \, \mathrm{mag\,arcsec^{-2}}$ and the blue dashed circle
  has radius \re. Masked regions are highlighted in yellow, the diameter of
  the grey circle in the bottom-left corner is equal to
  the PSF FWHM. These two galaxies were randomly selected from the Fundamental
  Plane sample.
  }\label{f.ds.succfit}
\end{figure}

\begin{figure}
  \centering
  \includegraphics[type=pdf,ext=.pdf,read=.pdf,width=0.5\textwidth]{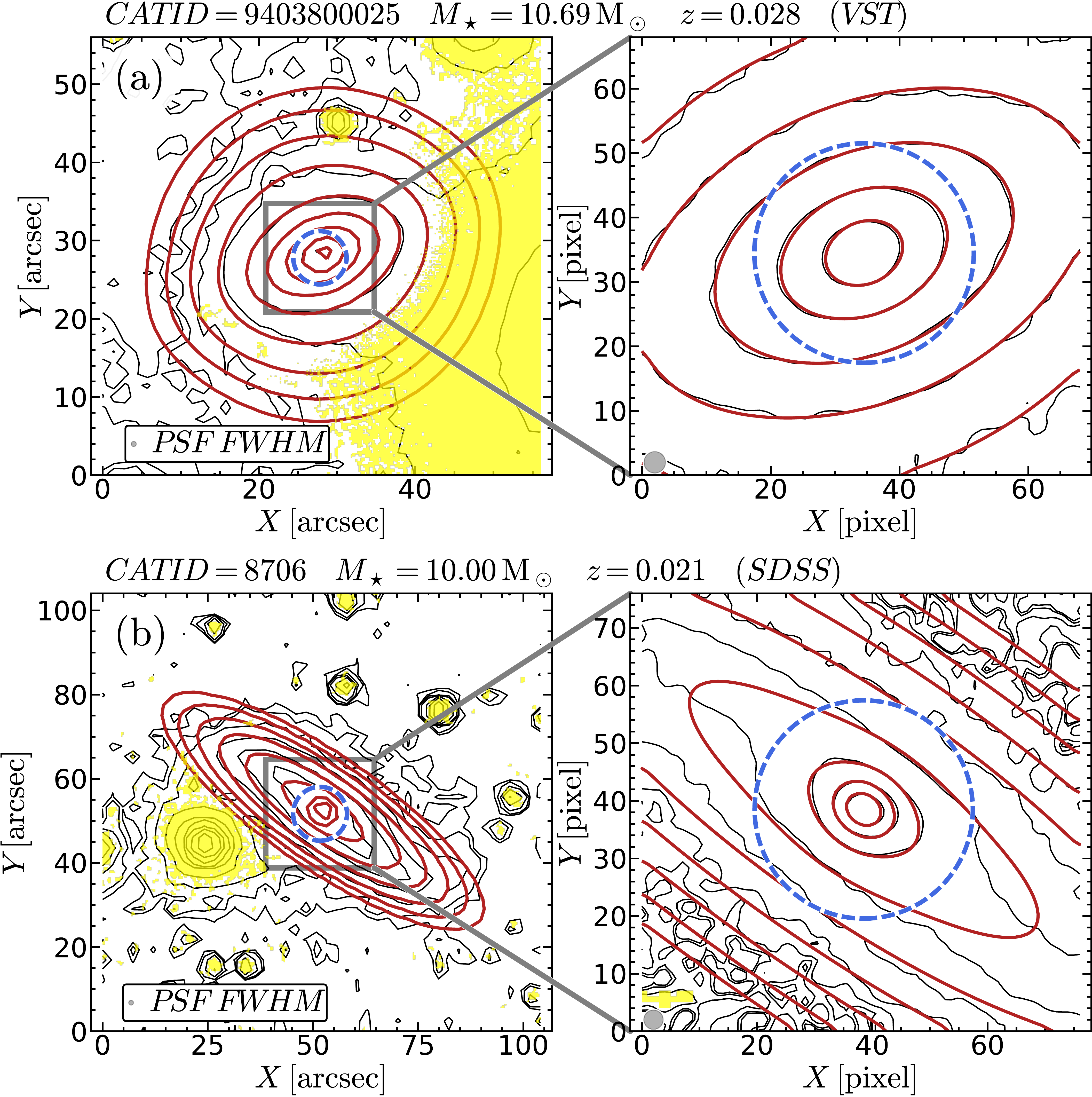}
  {\phantomsubcaption\label{f.ds.probfit.a}
   \phantomsubcaption\label{f.ds.probfit.b}}
  \caption{Example of two problematic $r-$band MGE fits. Galaxy 9403800025 from
  cluster Abell~4038, using VST photometry (panel~\subref{f.ds.probfit.a}) has
  substantial contamination from a neighbour (shaded yellow region). Galaxy 8706
  is a spiral galaxy from GAMA; the MGE model does not accurately describe its
  substructure (panel~\subref{f.ds.probfit.b}). Lines and symbols are the same
  as in \reffig{f.ds.succfit}.
  }\label{f.ds.probfit}
\end{figure}

\reffig{f.ds.succfit} shows two successful fits: observed galaxy isophotes and
best-fit model isophotes are traced by black and red solid lines, respectively
(isophotes are spaced by $1 \, \mathrm{mag\,arcsec^{-2}}$); the content of the grey
square in the left column is reproduced, enlarged, in the right column;
the blue dashed circle has radius $1\,\mathrm{R_e}$.
By contrast, \reffig{f.ds.probfit} shows two problematic fits. Even though MGE may not describe accurately the substructures common in
late-type galaxies (e.g.\ Fig.~\ref{f.ds.probfit.b}), single-S{\'e}rsic models do
not necessarily perform better, as highlighted by the fact that the two
measurements are generally in good agreement (\S~\ref{s.ds.ss.phot.sss.remag}).

From the best-fit MGE model we measure total magnitude $m$, galaxy ellipticity
$\epsilon$ and galaxy size \re. Total magnitude is given by the sum of the
fluxes of each Gaussian component. The circularised half-light radius \re is defined
as the radius enclosing half the total flux of the circularised MGE model (i.e.\ 
the model where all the Gaussian components have the same flux as the best-fit
model, but have isotropic dispersion equal to $\sqrt{a\,b}$, where $a$ and $b$
are the semi-major and semi-minor axes of the best-fit Gaussian component).
Finally, ellipticities are measured with the method of
moments as implemented in the algorithm {\sc find\_galaxy}\footnote{Part of
the {\sc mgefit} package available on \href{https://pypi.org/project/mgefit/}{PyPI}.};
we find the model isophote with area $A = \pi \mathrm{R_e^2}$, and use its
ellipticity as the galaxy ellipticity \citepalias{cappellari+2013a}.

Our magnitude measurements have been k-corrected to redshift $z = 0.05$, close
to the median value for the FP sample ($z=0.053$), using {\sc \href{http://kcorrect.org/}{kcorrect}} \citep{blanton+roweis2007}. For galaxies drawn from
GAMA, we find good agreement between our k-corrections $K$ (re-computed to
$z=0$ for this test), and previously published values \citep{loveday+2012}; we
find $K = (1.048 \pm 0.002) K_\mathrm{GAMA} + (0.004 \pm 0.001)$ and $rms=0.014$.
Neglecting the k-correction entirely does not change the fiducial parameters
or the slope of the FP.

We convert apparent sizes and magnitudes to proper sizes and luminosities
using the angular diameter and luminosity distance based on
the adopted cosmology (\S~\ref{s.i}) and on flow-corrected
spectroscopic redshifts \citetext{\citealp{tonry+2000}, \citealp{baldry+2012};
for cluster galaxies, we use the redshift of the cluster}. In order to convert
$r-$band luminosities to units of solar
luminosity ($\mathrm{L}_{\odot,r}$), we adopt an absolute magnitude of the Sun
of $\mathrm{M}_{\odot,r} = 4.64$~mag \citep{blanton+roweis2007}. The
uncertainty on \logl is therefore $\sigma_{\log L} = 0.4 \sigma_m = 0.03$~dex.
For $g-$ and $i-$band photometry we use $\mathrm{M}_{\odot,g} = 5.12$~mag and
$\mathrm{M}_{\odot,i} = 4.53$~mag respectively \citep{blanton+roweis2007}.

\subsubsection{Calibration between SDSS and VST photometry}\label{s.ds.ss.phot.sss.calib}

For a considerable fraction of our sample, photometry is available from VST only.
In order to remove possible bias due to systematic differences between the VST-
and SDSS-based measurements, we use a set of 436 galaxies from cluster Abell~85,
for which we have both VST and SDSS data. To compare the two measurements we use
the least-trimmed squares algorithm \citep[LTS,][]{rousseeuw+driessen2006}, in
the free implementation {\sc \href{https://pypi.org/project/ltsfit/}{lts\_linefit}}
of \citetalias{cappellari+2013a}.

\begin{figure}
  \centering
  \includegraphics[type=pdf,ext=.pdf,read=.pdf,width=0.5\textwidth]{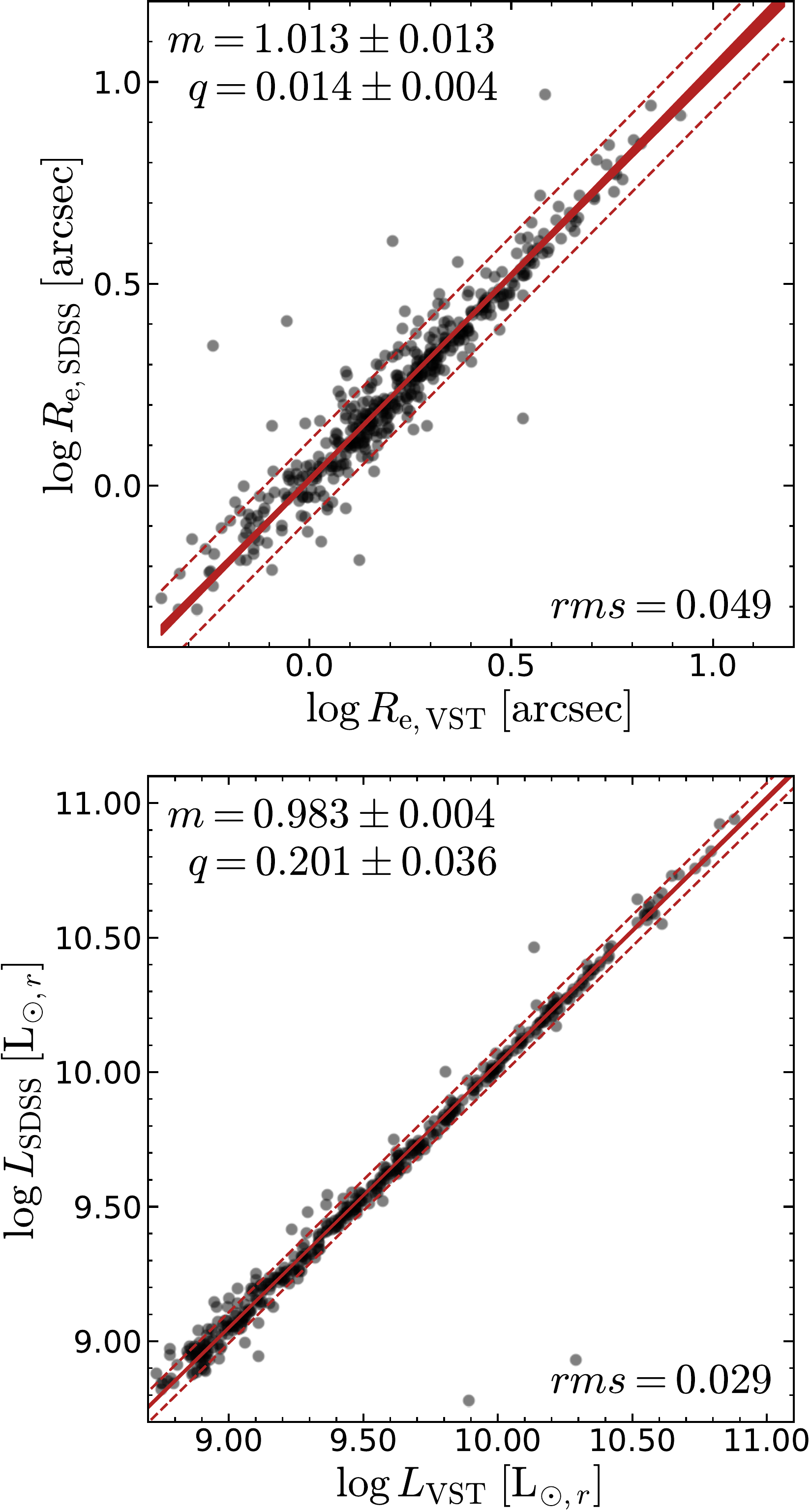}
  \caption{Comparison between SDSS-based and VST-based photometry. We correct
  the VST measurements to the SDSS standard using the best-fit relation. Circles
  represent individual SAMI galaxies, the shaded red region shows the 95\%
  confidence interval and the dashed red lines show the 95\% prediction
  interval. 
  }\label{f.ds.sdss.vs.vst}
\end{figure}

We set the data-point uncertainties to a uniform value of 0.045~dex
(estimated in \S~\ref{s.ds.ss.phot.sss.remag}), and the sigma-clipping keyword {\sc clip}
to a value of 3. The results for \re and $L$ are illustrated in
\reffig{f.ds.sdss.vs.vst}, where we show the data as black contours enclosing
the 50\textsuperscript{th}, 75\textsuperscript{th} and 90\textsuperscript{th}
percentiles; the shaded red region shows the 95\% confidence interval,
whereas the dashed red lines enclose the 95\% prediction interval. 
For \re, we find a relation consistent with no systematic trend (i.e.\ best-fit slope
consistent with 1)
\begin{equation}\label{eq.ds.re.vst}
\log R_\mathrm{e,SDSS} = (1.013 \pm 0.013) \log R_\mathrm{e,VST}
     + (0.014 \pm 0.004)
\end{equation}
but there is a small offset, with SDSS sizes $10^{0.014} \approx 3\%$ larger
than VST sizes (the typical uncertainty is $11\%$; see again
\S~\ref{s.ds.ss.phot.sss.remag}). For luminosity, we find
\begin{equation}\label{eq.ds.lum.vst}
\log L_\mathrm{SDSS} = (0.983 \pm 0.004) \log L_\mathrm{VST} + (0.201 \pm 0.036)
\end{equation}
which corresponds to an offset in the magnitude zero-point. This offset is
applied on top of the correction already derived by \citet{owers+2017}.

We use the two equations (\ref{eq.ds.re.vst}-\ref{eq.ds.lum.vst}) to convert
the VST-based measurements to their SDSS equivalent. Similar corrections are
applied to the other photometric variables derived from MGE.

\subsubsection{Measurement uncertainties}\label{s.ds.ss.phot.sss.remag}

In order to assess the uncertainties on our measurements of \re and $m$,
we compare our results to the corresponding values from the GAMA survey,
derived from single-S{\'e}rsic fits measured using {\sc sigma}
\citep{kelvin+2012}, a pipeline feeding
{\sc \href{https://users.obs.carnegiescience.edu/peng/work/galfit/galfit.html}{galfit}}
\citep{peng+2002}. We describe only the
procedure adopted for \re, because the procedure for $m$ is analogous.

\begin{figure}
  \centering
  \includegraphics[type=pdf,ext=.pdf,read=.pdf,width=0.5\textwidth]{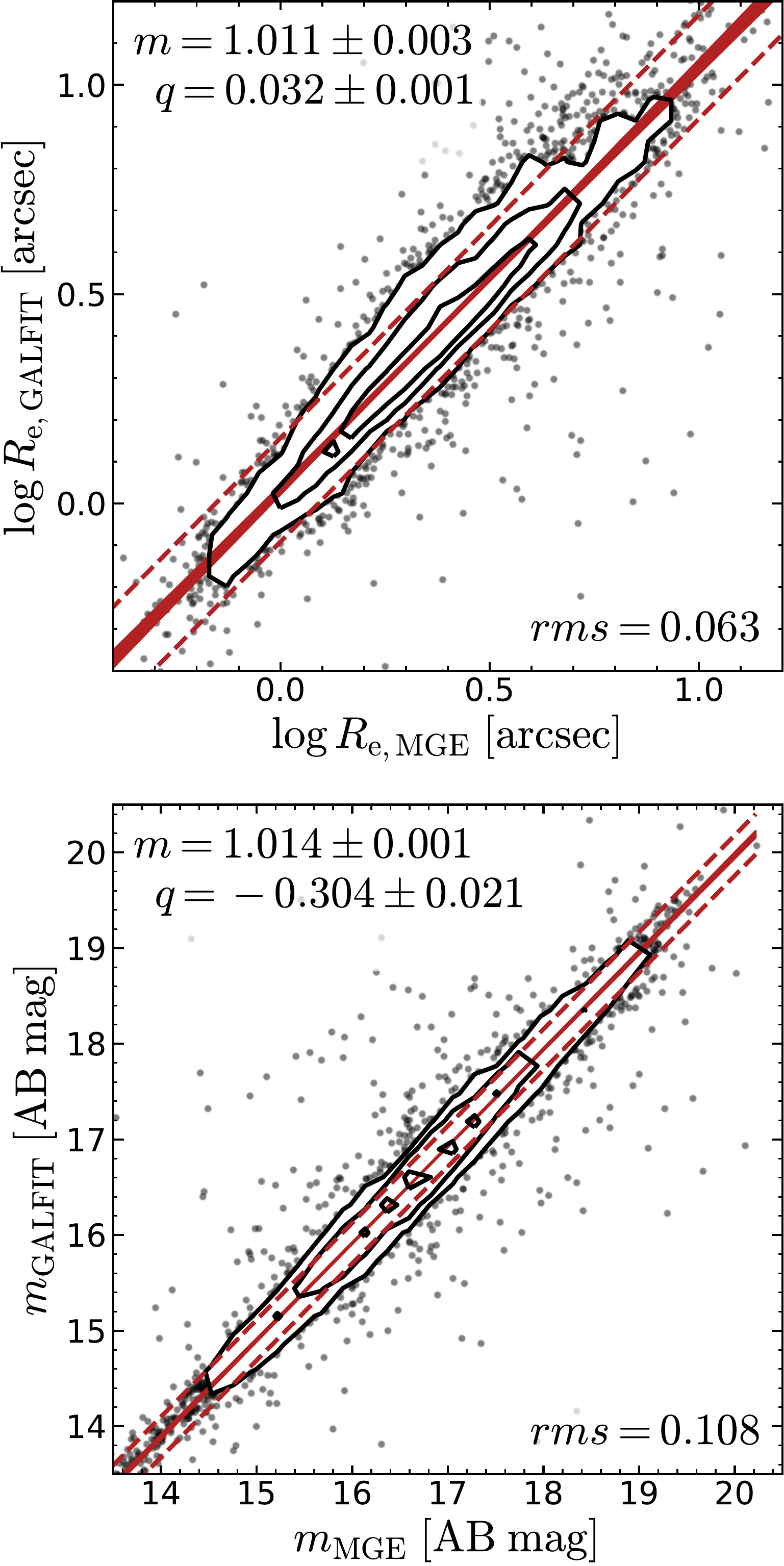}
  {\phantomsubcaption\label{f.ds.mge.vs.gama.a}
   \phantomsubcaption\label{f.ds.mge.vs.gama.b}}
  \caption{Comparison between our MGE measurements and the
  corresponding values from the GAMA survey, for both effective radius \re
  (panel~\subref{f.ds.mge.vs.gama.a}) and apparent magnitude $m$
  (panel~\subref{f.ds.mge.vs.gama.b}). The solid black contour
  lines enclose the 90\textsuperscript{th}, 75\textsuperscript{th} and
  50\textsuperscript{th} percentiles of the data, the shaded red region shows
  the 95\% confidence interval, and the dashed red lines show the 95\% prediction
  interval. We estimate the observational uncertainties on both \re and $m$
  using the observed root-mean square scatter about the best-fit relation (reported in
  the bottom right corner of each panel). The values $m$, $q$ and $x_0$ are the best-fit
  slope and zero-point of the linear relation
  (equations~(\ref{eq.ds.regama}) and (\ref{eq.ds.maggama}).
  }\label{f.ds.mge.vs.gama}
\end{figure}

We cross-match the SAMI parent sample with version 7 of the GAMA S{\'e}rsic
catalogue, finding 5496
galaxies in common. Of these, 211 have no size measurements in GAMA, 23 have no
MGE measurements and 26 have neither (the pipeline did not converge). The
effective overlap is therefore 5236
galaxies. To fit the GAMA sizes ($R_\mathrm{e,GALFIT}$) as a function of the
MGE sizes ($R_\mathrm{e,MGE}$) we use the same method
adopted to calibrate the VST and SDSS results (\S~\ref{s.ds.ss.phot.sss.calib}).
We set the data-point uncertainties to a negligible value of $10^{-5}$,
thereby assuming uniform observational uncertainties\footnote{
We find no evidence of a systematic dependence of the \re uncertainty
on either $m$ or \re itself. We model the distribution of \re as a bivariate
Gaussian, with standard deviation equal to $\sigma(m) = \alpha \, m + \beta$. The
best-fit values of $\alpha$ are always consistent with $0$, and entail a maximum
difference in the uncertainty $\sigma$ of $0.004$, negligible compared
to the average uncertainty of 0.045~dex.}.
The best-fit relation is
\begin{equation}\label{eq.ds.regama}
\log R_\mathrm{e,GALFIT} = (1.011 \pm 0.003) \log R_\mathrm{e,MGE}
                               + (0.032 \pm 0.001)
\end{equation}
with an observed root-mean square $rms = 0.063$. The data is shown
in \reffig{f.ds.mge.vs.gama.a}, where the meaning of the lines is the same as in
\reffig{f.ds.sdss.vs.vst}. We assume that the
uncertainties are equal between the {\sc galfit} and MGE measurements, so the
adopted uncertainty on \logre is equal to $\sigma_{\log R_\mathrm{e}} = rms/
\sqrt{2} \approx 0.045$~dex. This value is comparable to the uncertainties reported
by other authors \citepalias[e.g.][]{cappellari+2013a}. There is a systematic scaling
factor between the two measurements, as highlighted by the fact that the
best-fit linear coefficient of equation~(\ref{eq.ds.regama}) is not unity.

In order to understand the origin of this trend, we explore how the ratio
$R_\mathrm{e,MGE} / R_\mathrm{e,GALFIT}$ correlates with a number of
galaxy observables. The first three by statistical significance are stellar
mass, projected axis ratio $q$ and S{\'e}rsic index $n$. The most significant
correlation is between $R_\mathrm{e,MGE} / R_\mathrm{e,GALFIT}$ and $n$
(Spearman $\rho = -0.62$; the second most-significant correlation is with $q$
and has $\rho = -0.29$). 
We can remove the correlation between $R_\mathrm{e,MGE} /
R_\mathrm{e,GALFIT}$ and $n$ with an empirical correction based on the moving
median of $R_\mathrm{e,MGE} / R_\mathrm{e,GALFIT}$ as a function of $n$.
This correction also reduces the scatter about the best-fit relation
equation~(\ref{eq.ds.regama}) from $0.063$ to $0.048 \, \mathrm{dex}$, and further
removes the correlations with mass and shape (if we use a correction based on
$q$, the correlation with $n$ is also removed, but the scatter stays constant
at $0.063$).

Based on these tests, we believe that the most fundamental
correlation is with S{\'e}rsic index. The systematic trend between the two size
measurements is most likely due to the different
nature of the MGE and S{\'e}rsic fit: the first measures only detected light,
whereas the second attempts to extrapolate the total light based on the shape of
the detected profile \citep[in practice, in GAMA, the model is integrated only
within $R\leq5\,R_\mathrm{e}$,][]{kelvin+2012}. For a S{\'e}rsic profile, the fraction of
light at large radii increases with $n$, so the decreasing trend of
$R_\mathrm{e,MGE} / R_\mathrm{e,GALFIT}$ is qualitatively consistent with the fact
that, with increasing $n$, MGE misses more light and/or S{\'e}rsic models
overestimate missing light.

As for the scatter, it does not change between
the full overlap sample considered above and the subset of overlapping ETGs from
the FP sample (111 galaxies), suggesting that the systematic difference between
GAMA and SAMI photometry is not due to the inclusion of late-type galaxies (LTG) in the
comparison sample. Moreover, we find the same results between the full overlap
sample and the subset with $R_\mathrm{e} > 1.5 \, \mathrm{arcsec}$ (1111 galaxies),
suggesting the systematic slope is not caused by PSF modelling. This
hypothesis is confirmed by repeating the MGE measurements using the same PSF
reconstruction method as GAMA \citep[using the software
{\sc \href{https://www.astromatic.net/software/psfex}{PSFEx}},][]{
bertin2011}.

For magnitudes, we use the same procedure and find
\begin{equation}\label{eq.ds.maggama}
m_\mathrm{GALFIT} = (1.014 \pm 0.001) m_\mathrm{MGE} - (0.304 \pm 0.021)
\end{equation}
with an observed root-mean square scatter $rms = 0.108$ (\reffig{f.ds.mge.vs.gama.b}).
The uncertainty on $m$ is therefore $\sigma_m \approx 0.076$.

In summary, this comparison shows that circularised effective radii are
consistent between our MGE measurements and the values published in GAMA, but
there is a systematic factor such that $R_\mathrm{e,GALFIT}$ is $5-7\%$ larger
than $R_\mathrm{e,MGE}$ (the interval of the rescaling factor is the
16\textsuperscript{th}-84\textsuperscript{th} percentile of the distribution of
$R_\mathrm{e,MGE}$ for the FP sample, defined in \S~\ref{s.ds.ss.samp}).
Magnitudes, on the other hand, are consistent within $0.08 \, \mathrm{mag}$; the zero-point
$b = -0.304\pm0.021$ of the best-fit relation in equation~(\ref{eq.ds.maggama})
does not imply an equally large offset in magnitude zero-point, because the
best-fit slope is larger than unity. Over the magnitude range of the FP
sample, $m_\mathrm{GALFIT}$ is 0.09--0.06 $\mathrm{mag}$ brighter than
$m_\mathrm{MGE}$. For magnitudes, if we swap the measurements from
\citet{kelvin+2012} with the SDSS `model' values from SDSS Data Release 14
\citep{abolfathi+2018} or with the GAMA photometry from
{\sc \href{http://cutout.icrar.org/LAMBDAR.php}{lambdar}} \citep{wright+2016},
we find comparable (but better) agreement with our MGE measurements.

As a final remark, the adopted value of the sigma-clipping threshold
in {\sc lts\_linefit} (parameter {\sc clip}) does affect the derived uncertainties. For example,
by using a value of 4 the resulting uncertainties on \re are 30\% larger.
We tested the effect of adopting 50\% smaller or larger uncertainties on our FP
fit, and, although the best-fit FP parameters depend on the value of the
uncertainty, the key results of this work are qualitatively unchanged
(see \S~\ref{s.r.ss.galev.sss.corr} and Table~\ref{t.r.bestfit}, rows~11--15).

\subsubsection{Surface mass density}\label{s.ds.ss.phot.sss.sig}

We estimate the surface mass density in two elliptical apertures: an aperture of
fixed physical size, having circularised radius 1~kpc, and a galaxy-dependent
aperture having circularised radius 1~\re; in practice, the apertures we use are
the model isophotes with area closest to $\pi \mathrm{kpc}^2$ and
$\pi R_\mathrm{e}^2$, respectively. We measure $g$- and $i$-band magnitudes
inside these apertures, then estimate the enclosed stellar masses from
absolute $i-$band magnitude and $g-i$ colour \citep{taylor+2011}.
In practice, we implement the k-correction by defining
\begin{equation}\label{eq.mphot.bryant}
\begin{split}
  \log M_\star(<R)/\mathrm{M_\odot} \equiv & - 0.4 \; M_i(<R) - \log \, (1 + z)\\
       + & (1.2117 - 0.5893 z) \\
       + & (0.7106 - 0.1467z) (g -i)(<R)
\end{split}
\end{equation}
where $M_i(<R)$ is the enclosed absolute magnitude prior to the k-correction
\citep{bryant+2015}. We then define
\begin{equation}
\begin{split}
    \Sigma_\star(R<1\mathrm{kpc}) & \equiv M_\star(R<1\mathrm{kpc}) / (\pi \; 1 \, \mathrm{kpc}^2)\\
    \Sigma_\star(R<R_\mathrm{e}) & \equiv  M_\star(R<R_\mathrm{e}) / (\pi R_\mathrm{e}^2)
\end{split}
\end{equation}

\subsection{SAMI spectroscopy}\label{s.ds.ss.spec}

The main SAMI data consists of integral field spectra taken with the SAMI
instrument at the prime focus of the $3.9 \,\mathrm{m}$ Anglo-Australian
Telescope. The SAMI instrument uses 13 integral field units (IFUs), deployable
anywhere within a 1~degree diameter field of view. Each IFU is a fused fibre
bundle \citep[hexabundle;][]{bland-hawthorn+2011, bryant+2014}, containing 61
fibres of 1.6~arcsec diameter, for a total IFU diameter of 15~arcsec; the distinctive advantage of the SAMI instrument is that the
hexabundles have a higher fill factor than conventional fibre bundles
\citep[0.75 instead of $\sim$0.5,][]{croom+2012}.

The IFUs, as well as 26 single fibres used for sky measurements, are plugged into pre-drilled plates
using magnetic connectors. The fibres are fed to the double-beam AAOmega
spectrograph, which allows a range of different resolutions and wavelength
ranges \citep{sharp+2006}. For SAMI we use the 570V grating at 3750--5750~\AA\ (blue
arm) and the R1000 grating at 6300--7400~\AA\ (red arm). The spectral resolutions
for the blue and red arms are respectively $R$=1812 ($\sigma=70.3 \, \mathrm{km\;s^{-1}}$) and $R$=4263 ($\sigma=29.9 \, \mathrm{km\;s^{-1}}$); the reference wavelengths are $\lambda_\mathrm{blue}$ = 4800~\AA\ and $\lambda_\mathrm{red}$ = 6850~\AA\ \citep[][their Table~1]{vandesande+2017a}.
Each galaxy was exposed for approximately 3.5~hours, following a hexagonal
dither pattern of seven equal-length integrations \citep{sharp+2015}. The median
full-width-at-half-maximum seeing was $2.06 \pm 0.40$~arcsec. The
basic data reduction process is described in \citet{sharp+2015} and
\citet{allen+2015}; the data quality is illustrated in the public data release
papers, alongside a number of improvements in the data reduction
\citep{green+2018, scott+2018, croom+2021}.

\subsubsection{Spectroscopic measurements}\label{s.ds.ss.spec.sss.sigma}

\begin{figure*}
  \centering
  \includegraphics[type=pdf,ext=.pdf,read=.pdf,width=\textwidth]{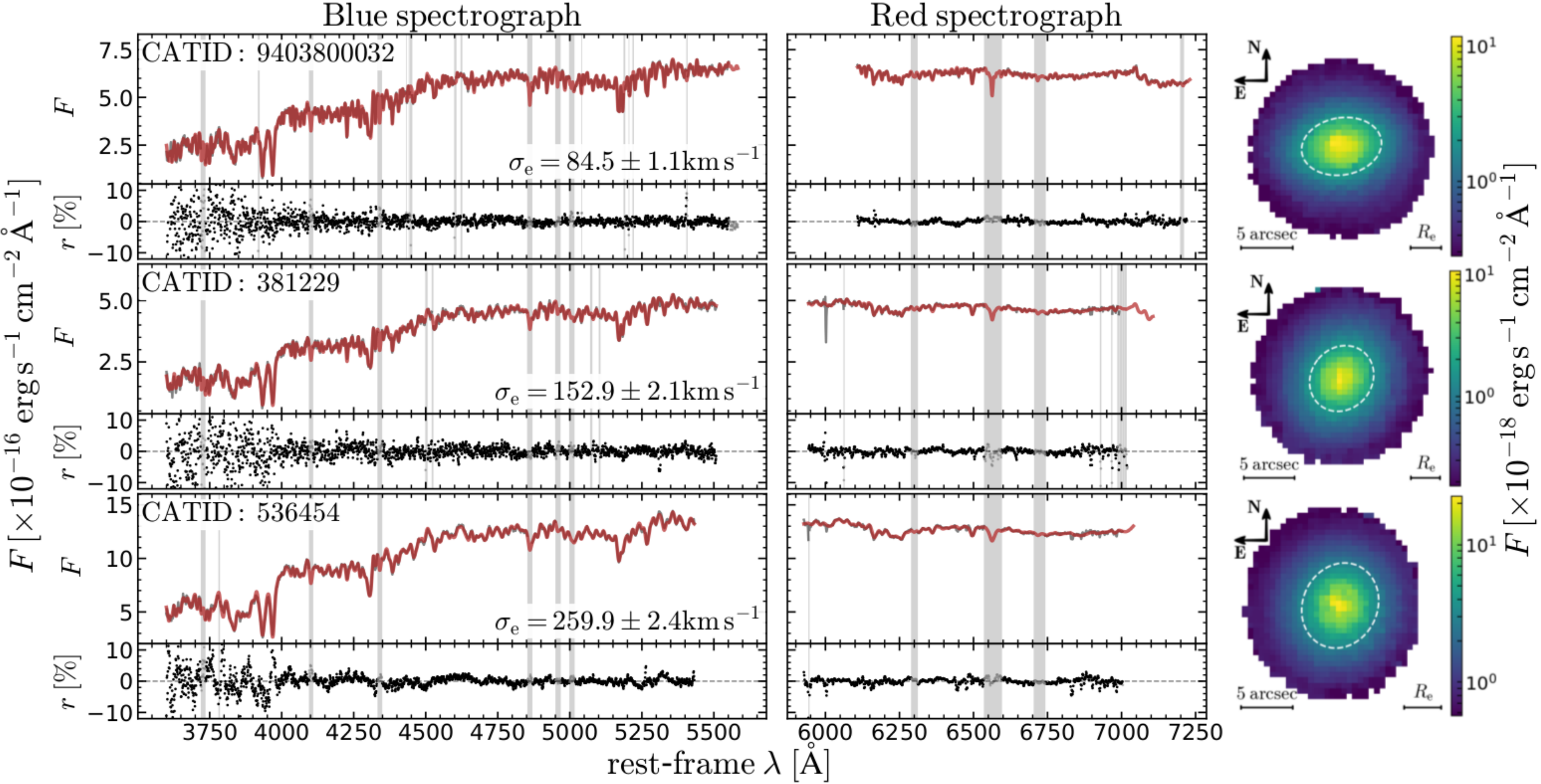}
  \caption{
  Three example {\sc pPXF} fits. From top to bottom, the three galaxies
  have \sige closest to the 5\textsuperscript{th}, 50\textsuperscript{th} and
  95\textsuperscript{th} percentiles of the \sige distribution. The grey lines
  show the SAMI blue-arm (left column) and red-arm spectra (central column); the
  red lines are the best-fit spectra. These aperture spectra are unweighted sums of
  the flux inside the MGE model isophote of area $\pi \, \mathrm{R_e}^2$ (right
  column). Gray vertical regions in spectra were not fit, either
  because they may contain possible (weak) emission lines or because they were
  sigma-clipped. Below each fit, we show the relative residuals. The right column
  shows the SAMI view of the three galaxies, with superimposed the elliptical
  aperture of area $\pi\,R_\mathrm{e}^2$.
  }\label{f.ds.ppxf}
\end{figure*}

For each galaxy, we constructed a synthetic elliptical aperture of equivalent
radius 1~$\mathrm{R_e}$, i.e.\ the $r-$band elliptical isophote with area $A = \pi \,
\mathrm{R_e^2}$ (see \S~\ref{s.ds.ss.phot.sss.mge}). This aperture size, like
all adaptive apertures, is problematic for both large galaxies (where the
aperture is larger than the IFU) and small galaxies (where the aperture is
smaller than the SAMI PSF); we discuss this issue in relation to the sample selection
(\S~\ref{s.ds.ss.samp}) as well as in relation to the FP determination
(\S~\ref{s.r.ss.galev.sss.sample}).
We derive the aperture spectra as a weighted sum of the IFU
spectra, where the spectrum in each spatial pixel (spaxel) is weighted by the
fraction of its area falling inside the ellipse. In order to mimic as closely
as possible the behaviour of an elliptical aperture, we do not use statistical
weights in the sum; because the SAMI spectra are flux calibrated \citep{green+2018};
applying, e.g., inverse-variance weighting would create spectra that
are more weighted towards the central spaxels than in a large, physical aperture.
Three example apertures are illustrated by the white dashed ellipses in the
right column of \reffig{f.ds.ppxf}.
Because SAMI has different spectral resolutions in the blue and red arm, we
convolve the red spectrum to the same spectral resolution as the blue spectrum
\citep[as described in][]{vandesande+2017a}. The resulting spectrum covers the
full spectral range of SAMI, with a $\sim$5500\,\AA gap between 5750\,\AA and
6300\,\AA.

We used these spectra to measure aperture kinematics. We obtained
the second moment of the velocity distribution (\sige) using the penalised pixel
fitting algorithm {\sc \href{https://pypi.org/project/ppxf/}{pPXF}}
\citep{cappellari+emsellem2004, cappellari2017} and the MILES stellar template library
\citep{sanchez-blazquez+2006, falcon-barroso+2011}. The process is identical
to that used to measure the kinematics in each spaxel \citep{vandesande+2017a}
and to measure the DR3 aperture kinematics \citep{croom+2021}.
In brief, we fit a Gaussian line-of-sight velocity distribution as well as
an additive 12\textsuperscript{th} order Legendre polynomial. The fit is iterated
three times. In the first iteration, we estimate (if necessary) a scaling for
the noise spectrum. In the second
iteration we reject bad pixels using an iterative sigma-clipping algorithm,
(the {\sc clean} keyword in {\sc pPXF}). The third iteration yields the
measurement of \sige. We estimate the uncertainties using a Monte Carlo approach:
we create an ensemble of \sige measurements from one hundred random-noise
realisations of the best-fit spectrum, and define the uncertainty to be the
standard deviation of the ensemble. The random noise was obtained by shuffling
the noise in 15 equal-width spectral intervals.
We show three example fits in
\reffig{f.ds.ppxf}: these galaxies were chosen to have \sige closest to the
5\textsuperscript{th}, 50\textsuperscript{th} and 95\textsuperscript{th}
percentiles of the \sige distribution for our final sample
(\S~\ref{s.ds.ss.samp}).
Compared to DR3 aperture kinematics, the values used in this work present
two differences: first, our apertures use regularised MGE fits (whereas DR3
uses unregularised fits). Second, our apertures are elliptical, whereas DR3
apertures are circular. Nevertheless, we find excellent agreement between the
two measurements
\begin{equation}
        \log \, \sigma_\mathrm{DR3} = (0.975 \pm 0.003) \log \sigma_\mathrm{e}
                                 + (0.054 \pm 0.007)
\end{equation}
with a scatter of 0.04~dex. Although \sige is systematically larger than
$\sigma_\mathrm{DR3}$, the difference is small (1\% at $\sigma_\mathrm{e} = 200
\, \mathrm{km\,s^{-1}}$) and can be intuitively explained as follows. Compared to
the circular aperture of the same area, an elliptical aperture along the galaxy
major axis includes more spaxels along the major axis and fewer spaxels along
the minor axis.
Most ($\sim$85\%) ETGs are `fast rotators' \citep{krajnovic+2011,
emsellem+2011}, i.e. galaxies with considerable rotation support \citep[the
exact definition of fast rotator varies, but they are identified as galaxies
that are intrinsically flat $\epsilon>0.4$ and/or having $(V/\sigma)_\mathrm{e} \gtrsim 0.1$;][]{
emsellem+2011, cappellari2016, vandesande+2017b, vandesande+2021}
Therefore, for most galaxies in our ETG sample, major-axis spaxels have
larger line-of-sight velocity offsets than minor-axis spaxels, and contribute
more to the aperture dispersion. Measurement uncertainties are estimated by
comparing repeat observations. We use a uniform value of 0.022~dex for all
galaxies.

The stellar-population parameters age, \ZH and \aFe were determined as part
of the SAMI survey. \citet{scott+2017} measured a set of twenty Lick
absorption indices \citep{worthey+1994, trager+1998} and used
simple-stellar-population (SSP) models to convert the empirical index values to
luminosity-weighted SSP-equivalent stellar-population parameters
\citep{schiavon2007, thomas+2010};
here \ZH is the
(logarithmic) metal-to-hydrogen mass fraction relative to the solar value
\begin{equation}
    \mathrm{[Z/H]} \equiv \log \mathrm{Z/X} - \log (\mathrm{Z/X})_\odot
\end{equation}
where Z and X are the metal and hydrogen mass fractions, respectively.
For an in-depth
discussion of the SAMI stellar-population parameters, see \citet{scott+2017}.
Here we stress that for any given galaxy, each SSP parameter captures only one
aspect (the light-weighted mean) of what is in fact a distribution of stellar
population properties.
We then infer $r-$band stellar mass-to-light
ratios from the best-fit age and metallicity measurements, interpolating on the
models of \citet{maraston2005} and using the same grid as \citet{scott+2017}.
Our mass-to-light ratios (\ups) assume a Chabrier IMF \citep{chabrier2003},
whereas the models of \citet{maraston2005} assume a Kroupa IMF
\citep{kroupa2001}. We therefore divided the Kroupa mass-to-light ratios
by a factor of $1.12$ to obtain \ups \citep{speagle+2014}.

\subsection{Correlated noise}\label{s.ds.ss.corr}

Apart from the measurement uncertainties on \logsige, \logre and \logl, the
covariance matrix also contains three off-diagonal elements. The largest of these
(by absolute value) is the covariance between the measurement uncertainties on
\logl and \logre. We estimate this entry by comparing the SDSS luminosity and
size measurements to the corresponding VST measurements, in all the three bands.
In order to obtain the measurement errors, as well as to remove the physical
correlation between size and luminosity
(Pearson correlation coefficient $\rho \approx 0.6$), we subtract from each set
of size and luminosity measurements, the median over all six measurements (three
bands for SDSS photometry and three bands for VST photometry). Furthermore, to
eliminate the effect of systematic changes between the three bands, we subtract
the median from each set of measurements in any given photometry and band.
After rejecting outliers (defined as lying outside the contour enclosing the
95\textsuperscript{th} percentile of the data) we find that the correlation
coefficient is $\rho = 0.235$ (the covariance is $2.8\times 10^{-4}$). Using a
stricter rejection threshold yields even lower correlation ($\rho = 0.202$ if we
reject data outside the 80\textsuperscript{th} percentile).
If we repeat the test to measure the correlation between effective radius and
the average surface brightness within \re, we find $\rho = -0.770$, a number
considerably smaller (in absolute value) than reported in the
literature \citepalias[$\rho =-0.95$;][]{magoulas+2012}. It is reasonable to
assume that the difference
is due to the flexibility of MGE photometry compared to the rigid functional
form of S\'{e}rsic photometry.
As a validation test, we repeat the MGE fits after rotating all galaxy images
in steps of 5 degrees, thus building a set of size and luminosity measurements
for each galaxy. With this method, we derive a correlation coefficient $\rho =
0.650$. Although this is much larger than our measured value of $\rho =
0.235$, these measurements are not independent and so a
larger correlation coefficient is expected. 

The second largest entry in the covariance matrix is the covariance between the
measurement uncertainties on \logsige and \logre. This correlation
arises because we use \re to create the aperture spectra (from which we measure
\sige) and because the second moment of the velocity distribution depends on
the radius of the aperture inside which it is measured \citep{jorgensen+1996}. We
estimate this covariance as $\alpha \,
\sigma^2_{\log R_\mathrm{e}} = -0.00015$ \citep[following][we assumed $\sigma_\mathrm{e} \propto
R_\mathrm{e}^{-0.066}$]{cappellari+2006}. Finally, we neglected the last component of the
covariance matrix and assumed that measurement uncertainties in \logsige and
\logl are uncorrelated.

\subsection{Ancillary data}\label{s.ds.ss.ancillary}

For each galaxy, we use $g-$,~$r-$ and $i-$band photometry, derived either from
SDSS (if available) or from VST \citep[for the clusters APMCC0917, EDCC442,
Abell~3880 and Abell~4038; ][]{owers+2017}.

Magnitudes in the $i-$band and $g-i$ colours were measured consistently across SAMI
\citep{owers+2017}. Apparent magnitudes were converted to absolute magnitudes
$M_i$ using the luminosity distance in the adopted cosmology.
The input redshifts for the luminosity distance are
individual spectroscopic redshifts from GAMA for the field and group
galaxies \citep{bryant+2015} or cluster redshifts for the cluster
galaxies \citep{owers+2017}.

We use photometric stellar masses ($M_\star$) from the SAMI catalogue
\citep{bryant+2015, owers+2017}. These masses (and mass densities) were derived
by combining absolute $i-$band magnitude and $g-i$ colour
\citetext{\citealp{taylor+2011}; \S~\ref{s.ds.ss.phot.sss.sig}}.

When available, we use $r-$band S\'{e}rsic indices $n$ \citetext{from
\citealp{kelvin+2012} or \citealp{owers+2019} for GAMA and cluster galaxies,
respectively}. To estimate the uncertainties on $n$, we compare $r-$ and
$i-$band indices for ETG galaxies, finding a scatter of 0.03~dex
or 6\% (after dividing by $\sqrt{2}$). This value is a compromise between
low- to intermediate-index galaxies, where the scatter is smaller, and large-index
galaxies ($n \gtrsim 5$), where it is larger.

Optical morphologies were assigned by twelve SAMI team members, using RGB
cutouts and following the classification scheme adopted by GAMA
\citep{kelvin+2014a}. Firstly, galaxies are divided into late- and early-types by
the presence or absence of spiral arms. Late-types are subdivided into early-
and late-spirals by the presence or absence of a bulge, whereas early-types are subdivided
into lenticulars and ellipticals by the presence or absence of a stellar disc. Whenever
the image quality was deemed insufficient, or no consensus ($>67\%$) was reached between
the classifications, galaxies were classified as uncertain. The SAMI
morphological classification was presented in \citet{cortese+2016}: ellipticals
have mtype=0, lenticulars (S0) have mtype=1 and intermediate
types have mtype=0.5. In this paper, we define early-type galaxies (ETGs) as having
mtype$\leq1$; this conservative definition ensures minimal contamination
from late-type galaxies (LTGs), even though adding galaxies with mtype=1.5 does not
change our results.

In addition, we use $(V/\sigma)_\mathrm{e}$ ratios measured on the SAMI kinematic maps.
These values have been corrected to an aperture of 1\,\re using a statistical
correction based on galaxies with sufficient radial coverage
\citep{vandesande+2017b}. Numerical
simulations have shown that $(V/\sigma)_\mathrm{e}$ is in good agreement with more
sophisticated measures of the relative importance of streaming and random motions
\citep[e.g.][]{thob+2019}.

\subsection{Sample selection}\label{s.ds.ss.samp}

\begin{figure}
  \centering
  \includegraphics[type=pdf,ext=.pdf,read=.pdf,width=1.0\columnwidth]{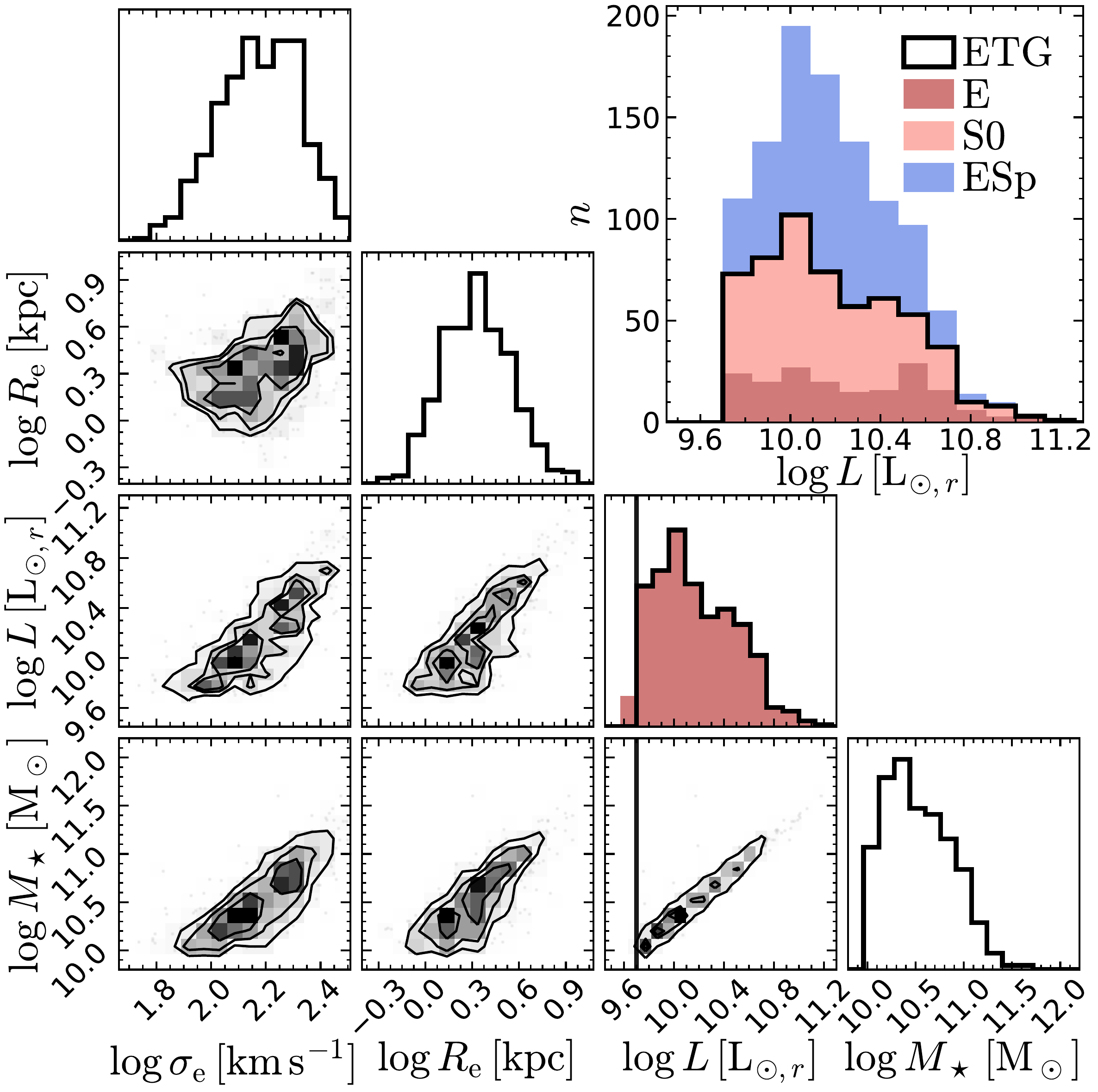}
  \caption{The properties of our volume- and mass-limited ETG sample.
  Notice the non-Gaussian nature of the \logl distribution (filled red
  histogram); the cut in \logl (solid black vertical line) enables us to
  remove the tail of $L<10^{9.7}\,\mathrm{L}_{\odot,r}$ galaxies and to
  use a simpler probability model for the FP (\S~\ref{s.da.ss.3dg});
  alternative optimisation methods also assume (implicitly) some
  distribution.
  The top-right histogram shows the relative contribution of
  ellipticals (E; red) and lenticulars (S0; salmon) to our sample of ETGs (solid
  line). Blue bars show the luminosity distribution of early spirals
  (ESp; notice the bars are stacked).
  }\label{f.ds.sample}
\end{figure}

We use a volume-limited subset of the SAMI
sample consisting of 1461 galaxies with $M_\star \geq 10^{10} \, \mathrm{M_\odot}$
and $z \leq 0.065$ \citep[cf.\ ][their Fig.~4; for cluster members
we used the redshift of the cluster instead of the redshifts of individual
galaxies; the mass cut effectively removes only three galaxies, which do not
affect our analysis]{bryant+2015}. From this initial
selection, 1460 have optical
morphologies, and we further select 642 ETGs by requiring $0 \leq \mathrm{mtype}
\leq 1$ (we prioritised sample purity over sample size, see
\S~\ref{s.ds.ss.ancillary}). Note that if we limited the selection to
elliptical galaxies only ($\mathrm{mtype}=0$) the results presented here would
be qualitatively unchanged, but the smaller sample size (216 galaxies) means
that the significance of some results would be lower
(top right histogram in \reffig{f.ds.sample}).
We further introduce censoring in \logl by requiring $L \geq 10^{9.7} \,
\mathrm{L}_{\odot,r}$; this cut removes the asymmetric tail of 26 low-luminosity
galaxies in Fig.~\ref{f.ds.sample} and allows an accurate yet relatively
simple probabilistic model. From this sample of 616 galaxies, we further remove
24 galaxies that are too large relative to the SAMI IFU and so have no
measurement of \sige. Even though removing these 24 galaxies introduces a
bias against large galaxies, their inclusion does not affect the results of the
subsequent analysis (\S~\ref{s.r.ss.galev.sss.sample}).
After visual inspection of the aperture spectra, we
also remove 31 galaxies with Balmer emission lines, thus bringing the sample
down to 561 galaxies. 

In general an accurate determination
of the FP requires removing low-quality measurements, e.g.\ galaxies with large
measurement uncertainties in
\sige or with \sige below the instrument spectral resolution.
However, this additional quality selection is not necessary, because the mass and
redshift cuts already remove low-quality data. Comparing the S\'{e}rsic and
MGE photometry, we find just 14 galaxies that have measurement discrepancies larger
than three standard deviations.
We chose not to remove these galaxies either, but use instead the robust fitting
algorithm to identify the most likely outliers. In fact, even
though some of these 14 galaxies have neighbours that could have affected the
photometry, others
appear regular galaxies where either the S\'{e}rsic or MGE fit performed poorly.
Moreover, that approach does not reject unreliable measurements where systematic
errors bias both the S\'{e}rsic and MGE fits in the same direction; we manually remove
9016800416, as it is too close to the edge of the image frame to measure \re, even
though there is good agreement between the two fitting methods.

We are left with a final sample of 560 ETGs. Within the mass and volume limits of our
selection, this sample has the same completeness as the original SAMI ETG
sample. The FP sample is not
representative of the SAMI survey volume, however, because cluster galaxies are
over-represented \citep{vandesande+2021}.

\section{Data Analysis}\label{s.da}

The value of the best-fit FP parameters depends on the model adopted to
describe the data. For this reason, the model choice is of central importance
in comparing and interpreting the results. In this section we describe the
FP model adopted throughout this paper (\S~\ref{s.da.ss.3dg}) and motivate
our choice using a comparison to an alternative optimisation approach
popular in the literature (\S~\ref{s.da.ss.comp} and \S~\ref{s.da.ss.outl}).
We finally illustrate the expected residual correlations for our model
(\S~\ref{s.da.ss.rescorr}).

\subsection{3dG: a 3-d Gaussian algorithm}\label{s.da.ss.3dg}

We adopt a Bayesian approach, and model the FP as a three-dimensional (3-d)
Gaussian with a cut in \logl (censoring).
This choice is motivated by a visual inspection of the distribution of the SAMI
observables (histograms of \logsige, \logre and \logl in \reffig{f.ds.sample}),
as well as from previous experience fitting the 6dFGS FP \citepalias{magoulas+2012}.
Even though the assumption of a Gaussian distribution might appear crude, it is far more accurate
than the assumption of a uniform distribution that implicitly underlies algorithms minimising the
distance to the plane. Censoring (equation~\ref{eq.ds.cens} below) is introduced to
account for truncation of the distribution in \logl, which is a consequence of our sample selection criteria
(see the vertical line in \reffig{f.ds.sample}).

The most general 3-d Gaussian has nine parameters: three for the centroid and
six for the symmetric covariance matrix. Another parameter is needed to
implement censoring, so the model has ten parameters in total.

In order to preserve a straightforward geometric interpretation, we factorise
the generic covariance matrix into a diagonal and an orthogonal factor:
the former ($\bm{\Sigma}$) represents the uncorrelated covariance of the
FP in the eigenvector basis, while the latter ($\bm{R}$) represents the rotation transforming the observable basis to the eigenvector basis.

To write down this model in a concise manner, we first introduce some
definitions: $\bm{x}\equiv (\log \sigma_\mathrm{e}, \log R_\mathrm{e}, \log L$) are the
coordinates in the observable space whereas
$\bm{v}$ are the coordinates in the reference frame of the 3-d Gaussian. The
most generic multivariate Gaussian is given by
\begin{equation}
    \mathcal{N}_{\bm{m},\bm{A}}(\bm{x}) \equiv 
    \frac{1}{\sqrt{\det{\left(2\pi \bm{A}\right)}}} \exp \left\{-\frac{1}{2}(\bm{x}-\bm{m})^T 
    \bm{A}^{-1} (\bm{x}-\bm{m})\right\}
\end{equation}
where $\bm{A}$ is the covariance matrix and $\bm{m}$ is the centroid. In the
reference frame of $\bm{v}$, the model probability is given by
\begin{equation}
    p(\bm{v}) = \mathcal{N}_{\bm{0},\bm{\Sigma}}(\bm{v}) 
\end{equation}
where $\bm{\Sigma}$ is the diagonal correlation matrix. We further assume that
the $\bm{v}$ coordinates are sorted according to the scatter of the
3-d Gaussian (from largest to smallest) and that the $\bm{v}$ reference frame
has the same chirality as the $\bm{x}$ reference frame. With these assumptions,
there exists a rotation $R$ by an angle $\theta$ about an axis $\bm{\hat{u}}$, such
that $\bm{v} = R(\bm{x})$.
In the base of $\bm{x}$, $R$ can be represented by an orthogonal matrix $\bm{R}$,
such that $\forall \bm{x}$, $R(\bm{x}) = \bm{R \, x}$.
If we call the mean of the observables $\bm{\mu}$, we can write the probability
in the $\bm{x}$ reference frame as
\begin{equation}\label{eq.da.multinorm}
p(\bm{x}|\mathrm{model}) = \mathcal{N}_{\bm{\mu}, \bm{R}^T \bm{\Sigma} \bm{R}}(\bm{x})
\end{equation}
which expresses the probability of the observables given the model. So far, the
model has nine free parameters: three Gaussian centroids ($\bm{\mu}$), three
standard deviations (the diagonal matrix $\bm{\Sigma}$) and three parameters to
identify the rotation $R$. It is useful to recall that, because
$\bm{R}$ is orthogonal, the inverse of the covariance matrix $\bm{R}^T \bm{\Sigma}
\bm{R}$ is equal to $\bm{R}^T \bm{\Sigma}^{-1} \bm{R}$ and
$\det{\left(2\pi \bm{R}^T \bm{\Sigma} \bm{R}\right)}$ is
equal to $\det{\left(2\pi \bm{\Sigma}\right)}$.

We then assume Gaussian noise for each of the $N$ measurements $\bm{x}_i$, with
covariance matrix $\bm{E}_i$, so that given a true value of $\bm{x}$, a
measurement $\bm{x}_i$ has probability
\begin{equation}
    p(\bm{x}_i|\bm{x}) = \mathcal{N}_{\bm{x}, \bm{E}_i}(\bm{x}_i)
\end{equation}
and the probability of $\bm{x}_i$ given the model is given by the definite
integral
\begin{equation}
\begin{split}
    p(\bm{x}_i|\mathrm{model}) & = \iiint d\bm{x} \, p(\bm{x_i}|\bm{x}) \, p(\bm{x}|\mathrm{model})\\
    & = \mathcal{N}_{\bm{\mu}, \bm{R}^T \bm{\Sigma} \bm{R} + \bm{E}_i}(\bm{x}_i)
\end{split}
\end{equation}
where we have integrated the convolution of two Gaussians by completing the
squares. We introduce a cut in \logl, so that the new probability
$p(\bm{x}_i|\mathrm{model})$ is given by
\begin{equation}\label{eq.ds.cens}
    p(\bm{x}_i|\mathrm{model}) =
\begin{dcases}
    f_i(L_\mathrm{min}) \mathcal{N}_{\bm{\mu}, \bm{R}^T \bm{\Sigma} \bm{R} + \bm{E}_i}(\bm{x}_i),& \text{if } L \geq L_\mathrm{min}\\
    0,              & \text{otherwise}
\end{dcases}
\end{equation}
where for each $i$, $f_i(L_\mathrm{min})$ is a constant such that
\begin{equation}\label{eq.app.fi}
    \int_{-\infty}^{\infty} \hspace{-1.0em} dx_1
    \int_{-\infty}^{\infty} \hspace{-1.0em} dx_2
    \int_{\log L_\mathrm{min}}^{\infty} \hspace{-1em} dx_3 f_i(L_\mathrm{min}) \mathcal{N}_{\bm{\mu}, \bm{R}^T \bm{\Sigma} \bm{R} + \bm{E}_i}(\bm{x}) = 1
\end{equation}
An expression for $f_i(L_\mathrm{min})$ is given in {Appendix~\ref{a.s.fi}}.

Using Bayes' theorem, the probability of the model given all the data is then
\begin{equation}\label{eq.ds.pmod}
    p(\mathrm{model}|\mathrm{data}) = \dfrac{\prod_{i=1}^N p(\bm{x}_i|\mathrm{model}) p(\mathrm{model})}{p(\mathrm{data})}
\end{equation}
where $p(\mathrm{model})$ is the prior. Unfortunately, integrating this
expression is complicated by the cut in \logl, so that, even assuming conjugated
priors, the evidence would \textit{not} be Gaussian-inverse-Wishart
\citep[e.g.][]{degroot1970}. For this reason, we assume flat, uninformative
priors for all the parameters and, to determine their value, we integrate
equation~(\ref{eq.ds.pmod}) with {\sc \href{https://pypi.org/project/emcee/}{emcee}}
\citep{foreman-mackey+2013}, an implementation of the Markov Chain Monte Carlo
algorithm \citep[MCMC,][]{metropolis+1953} proposed by \citep{goodman+weare2010}.
The implementation of the 3-d Gaussian model (3dG) is a generalisation
of the algorithm already presented in \citetalias{barone+2018} and
\citet{barone+2020}, with the addition of censoring
in \logl. To verify that the method is correct, we also use a nested sampler
and find the same results as with the MCMC integrator \citep[we used the python
package {\sc \href{https://pypi.org/project/dynesty/}{dynesty}},][]{
skilling2004, skilling2006, feroz+2009, higson+2019, speagle2020}.

\subsection{Model bias}\label{s.da.ss.comp}

\begin{figure}
  \centering
  \includegraphics[type=pdf,ext=.pdf,read=.pdf,width=0.46\textwidth]{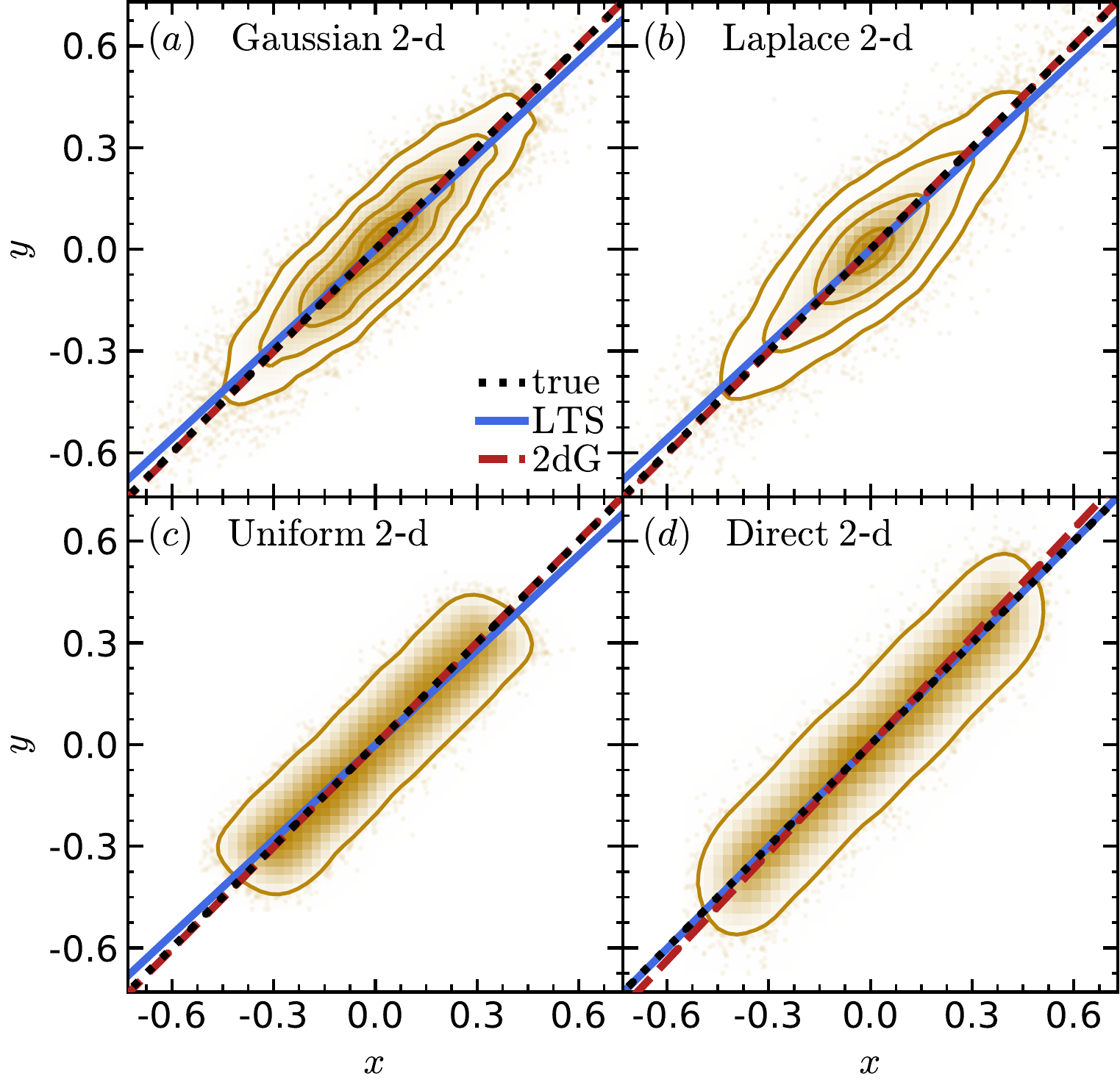}
  {\phantomsubcaption\label{f.ds.2dg.a}
   \phantomsubcaption\label{f.ds.2dg.b}
   \phantomsubcaption\label{f.ds.2dg.c}
   \phantomsubcaption\label{f.ds.2dg.d}}
  \caption{
  Two-dimensional mock datasets illustrating the bias inherent to
  different models. The panels compare the results from the 
  {\sc lts\_linefit} (solid blue line) and 2dG (dashed red line) algorithms to the
  true model $y=x$ (dotted black line). The algorithms have been tested with four
  different generative models: a 2-d Gaussian (panel~\subref{f.ds.2dg.a}),
  a 2-d Laplace distribution (panel~\subref{f.ds.2dg.b}), a uniform distribution with intrinsic orthogonal scatter
  (panel~\subref{f.ds.2dg.c}) and a uniform
  distribution with intrinsic scatter along the $y$ axis (panel~\subref{f.ds.2dg.d}). Each point also includes
  Gaussian measurement uncertainties. For panels~\subref{f.ds.2dg.a} and
  \subref{f.ds.2dg.b}, the contours trace the 12\textsuperscript{th},
  40\textsuperscript{th}, 68\textsuperscript{th} and 86\textsuperscript{th}
  percentiles of the data (roughly corresponding to $0.5$, $1$, $1.5$ and $2$
  $\sigma$); for panels~\subref{f.ds.2dg.c} and \subref{f.ds.2dg.d}, the single
  contour line traces the 90\textsuperscript{th} percentile of the data.
  For the models with orthogonal scatter
  (panels~\subref{f.ds.2dg.a}-\subref{f.ds.2dg.c}), {\sc lts\_linefit} infers a
  systematically smaller $rms$, at the price of flatter slope $m$ (see
  Table~\ref{f.ds.2dg}). The opposite is true for the model with scatter
  along $y$ (panel~\subref{f.ds.2dg.d}): here it is 2dG that infers a
  systematically steeper slope.
  This model bias greatly affects the strength of the correlation
  between the fit residuals $\Delta$ and the variables $x$ and $y$ (see
  Table~\ref{t.ds.2dg}).
  }\label{f.ds.2dg}
\end{figure}

In order to assess our model, we use the well-established
{\sc \href{https://pypi.org/project/ltsfit/}{lts\_planefit}}
algorithm as a comparison. {\sc lts\_planefit} \citepalias{cappellari+2013a},
is a 3-d extension of the {\sc lts\_linefit} algorithm which we have briefly
introduced in \S~\ref{s.ds.ss.phot.sss.remag}. The core of {\sc lts\_planefit}
is a least-squares minimisation, where the quantity being minimised is the sum
of the squares of the distance of each data point to the plane. The distance
is calculated along the $z$ axis
\citetext{\citetalias{cappellari+2013a}, equation~7 and \citealp{press+2007},
section~15.3}.
Because the quantity minimised by these methods is directly related to the
observed $rms$, the best-fit {\sc lts\_planefit} models have, by construction, lower
$rms$ compared to the best-fit 3dG models. Notice, however, that lower $rms$ is \emph{not}
necessarily
an indication of a more likely model: as we will see (\S~\ref{s.da.ss.rescorr},
Table~\ref{t.ds.2dg} and Fig.~\ref{f.ds.2dg}),
depending on the underlying model, the price for this lower $rms$ may be a bias
on the value of the inferred model parameters.
The probability distribution underlying direct-fit methods like
{\sc lts\_planefit} assumes an infinite plane with Gaussian scatter along the
$z$ axis
\begin{equation}\label{eq.ds.inforth}
    p(x,y,z) \propto \frac{1}{\sqrt{2\pi \sigma_\mathrm{int}^2}} \exp \left\{-\frac{(z - a \, x - b \, y - c)^2}{2 \sigma_\mathrm{int}^2} \right\}
\end{equation}
where, in order for the probability to be integrable, some truncation is
required in both $x$ and $y$. The limits of this truncation determine the missing
multiplicative factor.
In equation~(\ref{eq.ds.inforth}) the probability distribution of the
independent variables is not constrained.
Formally, this model can be thought of as the limit for an infinitely extended
plane \citetext{see Appendix~\ref{a.s.lq} and \citealp{kelly2007}}, which clearly
does not apply to our sample. In
fact, by inspecting \reffig{f.ds.sample}, we can see that the range of the FP
observables is larger, but not much larger, than the FP thickness (the
narrowest histogram has a FWHM of 0.36~dex, roughly four times larger than the
typical FP scatter of $\approx$0.05--0.1~dex).
This is a critical aspect of our model choice. In fact, for an infinitely extended
underlying distribution, the direction of scatter does not matter, as the normal
and axial scatters are just projections of each other. However, because our data
is not infinitely extended and uniform (Fig.~\ref{f.ds.sample}), then the varying
underlying distribution combined with the direction of the scatter modifies the
observed distribution (e.g. consider the ends of a \emph{finite} uniform
distribution with normal and axial scatter, as in panels~\subref{f.ds.2dg.c} and
\subref{f.ds.2dg.d} of Fig.~\ref{f.ds.2dg}). In the
realm of finite distributions, we find that the precise shape of the 1-d
distribution of the observables does not appear to bias the FP coefficients (as
long as the distribution stays symmetric, Fig.~\ref{f.ds.2dg.a}-\subref{f.ds.2dg.c}).
What matters most for these finite distributions is the direction of the
intrinsic scatter, whether it is orthogonal to the FP (as assumed by our
algorithm) or along the $z$ axis (as assumed by direct-fit methods).
To quantify the effect of model mismatch, we conduct two sets of tests.
In the first set, we test the 3dG and {\sc lts\_planefit} algorithms on a mock
dataset based on the 3-d Gaussian generative model. For the second set, we use
mock datasets based on a probability distribution close to the generative model
of {\sc lts\_planefit}.

For the first set of tests, we create a range of mock datasets based on the
fiducial SAMI FP (\S~\ref{s.r}) and attempt to recover the input FP parameters
with {\sc lts\_planefit} and with 3dG. For a 3-d Gaussian generative model, we
generate one hundred mock samples from the multivariate Gaussian with covariance
matrix
\begin{equation}
    \bm{R}^T \bm{\Sigma} \bm{R} + \bm{E} = 
    \begin{bmatrix}
        0.0316 & 0.0263 & 0.0614 \\
        0.0263 & 0.0704 & 0.0946 \\
        0.0614 & 0.0946 & 0.1686 \\
    \end{bmatrix}
\end{equation}
obtained from the fiducial FP (\S~\ref{s.r.ss.galev}, this matrix includes both
the intrinsic covariance matrix and homoscedastic measurement uncertainties).
Each mock
consists of 560 triplets $(\log \sigma_\mathrm{e}, \log R_\mathrm{e}, \log L)$,
where the number 560 is chosen to match the size of the FP sample (see
\S~\ref{s.ds.ss.samp}). We run the fit one hundred times, and determine the
bias on each model parameter as the mean of the inferred values; the
significance is given by the uncertainty on the mean. For the
{\sc lts\_planefit} algorithm, the bias is $-8\%$, $+5\%$ and $+3\%$ in $a$,
$b$ and $c$ (all offsets are statistically significant). If we introduce
censoring in \logl, the bias is $-12\%$, $-2\%$ and $+5\%$, consistent with the
findings of \citetalias{magoulas+2012}.
In contrast, the same tests show that the 3dG algorithm recovers $a$, $b$ and
$c$ without measurable bias. For the 3-d Gaussian generative model, the bias in
$a$, $b$ and $c$ is $+0.5\%$, $+0.02\%$ and $-0.2\%$, respectively; for the
censored 3-d Gaussian, the bias is $-0.06\%$, $+0.1\%$ and $+0.1\%$. None of
these values are statistically significant.

For the second set of tests, we aim to assess the bias of the 3dG algorithm
when fitting mock data drawn from the model underlying direct-fit methods.
However, because this probability distribution is not integrable
(equation~\ref{eq.ds.inforth}), and because SAMI data is not uniformly
distributed in \logsige and \logre, we adopt the following approach. First, we
take the median and standard deviation of the measured \logsige and \logre;
we correct the standard deviations for observational uncertainties by
subtracting in quadrature the median measurement uncertainty in \logsige and
\logre. We then sample 560 values of \logsige and \logre from the Gaussian
distributions with mean equal to the observed mean and standard deviation
equal to the corrected standard deviation. \logl is obtained drawing randomly
from the Gaussian distribution with mean $a \, \log \sigma_\mathrm{e} + b \,
\log R_\mathrm{e} + c$ and standard deviation $\sigma_{\log \, L} =
0.082$~dex.
Notice this Gaussian scatter is added along the direction of \logl only. We
further add uniform Gaussian scatter to the mock values of \logsige, \logre
and \logl, to represent (uncorrelated) observational uncertainties. The process
is repeated one hundred times, to create one hundred independent realisations
of the mock sample. The values of the FP coefficients $a$, $b$ and $c$ and the 
FP intrinsic scatter $\sigma_{\log \, L}$ are taken from fitting
{\sc lts\_planefit} to the SAMI FP data (Appendix~\ref{a.s.lts}).
As expected, for this test {\sc lts\_planefit} recovers the input parameters
with no detectable bias (the bias in $a$, $b$ and $c$ is $-0.2\%$, $+0.8\%$
and $+0.3\%$, respectively; none of these is statistically significant). In
contrast, for 3dG the bias in $a$, $b$ and $c$ is $+11$\%, $+5$\% and $-4$\%,
with significance of 3, 2 and 4 standard deviations, respectively.

These tests show that each of the two algorithms suffers from considerable
bias when fitting data taken from a different model (i.e. bias results from
using an incorrect model). For this reason, the
choice of algorithm is dictated by other considerations. We prefer the 3dG
algorithm because a Gaussian model appears to be closer to the
distribution of the real data - at least for the SAMI sample
(Fig.~\ref{f.ds.sample}). Whether the scatter should be orthogonal to the
plane or along \logl could be explored with a dedicated model, but we defer
this analysis to future work. In this work, we address this model uncertainty
by marginalising over it, i.e. repeating the analysis with the
{\sc lts\_planefit} algorithm (Appendix~\ref{a.s.lts}) and showing that we
find a different FP, but the same qualitative results as the 3dG method.

\subsection{Outlier rejection}\label{s.da.ss.outl}

An additional source of bias is represented by outliers.
{\sc lts\_planefit} is designed to deal with outliers through
the eponymous
inside-out sigma clipping \citetext{\citealp{rousseeuw+driessen2006},
\citetalias{cappellari+2013a}}.
The number of standard deviations
beyond which the sample is clipped is set by the value of the keyword
{\sc clip}. The optimal value of {\sc clip} depends on
the probability distributions of both the sample and of the outliers, which are
not known \textit{a priori}. If we assume Gaussian scatter and no outliers
({\sc clip}=0), given our sample size we expect two
valid galaxies to lie beyond three standard deviations from the best-fit plane
and no galaxy to lie beyond four standard deviations; in fact we find eight and
two respectively, and so need to apply sigma-clipping. As the best-fit
parameters are statistically consistent between these two choices, we use 
the more conservative value {\sc clip=3}.

For the 3dG algorithm, we proceed as follows. Even
though outliers can in principle be modelled, in practice this is not required
because of our good data quality. In fact, we find the same FP by either 
rejecting the points outside of the contours enclosing the 95\textsuperscript{th}
percentile of the data \citepalias[see e.g.][]{barone+2018} or by using the 
posterior probability to reject the 1\textsuperscript{st} percentile of the data
and repeating the optimisation on the pruned sample \citepalias{magoulas+2012}.
These two post-optimisation methods find largely the same outliers as the
robust {\sc lts\_planefit} with {\sc clip=3} (\S~\ref{s.r.ss.galev.sss.outl}). Even though
outliers do not affect the FP parameters $a$, $b$ and $c$ (see
Table~\ref{t.r.bestfit}, rows 3--4, columns 3--5) the strictness of the outlier
rejection threshold does directly affect the observed $rms$ and intrinsic scatter
(columns 6--8).

To test whether our results depend on the model or algorithm used, we
repeated the FP analysis using the
{\sc lts\_planefit} algorithm {(Appendix~\ref{a.s.lts})}. Even though the FP
parameters differ between the two algorithms (Table~\ref{t.r.bestfit}, rows 1
and 18--19), the main results of this paper are the same for the 3dG and
{\sc lts\_planefit} algorithms, in that the ranking by significance of the
residual correlations is the same.

\subsection{Residual correlations}\label{s.da.ss.rescorr}

\begin{table}
  \begin{center}
  \setlength{\tabcolsep}{3pt}
  \caption{
  True and inferred model parameters for 2-dimensional mock
  datasets, illustrating the bias inherent to different models and algorithms
  (see also Fig.~\ref{f.ds.2dg}).
  For models with orthogonal intrinsic scatter, direct-fit least-squares minimisation
  methods (column 5), obtain consistently flatter slopes and smaller $rms$
  compared to both 2dG (column 4) and to the input model value (column 3).
  Conversely, for models with intrinsic scatter along $y$ (labelled `Direct 2-d'),
  direct-fit methods recover the input parameters with no bias whereas 2dG
  infers systematically steeper slope $m$.
  The bias in the slope $m$ propagates to the correlation between
  the residuals $\Delta$ and the data (measured here using both covariance
  $\mathrm{Cov}$ and the Spearman rank correlation coefficient $\rho$). The
  bias on the residual correlations is both large (50--100\%) and statistically significant
  ($>10 \sigma$).
  }\label{t.ds.2dg}
  \begin{tabular}{ccccc}
  \hline
     model & & $\mathrm{true}$ & $\mathrm{2dG}$ & $\mathrm{LTS}$ \\
       \hspace{-4mm} (1) & \hspace{-4mm} (2) & (3) & (4) & (5) \\
 \hline
 \parbox[t]{6mm}{\multirow{6}{*}{\rotatebox[origin=c]{90}{Gaussian 2-d}}}  & \hspace{-4mm} $m$                                           & $1$                   & $\hphantom{-}1.003\pm0.005$ & $0.930\pm0.004$ \\
                                                                           & \hspace{-4mm} $rms$                                         & $0.102$               & $\hphantom{-}0.103\pm0.001$ & $0.100$  \\
                                                                           & \hspace{-4mm} $10^{3} \! \times \! \mathrm{Cov}(\Delta, x)$ & $-6.01$               & $            -6.17\pm0.26$  & $-2.42$  \\
                                                                           & \hspace{-4mm} $10^{3} \! \times \! \mathrm{Cov}(\Delta, y)$ & $\hphantom{-}4.49$    & $ \hphantom{-}4.34\pm0.23$  & $\hphantom{-}7.66$   \\
                                                                           & \hspace{-4mm} $\rho(\Delta, x)$                             & $-0.247$              & $           -0.253\pm0.010$ & $-0.102$ \\
                                                                           & \hspace{-4mm} $\rho(\Delta, y)$                             & $\hphantom{-}0.191$   & $\hphantom{-}0.185\pm0.010$ & $\hphantom{-}0.334$  \\
 \hline                                                                                                                                                                           
 \parbox[t]{6mm}{\multirow{6}{*}{\rotatebox[origin=c]{90}{Laplace 2-d}}}   & \hspace{-4mm} $m$                                           & $1$                   & $\hphantom{-}1.005\pm0.005$ & $0.932\pm0.004$ \\
                                                                           & \hspace{-4mm} $rms$                                         & $0.103$               & $\hphantom{-}0.103\pm0.001$ & $0.100$  \\
                                                                           & \hspace{-4mm} $10^{3} \! \times \! \mathrm{Cov}(\Delta, x)$ & $-5.99$               &             $-6.25\pm0.26$  & $-2.42$  \\
                                                                           & \hspace{-4mm} $10^{3} \! \times \! \mathrm{Cov}(\Delta, y)$ & $\hphantom{-}4.65$    &  $\hphantom{-}4.41\pm0.23$  & $\hphantom{-}7.80$   \\
                                                                           & \hspace{-4mm} $\rho(\Delta, x)$                             & $-0.285$              &            $-0.295\pm0.010$ & $-0.143$ \\
                                                                           & \hspace{-4mm} $\rho(\Delta, y)$                             & $\hphantom{-}0.209$   & $\hphantom{-}0.198\pm0.010$ & $\hphantom{-}0.349$  \\
 \hline                                                                                                                                                                           
 \parbox[t]{6mm}{\multirow{6}{*}{\rotatebox[origin=c]{90}{Uniform 2-d}}}   & \hspace{-4mm} $m$                                           & $1$                   & $\hphantom{-}1.005\pm0.005$ & $0.933\pm0.004$ \\
                                                                           & \hspace{-4mm} $rms$                                         & $0.102$               & $\hphantom{-}0.102\pm0.001$ & $0.099$  \\
                                                                           & \hspace{-4mm} $10^{3} \! \times \! \mathrm{Cov}(\Delta, x)$ & $-5.82$               &             $-6.10\pm0.25$  & $-2.43$  \\
                                                                           & \hspace{-4mm} $10^{3} \! \times \! \mathrm{Cov}(\Delta, y)$ & $\hphantom{-}4.49$    & $ \hphantom{-}4.24\pm0.22$  & $\hphantom{-}7.49$   \\
                                                                           & \hspace{-4mm} $\rho(\Delta, x)$                             & $           -0.239$   &            $-0.250\pm0.010$ & $-0.101$ \\
                                                                           & \hspace{-4mm} $\rho(\Delta, y)$                             & $\hphantom{-}0.187$   & $\hphantom{-}0.176\pm0.010$ & $\hphantom{-}0.322$  \\
 \hline                                                                                                                                                                           
 \parbox[t]{6mm}{\multirow{6}{*}{\rotatebox[origin=c]{90}{Direct 2-d}}}    & \hspace{-4mm} $m$                                           & $1$                   & $\hphantom{-}1.062\pm0.004$ & $1.003\pm0.004$ \\
                                                                           & \hspace{-4mm} $rms$                                         & $0.104$               & $\hphantom{-}0.107\pm0.001$ & $0.104$  \\
                                                                           & \hspace{-4mm} $10^{3} \! \times \! \mathrm{Cov}(\Delta, x)$ & $-2.41$               &             $-6.59\pm0.29$  & $-2.61$  \\
                                                                           & \hspace{-4mm} $10^{3} \! \times \! \mathrm{Cov}(\Delta, y)$ & $\hphantom{-}8.41$    & $ \hphantom{-}4.39\pm0.28$  & $\hphantom{-}8.23$   \\
                                                                           & \hspace{-4mm} $\rho(\Delta, x)$                             & $           -0.084$   &            $-0.229\pm0.010$ & $-0.091$ \\
                                                                           & \hspace{-4mm} $\rho(\Delta, y)$                             & $\hphantom{-}0.273$   & $\hphantom{-}0.128\pm0.010$ & $\hphantom{-}0.266$  \\

  \end{tabular}
  \end{center}Columns: (1)~probabilistic model used to generate the data; (2)~parameter; (3)~input value of parameter; (4)~parameter values from the 2dG; (5)~parameter value from
  {\sc lts\_linefit} algorithm. The mock data and models are illustrated in Fig.~\ref{f.ds.2dg}
\end{table}

The main scientific goal of this work is to compare various structural and
stellar-population parameters to the FP scatter, by studying how these
observables correlate with the FP residuals: the presence and strength of
correlations \textit{may} point to underlying physical trends. For this reason,
we discuss here the expected residual correlations between the plane-fit
residuals and the data.
In \S~\ref{s.r} we quantify the strength and significance of residual trends
using the Spearman rank correlation coefficient $\rho$, because it does not
penalise non-linear correlations. However, $\rho$ is difficult to relate
analytically to the assumed probability distribution; for this reason, in this
section we focus on covariance, which is directly related to the \emph{Pearson}
correlation coefficient by
\begin{equation}
    \rho_\mathrm{Pearson} \equiv \dfrac{\mathrm{Cov}(x, y)}{
        \displaystyle\sqrt{{\mathrm{Cov}(x, x) \mathrm{Cov}(y, y)}}
    }
\end{equation}
As we shall see, it is in the residual trends with respect to the FP observables
that the 3dG and {\sc lts\_planefit} algorithms differ the most.
This difference is due to the different probability distribution between the two
models. For the infinite plane model, the residuals with respect to the
\textit{true} solution correlate with $z$, but not with $x$ and $y$.
This can readily be ascertained by calculating the covariance explicitly. We
define the residuals as
\begin{equation}
    \Delta \equiv z - a \, x - b \, y - c
\end{equation}
and we integrate the moments of the probability model underlying
{\sc lts\_planefit}, equation~(\ref{eq.ds.inforth}), obtaining
\begin{equation}
  \begin{split}
     \mathrm{Cov}(\Delta,x) & = 0\\
     \mathrm{Cov}(\Delta,y) & = 0\\
     \mathrm{Cov}(\Delta,z) & = \mathrm{Cov}(z,z) \hspace{-0.15em} - a \mathrm{Cov}(x,z) \hspace{-0.15em} - b \mathrm{Cov}(y,z) \hspace{-0.1em} = \sigma_\mathrm{int}^2
  \end{split}
\end{equation}
where we have assumed that the integrals
\begin{equation}
    \int dx \, x, \int dx \, x^2, \int dy \, y, \int dy \, y^2
\end{equation}
are all finite (i.e.\ that the probabilities of $x$ and $y$ are uniform over a
finite interval that is `large' compared to the thickness of the plane,
$\sigma_\mathrm{int}$). For the uncensored 3dG model, we consider the
probability distribution equation~(\ref{eq.da.multinorm}) and find
\begin{equation}\label{eq.ds.3dg.cov}
  \begin{split}
     \mathrm{Cov}(\Delta,x) & = \left[\hspace{-0.13em}\bm{R}^T  \hspace{-0.1em}\bm{\Sigma} \bm{R}\hspace{-0.09em}\right]_{xz}\hspace{-0.5em} - a  \left[\hspace{-0.13em}\bm{R}^T  \hspace{-0.1em}\bm{\Sigma} \bm{R}\hspace{-0.09em}\right]_{xx}\hspace{-0.5em} - b  \left[\hspace{-0.13em}\bm{R}^T  \hspace{-0.1em}\bm{\Sigma} \bm{R}\hspace{-0.09em}\right]_{xy}\\
     \mathrm{Cov}(\Delta,y) & = \left[\hspace{-0.13em}\bm{R}^T  \hspace{-0.1em}\bm{\Sigma} \bm{R}\hspace{-0.09em}\right]_{yz}\hspace{-0.5em} - a  \left[\hspace{-0.13em}\bm{R}^T  \hspace{-0.1em}\bm{\Sigma} \bm{R}\hspace{-0.09em}\right]_{xy}\hspace{-0.5em} - b  \left[\hspace{-0.13em}\bm{R}^T  \hspace{-0.1em}\bm{\Sigma} \bm{R}\hspace{-0.09em}\right]_{yy}\\
     \mathrm{Cov}(\Delta,z) & = \left[\hspace{-0.13em}\bm{R}^T  \hspace{-0.1em}\bm{\Sigma} \bm{R}\hspace{-0.09em}\right]_{zz}\hspace{-0.5em} - a  \left[\hspace{-0.13em}\bm{R}^T  \hspace{-0.1em}\bm{\Sigma} \bm{R}\hspace{-0.09em}\right]_{xz}\hspace{-0.5em} - b  \left[\hspace{-0.13em}\bm{R}^T  \hspace{-0.1em}\bm{\Sigma} \bm{R}\hspace{-0.09em}\right]_{yz}
  \end{split}
\end{equation}
none of which is, in general, zero (censoring further complicates
these expressions). Direct fit methods such as {\sc lts\_planefit} tend to
bias the optimisation towards the assumed model, artificially lowering
$\vert\mathrm{Cov}(\Delta,x)\vert$ and $\vert\mathrm{Cov}(\Delta,x)\vert$ and
increasing $\vert\mathrm{Cov}(\Delta,z)\vert$. The conclusion is that, whenever residual correlations
are important, direct-fit methods are not a good solution.

To better illustrate the implications, we use a two-dimensional
example. In Fig.~\ref{f.ds.2dg} we present three mock datasets, consisting of
10,000 points generated from a probability distribution with mean $\mathbf{0}$
and symmetry axes $x=y$ and $x=-y$. Along these axes, the standard deviations
were chosen to match the values of the fiducial FP (Fig.~\ref{f.r.fpbench}).
The generative probability distributions are a 2-d Gaussian
(panel~\subref{f.ds.2dg.a}), a 2-d Laplacian distribution
(panel~\subref{f.ds.2dg.b}), a uniform distribution with orthogonal scatter
(labelled `Uniform 2-d', panel~\subref{f.ds.2dg.c}), and a uniform distribution
with scatter along the $y$ axis (labelled `Direct 2-d', panel~\subref{f.ds.2dg.d}).
Each point was convolved with a 2-d Gaussian reproducing the measurement
uncertainties on the FP observables. The dotted black line
has equation $y=x$ and represents the true model, with slope $m=1$ and
zero-point $q=0$. The dashed red and solid blue lines are the models inferred
with the 2dG and {\sc lts\_linefit} algorithms,
respectively.
For the models with orthogonal scatter (panels~\subref{f.ds.2dg.a}-\subref{f.ds.2dg.c}),
the LTS algorithm has consistently shallower slope $m$ and lower $rms$
(Table~\ref{t.ds.2dg}, column 5) compared to both the 2dG model (column 4) and,
revealingly,  even the true value (column 3).
Within the realm of plausible FP models, the bias does not depend on the
functional form of the true model, because it is identical between the three
mock datasets. Conversely, the 2dG algorithm consistently finds the true solution
and the true $rms$ (Table~\ref{t.ds.2dg}, column 4).
Crucially, the bias of the LTS solution propagates to the correlations between
the residuals $\Delta \equiv y - (m x + q)$ and the variables $x$ and $y$. As
predicted for the 3-d case, the correlation with the independent variable decreases
(in absolute value) and the correlation with the dependent variable increases.
Unsurprisingly, this bias becomes more/less severe with decreasing/increasing
aspect ratio of the data: the more the data tends to the infinite line,
the more accurate direct-fit solutions become.
In contrast, the 2dG algorithm recovers the underlying model and the correct
residual correlations even for `thick' (high-scatter) correlations and for
non-Gaussian scatter (Table~\ref{t.ds.2dg}, column~4).
On the other hand, for the model with scatter along the $y$ axis, it is the
LTS algorithm that recovers the true model parameters and residual
correlations, with 3dG finding steeper slope $m$ and stronger (weaker)
correlations between the residuals $\Delta$ and x (y).
The fact that the 3dG and {\sc lts\_planefit} algorithms have different
correlations between the FP
residuals and the FP variables $L$, \re and \sige (cf. \S~\ref{s.r.ss.nnl}),
means that our results depend on the model (and hence method) that we adopt.
This undesirable dependency
also extends to the correlations between the FP residuals and the other
structural and stellar-population observables, because these observables
correlate in turn with the FP variables (e.g.\ $age$ correlates with $L$ and
\sige).
For the reasons discussed above (\S~\ref{s.da.ss.comp}), we adopt the 3-d
Gaussian as the model of choice and the 3dG algorithm as the default method.
Therefore, in the following, we adopt the 3dG FP as the fiducial FP.
Nevertheless, it is important to test our results against other
model/algorithms, with {\sc lts\_planefit} representing the most
different alternative.
We find that for most of the residual correlations (including those
with the highest significance) there is no qualitative difference between the
results of the 3dG and {\sc lts\_planefit} algorithms. Consequently, the
interpretation of our results does not, in this case, depend on the adopted model.

\section{Results}\label{s.r}

We present the fiducial FP for SAMI (\S~\ref{s.r.ss.galev}), including its
dependence on a number of assumptions
(\S\S~\ref{s.r.ss.galev.sss.sample}--\ref{s.r.ss.galev.sss.model}).
We then study the residuals of the FP and compare them: (i)~to the FP variables
(\S~\ref{s.r.ss.nnl}); (ii)~to a set of structural parameters (\S~\ref{s.r.ss.str});
and (iii)~to a set of stellar-population properties
(\S~\ref{s.r.ss.ssp}). The most significant correlation is with
stellar-population age. We show that this
correlation arises from the large scatter in the relation between
stellar mass-to-light ratio and surface mass density at any given position on
the FP (\S~\ref{s.r.ss.ml}).
We then use mock data to isolate the effect of stellar-population trends with
velocity dispersion and size on the tilt and scatter of the FP
(\S~\ref{s.r.ss.mock}).

\subsection{The fiducial SAMI Fundamental Plane}\label{s.r.ss.galev}

\begin{figure}
  \includegraphics[type=pdf,ext=.pdf,read=.pdf,width=1.0\columnwidth]{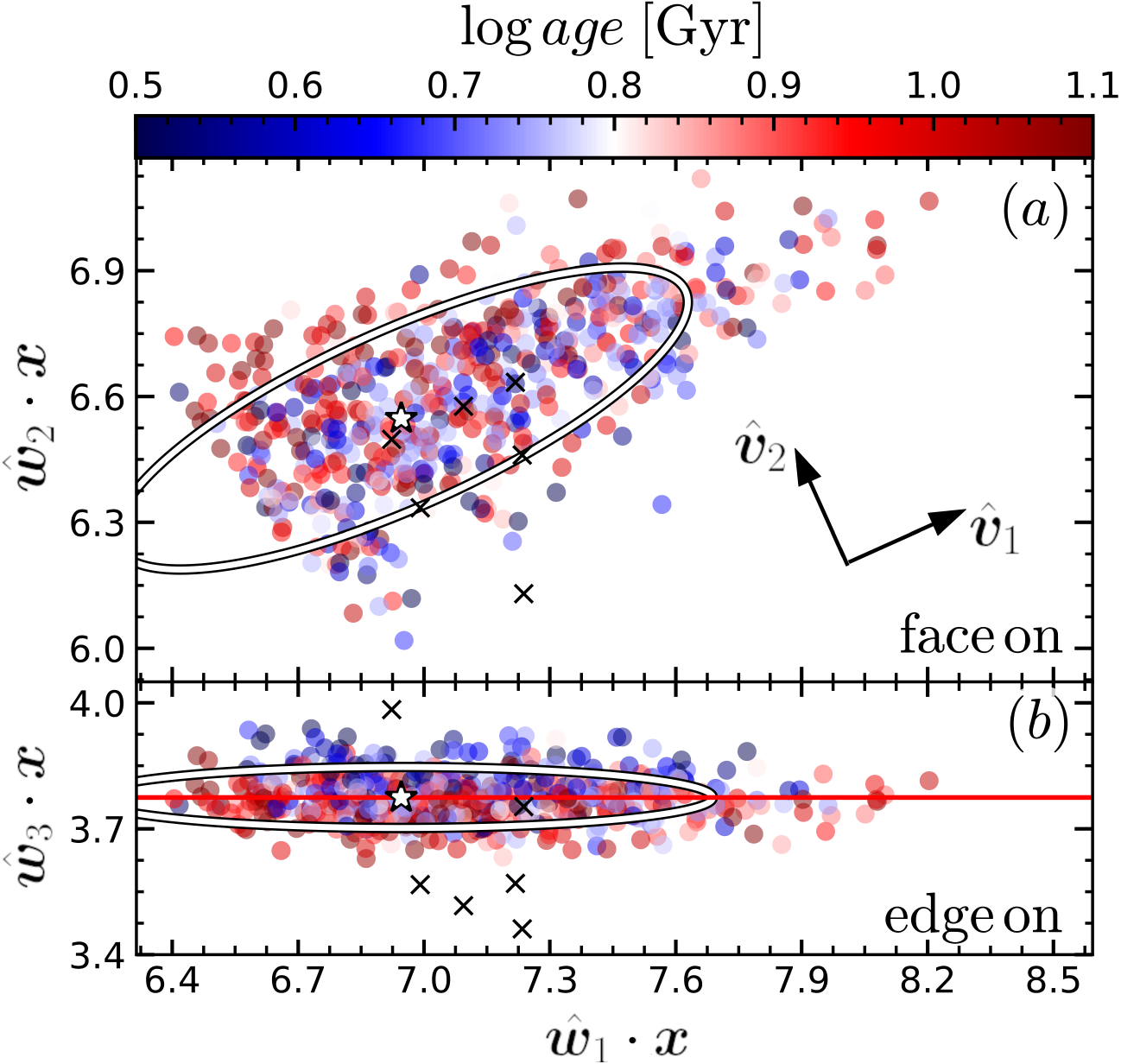}
  {\phantomsubcaption\label{f.r.fpproj.a}
   \phantomsubcaption\label{f.r.fpproj.b}}
  \caption{
  The fiducial FP for the SAMI ETGs, shown face on
  (panel~\subref{f.r.fpproj.a}) and edge on (panel~\subref{f.r.fpproj.b}).
  Each circle represents a SAMI galaxy, colour-coded by light-weighted SSP age
  (the six crosses are outlier galaxies). For ETGs, there is a clear age trend
  across the FP. The white ellipses centred on the
  white star show the projection of the fiducial model. The solid red line in
  panel~\subref{f.r.fpproj.b} represents the edge-on FP. The axes of the
  ellipse are equal to three times the \emph{intrinsic} dispersion of the
  3-d Gaussian.
  The adopted vector base $\{ \hat{\bm{w}}_1, \hat{\bm{w}}_2, \hat{\bm{w}}_3 \}$
  is expressed in terms of $a$ and $b$ only (equation~\ref{eq.r.eigen}); this
  base is closely related to the eigenvectors of the 3-d Gaussian model
  $\{ \hat{\bm{v}}_1, \hat{\bm{v}}_2, \hat{\bm{v}}_3 \}$:
  $\hat{\bm{w}}_3 \equiv \hat{\bm{v}}_3$, whilst $\hat{\bm{v}}_1$ and
  $\hat{\bm{v}}_2$ are illustrated by the arrows in panel~\subref{f.r.fpproj.a}
  (downscaled by a factor three for illustration purposes). The transformation
  matrix between the  $\hat{\bm{v}}_i$'s and
  $\hat{\bm{w}}_i$'s is reported in Table~\ref{t.r.eigen}.
  }\label{f.r.fpproj}
\end{figure}

\begin{figure}
  \includegraphics[type=pdf,ext=.pdf,read=.pdf,width=1.0\columnwidth]{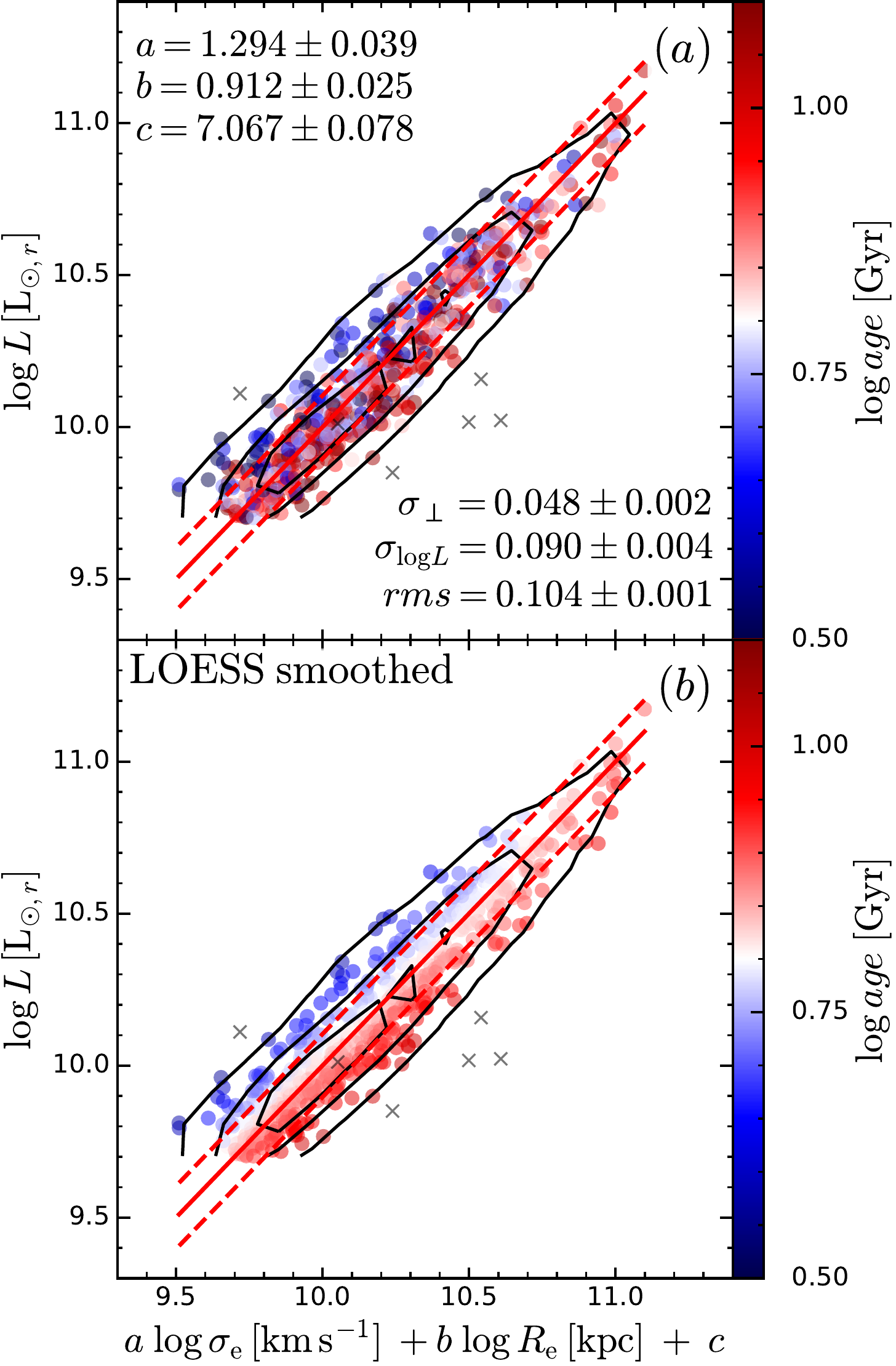}
  {\phantomsubcaption\label{f.r.fpbench.a}
   \phantomsubcaption\label{f.r.fpbench.b}}
  \caption{The fiducial FP for the SAMI ETGs, showing a clear age gradient
  across the plane. Each circle represents a SAMI galaxy, colour-coded by
  SSP age (panel~\subref{f.r.fpbench.a}) or by LOESS-smoothed SSP age
  (panel~\subref{f.r.fpbench.b}). Black crosses mark galaxies excluded
  from the fit; after an initial fit, they lie in the 1\textsuperscript{st}
  percentile of the posterior probability distribution (the cross
  that lies near the plane is excluded because it
  lies far from the galaxy locus within the plane). The best-fit FP
  is traced by
  the solid red line, whereas the dashed red lines encompass $\pm rms$.
  The solid black contours enclose the 40\textsuperscript{th},
  68\textsuperscript{th}, and 96\textsuperscript{th} percentiles of the data
  distribution. 
  There is a clear age gradient across the FP: at fixed \sige and \re, old
  galaxies (red hues) are under-luminous and lie preferentially below the
  best-fit plane, and conversely for young galaxies (blue hues).
  }\label{f.r.fpbench}
\end{figure}

\begin{figure}
  \includegraphics[type=pdf,ext=.pdf,read=.pdf,width=1.0\columnwidth]{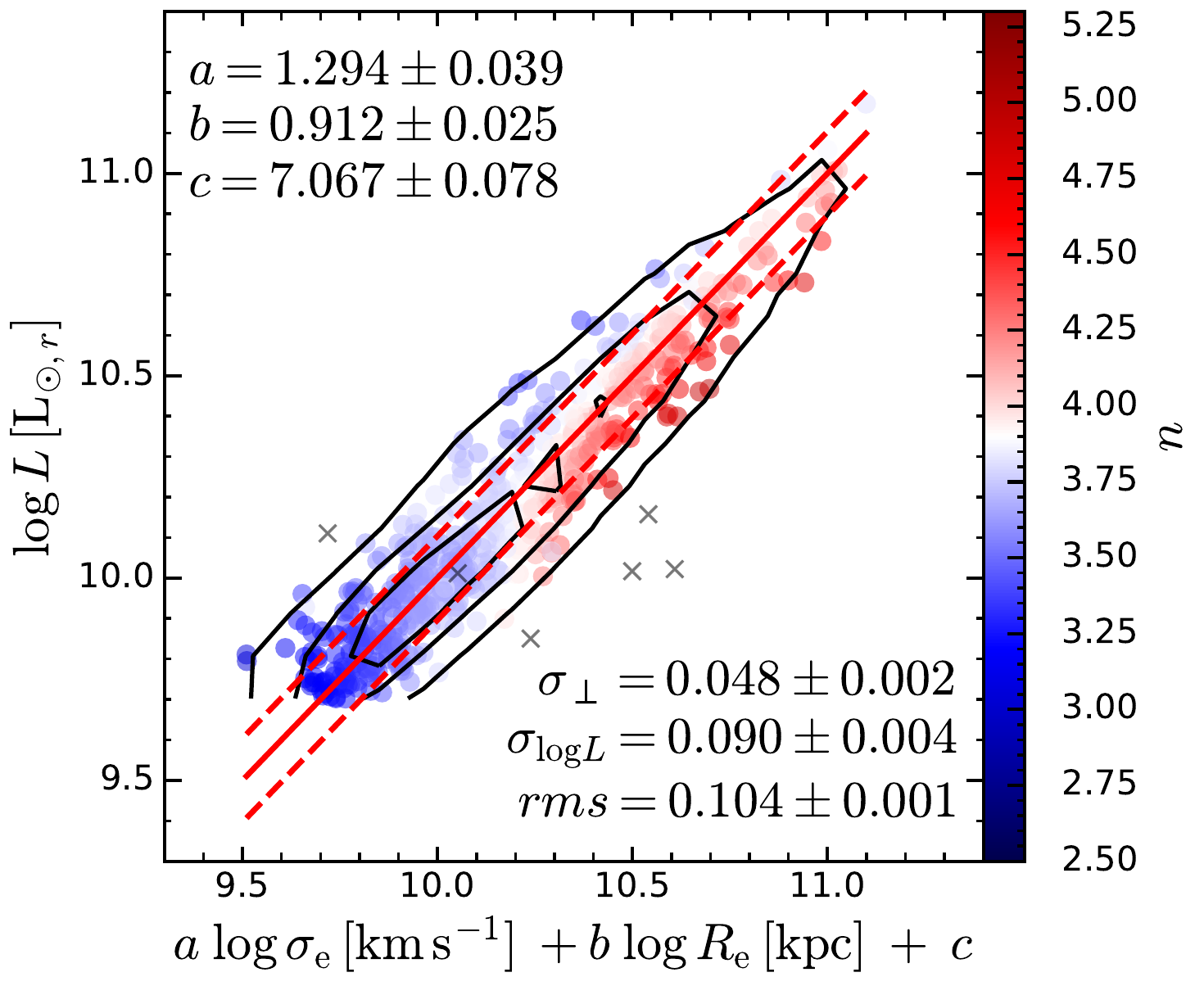}
  \caption{The fiducial FP, colour-coded with
  the LOESS-smoothed S{\'e}rsic index $n$; symbols are otherwise
  the same as in \reffig{f.r.fpbench}. Compared to SSP age, $n$ shows a trend 
  both along the FP and across the FP.
  }\label{f.r.fpsersic}
\end{figure}

Based on the considerations in \S~\ref{s.da}, we use the 3dG algorithm to find
the fiducial FP. We stress again that the fiducial plane is not the plane with
the least $rms$, but the plane with the \emph{least bias.} The fiducial FP for SAMI
ETGs has equation
\begin{equation}\label{eq.r.fiducialfp}
  \log \left( \dfrac{L}{\mathrm{L_{\odot,r}}}\right) = a \log \left( \dfrac{\sigma_\mathrm{e}}{\mathrm{km \, s^{-1}}} \right) + b \log \left( \dfrac{R_\mathrm{e}}{\mathrm{kpc}} \right) + c
\end{equation}
where $a=1.294\pm0.039$, $b=0.912\pm0.025$, and $c = 7.067\pm0.078$. Along the
direction of \logl, the FP has $rms=0.104\pm0.001$ and intrinsic scatter $\sigma_{\log L}
= 0.090\pm0.004$\,dex (the intrinsic orthogonal scatter is $\sigma_\perp = 0.048 \pm
0.002$\,dex, see Table~\ref{t.r.bestfit}). Notice that we report the $rms$ along
\logl because this value is easier to compare between different studies, which
may find different FP coefficients. For each model parameter we quote as
the fiducial value the 50\textsuperscript{th} percentile of the marginalised
posterior probability distribution, while the uncertainties are the
semi-difference between the 84\textsuperscript{th} and 16\textsuperscript{th}
percentiles. These uncertainties are of the order of 2--3\% and set the significance threshold to assess the effect of
the assumptions that we test below. With the fiducial values of $a$ and $b$, the
resulting tilt of the FP relative to the virial plane is $0.706$ and $0.088$ in
the directions of \logsige and \logre respectively\footnote{With the
Gaussian assumption we adopt, the usual practice of subtracting the FP
coefficients $a$ and $b$ from their virial equivalent introduces a bias; the
correct approach is to consider the difference between the random variables
$L$ and $M_\mathrm{vir}(\sigma_\mathrm{e}, R_\mathrm{e})$ and combine their
covariance matrices.}.

The face-on and edge-on projections of the fiducial plane are shown in
in \reffig{f.r.fpproj}. Rather than using the eigenvectors
$\{ \hat{\bm{v}}_1, \hat{\bm{v}}_2, \hat{\bm{v}}_3 \}$ of the fiducial 3-d
Gaussian, we use the base $\{ \hat{\bm{w}}_1, \hat{\bm{w}}_2, \hat{\bm{w}}_3 \}$
because, unlike the eigenvectors, all the $\hat{\bm{w}}_i$'s can be
expressed in the base of the observables
${\log \sigma_\mathrm{e}, \log R_\mathrm{e}, \log L}$, using only the FP
coefficients $a$ and $b$
\begin{equation}\label{eq.r.eigen}
\begin{array}{ccrrrrr}
    \bm{w}_1 & \equiv & ( & 0,              & 1/b, &  1 & ) \\
    \bm{w}_2 & \equiv & ( & (1+b^2)/(a\,b), &  -1, &1/b &) \\
    \bm{w}_3 & \equiv & ( & -a,             &  -b, &  1 & ) \\
\end{array}
\end{equation}
where the unit vectors are given by $\hat{\bm{w}}_i \equiv \bm{w}_i /
||\bm{w}_i||$.
Notice that $\hat{\bm{v}}_3 = \hat{\bm{w}}_3$, so this choice of
base preserves the edge-on view of the FP (\reffig{f.r.fpproj.b}) and
is equivalent to a coordinate rotation within the FP. The relation between
$(\hat{\bm{v}}_1, \hat{\bm{v}}_2)$ and $(\hat{\bm{w}}_1, \hat{\bm{w}}_2)$ is
illustrated by the arrows in panel~\subref{f.r.fpproj.b}: $\hat{\bm{w}}_1$ is
very close to $\hat{\bm{v}}_1$, but combines only photometric observables.
The transformation between our base $\hat{\bm{w}}_i$ and the eigenvector
base $\hat{\bm{v}}_i$ is given in columns 1--3 of Table~\ref{t.r.eigen}.

In Fig.~\ref{f.r.fpproj}, the white stars and ellipses trace the centroid
and the 2-d projections of the fiducial 3-d Gaussian; the axes of the
ellipses are equal to three times the \emph{intrinsic} dispersion. Because
of observational uncertainties, our data extends further than the ellipses
(which should otherwise enclose 99\% of the data). Each circle represents a
galaxy (black crosses are outliers), colour-coded by its light-weighted SSP age: 
there is a clear age trend across the plane (panel~\subref{f.r.fpproj.b});
this trend is also visible in \reffig{f.r.fpbench}, where we show
predicted vs measured \logl. Here the solid red line is the 1:1 relation
and the red dashed lines are offset by the observed $rms$.
The solid black contours enclose respectively the
40\textsuperscript{th}, 68\textsuperscript{th}, and 96\textsuperscript{th}
percentiles of the data (corresponding roughly to 1, 1.5 and 2.5 standard
deviations of the 2-d data distribution).
In panel~\subref{f.r.fpbench.b}, the age distribution has been
smoothed using the {\sc \href{https://pypi.org/project/loess/}{LOESS}}
algorithm \citep{cappellari+2013b}, highlighting the strong age gradient
across the plane: at fixed \sige and \re, the oldest galaxies (red hues in
Figs~\ref{f.r.fpproj} and \ref{f.r.fpbench}) are
under-luminous compared to the FP
prediction, and lie preferentially below the fiducial plane; the opposite is
true for the youngest galaxies (blue hues). In contrast, when colouring the
FP with S{\'e}rsic index $n$, the colour trend is both \textit{across} and \textit{along}
the FP (\reffig{f.r.fpsersic}).

\begin{table}
  \begin{center}
  \setlength{\tabcolsep}{4pt}
  \caption{
  Transformation between the eigenvectors of the fiducial 3-d Gaussian
  $\{ \hat{\bm{v}}_1, \hat{\bm{v}}_2, \hat{\bm{v}}_3 \}$ and the vector base
  $\{ \hat{\bm{w}}_1, \hat{\bm{w}}_2, \hat{\bm{w}}_3 \}$ from
  Fig.~\ref{f.r.fpproj} and equation~(\ref{eq.r.eigen}). $\hat{\bm{l}}_3$ is
  the normal to the LTS best-fit model (for this model, the distribution within
  the FP is not constrained, therefore there is only a single vector).
  }\label{t.r.eigen}
  \begin{tabular}{cccccc}
  \hline
                    & $\hat{\bm{v}}_i \cdot \dfrac{\hat{\bm{w}}_1}{||\hat{\bm{w}}_1||}$
                    & $\hat{\bm{v}}_i \cdot \dfrac{\hat{\bm{w}}_2}{||\hat{\bm{w}}_2||}$
                    & $\hat{\bm{v}}_i \cdot \dfrac{\hat{\bm{w}}_3}{||\hat{\bm{w}}_3||}$
                    & $\hat{\bm{v}}_i \cdot \dfrac{\hat{\bm{l}}_3}{||\hat{\bm{l}}_3||}$
                    & $\hat{\bm{v}}_i \sphericalangle \hat{\bm{l}}_3$ \\
                    & (1) & (2) & (3) & (4) & (5) \\
                    & dex & dex & dex & dex & \degree \\
  \hline
   $\hat{\bm{v}}_1$ & \phantom{-}0.912 & 0.410 & 0 & 0.029 &           88.3$\pm$0.6 \\
   $\hat{\bm{v}}_2$ &           -0.410 & 0.912 & 0 & 0.048 &           87.2$\pm$1.0 \\
   $\hat{\bm{v}}_3$ &                0 &     0 & 1 & 0.998 & \phantom{0}3.2$\pm$1.0 \\
  \hline
  \end{tabular}
  \end{center}
  Columns: (1--3)~scalar product between each eigenvector
  $\hat{\bm{v}}_i$ and $\hat{\bm{w}}_j$; (4)~scalar product between each
  eigenvector $\hat{\bm{v}}_i$ and the normal to the LTS best-fit plane
  $\hat{\bm{l}}_3$; (5)~angle between each eigenvector $\hat{\bm{v}}_i$ and the
  normal to the LTS best-fit plane.
\end{table}

In \S\S~\ref{s.r.ss.nnl}--\ref{s.r.ss.ssp}, we
quantify the correlation between the FP residuals and a number of galaxy
observables, finding SSP age to be the best predictor of the FP
residuals. However, before delving into the study of the residuals, we review
below how our assumptions affect the FP determination. We do so by changing
one assumption at a time and by comparing the resulting FP to the fiducial FP.
A summary of the results is reported in Table~\ref{t.r.bestfit}.

\subsubsection{Sample uncertainties}\label{s.r.ss.galev.sss.sample}

We estimate the sample random uncertainty by bootstrapping the FP sample one
thousand times and find the uncertainties about $a$, $b$, $c$, $rms$, $\sigma_{\log L}$ and
$\sigma_\perp$ to be respectively 0.036, 0.024, 0.077, 0.004, 0.005 and
0.003 (Table~\ref{t.r.bestfit}, row~2). Except for the $rms$, these
values are consistent with the uncertainties estimated by the 3dG algorithm
for the parameters of the fiducial plane.
The bootstrapping uncertainty about the $rms$ is larger than the
uncertainty calculated from the posterior distribution because the $rms$ is
sensitive to the presence and number of outliers. On the other hand, the
similarity between the posterior and bootstrapping uncertainties for the other
parameters implies that, for these parameters, the total uncertainty is
primarily due to measurement uncertainty, rather than sample variance. For
studies with significantly smaller sample size than considered here, the
relative contribution of sample uncertainty is likely to be higher.

As for systematic errors due to the sample, we run two tests. In
\S~\ref{s.ds.ss.samp}, we rejected 24 galaxies that, due to their large
apparent radius, are not fully covered by the SAMI IFU. This selection is worth
considering because these galaxies have larger than average physical radius,
so their inclusion/exclusion might bias the FP. We tested their effect in
two ways, using $\sigma_{\mathrm{e}/4}$ (the aperture velocity dispersion
measured inside the isophote of area $\pi (\mathrm{R_e}/4)^2$). First, we approximated
the 24 missing \sige values with their aperture-corrected version, based on
$\sigma_{\mathrm{e}/4}$
\citep[$\sigma_\mathrm{e} =
\sigma_{\mathrm{e}/4} \, 4^{-0.04}$,][]{jorgensen+1996}.
Secondly, we replaced \sige with $\sigma_{\mathrm{e}/4}$ for
all galaxies. Both tests return the same FP parameters within the uncertainties,
so we conclude that including/removing these galaxies does not affect our
conclusions (Table~\ref{t.r.bestfit}, row~3).
Likewise, we find that including or excluding the 22 galaxies that have
$R_\mathrm{e} < \sigma_\mathrm{PSF}$ and the 51 galaxies that have
$R_\mathrm{e} < 0.5$\,FWHM$_\mathrm{PSF}$
(which in principle could systematically affect \sige) also does not change
the FP parameters (Table~\ref{t.r.bestfit}, row~4).

\begin{table*}
  \begin{center}
  \caption{The parameters of the fiducial Fundamental Plane defined by
  equation~\ref{eq.r.fiducialfp} (first row) and
  their dependence on
  assumptions about the model and data (rows 2--21). Photometry (MGE or
  S{\'e}rsic), uncertainties on \logre, and the algorithm used have the largest
  effect on the FP parameters (rows 11--12, 14 \& 17, and 20--21 respectively).
  The last two rows give the parameters of the mock FPs.}\label{t.r.bestfit}
  \begin{tabular}{cccccccc}
  \hline
     \S                                    & description                                   & $a$ & $b$ & $c$ & $rms$ & $\sigma_{\log L}$ & $\sigma_\perp$ \\
  (1) & (2) & (3) & (4) & (5) & (6) & (7) & (8) \\
  \hline
   1$^*$       \hfill      \ref{s.r.ss.galev}        & \textbf{fiducial}                               & $1.294\pm0.039$ & $0.912\pm0.025$ & $7.067\pm0.078$ & $0.104\pm0.001$ & $0.090\pm0.004$ & $0.048\pm0.002$ \\
  \hline
   2           \hfill  \ref{s.r.ss.galev.sss.sample} & 1000 bootstrap.                                 & $\phantom{0.000}\pm0.036$ & $\phantom{0.000}\pm0.024$ & $\phantom{0.000}\pm0.077$ & $\phantom{0.000}\pm0.004$ & $\phantom{0.000}\pm0.005$ & $\phantom{0.000}\pm0.003$ \\
   3           \hfill  \ref{s.r.ss.galev.sss.sample} & Incl. large targets                             & $1.282\pm0.037$ & $0.880\pm0.023$ & $7.089\pm0.075$ & $0.105\pm0.001$ & $0.093\pm0.004$ & $0.050\pm0.002$ \\
   4           \hfill  \ref{s.r.ss.galev.sss.sample} & Excl. small targets                             & $1.297\pm0.042$ & $0.932\pm0.032$ & $7.050\pm0.085$ & $0.106\pm0.002$ & $0.092\pm0.004$ & $0.049\pm0.002$ \\
  \hline                                                                                              
   5           \hfill  \ref{s.r.ss.galev.sss.outl}   & No rejection                                    & $1.317\pm0.042$ & $0.900\pm0.027$ & $7.018\pm0.085$ & $0.112\pm0.002$ & $0.100\pm0.004$ & $0.053\pm0.002$ \\
   6           \hfill  \ref{s.r.ss.galev.sss.outl}   & 5\textsuperscript{th} \%-ile                    & $1.321\pm0.039$ & $0.904\pm0.026$ & $7.007\pm0.080$ & $0.098\pm0.001$ & $0.083\pm0.004$ & $0.044\pm0.002$ \\
  \hline
   7           \hfill  \ref{s.r.ss.galev.sss.aper}   & $\mathrm{\sigma_e \rightarrow \sigma_{e/4}}$    & $1.282\pm0.040$ & $0.892\pm0.026$ & $7.085\pm0.079$ & $0.106\pm0.001$ & $0.093\pm0.004$ & $0.050\pm0.002$ \\
   8           \hfill  \ref{s.r.ss.galev.sss.aper}   & $\mathrm{\sigma_e \rightarrow \sigma_e^{circ}}$ & $1.292\pm0.039$ & $0.917\pm0.025$ & $7.072\pm0.076$ & $0.104\pm0.001$ & $0.091\pm0.004$ & $0.048\pm0.002$ \\
   9           \hfill  \ref{s.r.ss.galev.sss.aper}   & $\mathrm{\sigma_e \rightarrow \sigma_{1.5''}}$  & $1.296\pm0.039$ & $0.877\pm0.026$ & $7.069\pm0.080$ & $0.105\pm0.001$ & $0.092\pm0.004$ & $0.049\pm0.002$ \\
  \hline
  10           \hfill  \ref{s.r.ss.galev.sss.phot}   & $R_{\rm e} \rightarrow R_{\rm e}^{\rm maj}$     & $1.311\pm0.040$ & $0.874\pm0.025$ & $6.998\pm0.081$ & $0.106\pm0.002$ & $0.094\pm0.004$ & $0.050\pm0.002$ \\
  11$^\dagger$ \hfill  \ref{s.r.ss.galev.sss.phot}   & MGE phot.                                       & $1.294\pm0.040$ & $0.922\pm0.028$ & $7.071\pm0.081$ & $0.103\pm0.002$ & $0.090\pm0.004$ & $0.048\pm0.002$ \\
  12$^\dagger$ \hfill  \ref{s.r.ss.galev.sss.phot}   & S{\'e}rsic phot.                                & $1.376\pm0.044$ & $0.825\pm0.025$ & $6.912\pm0.090$ & $0.113\pm0.002$ & $0.101\pm0.004$ & $0.054\pm0.002$ \\
  \hline
  13           \hfill  \ref{s.r.ss.galev.sss.corr}   & $0.5\times u_{\log \sigma_{\rm e}}$             & $1.274\pm0.038$ & $0.923\pm0.025$ & $7.107\pm0.077$ & $0.103\pm0.001$ & $0.093\pm0.004$ & $0.050\pm0.002$ \\
  14           \hfill  \ref{s.r.ss.galev.sss.corr}   & $0.5\times u_{\log R_{\rm e}}$                  & $1.336\pm0.039$ & $0.880\pm0.024$ & $6.985\pm0.078$ & $0.104\pm0.002$ & $0.096\pm0.004$ & $0.051\pm0.002$ \\
  15           \hfill  \ref{s.r.ss.galev.sss.corr}   & $0.5\times u_{\log L_{\phantom{e}}}$            & $1.294\pm0.038$ & $0.916\pm0.025$ & $7.066\pm0.078$ & $0.104\pm0.001$ & $0.092\pm0.004$ & $0.049\pm0.002$ \\
  16           \hfill  \ref{s.r.ss.galev.sss.corr}   & $1.5\times u_{\log \sigma_{\rm e}}$             & $1.322\pm0.039$ & $0.896\pm0.025$ & $7.010\pm0.079$ & $0.104\pm0.002$ & $0.086\pm0.004$ & $0.045\pm0.002$ \\
  17           \hfill  \ref{s.r.ss.galev.sss.corr}   & $1.5\times u_{\log R_{\rm e}}$                  & $1.210\pm0.038$ & $0.975\pm0.027$ & $7.230\pm0.078$ & $0.104\pm0.002$ & $0.077\pm0.004$ & $0.041\pm0.002$ \\
  18           \hfill  \ref{s.r.ss.galev.sss.corr}   & $1.5\times u_{\log L_{\phantom{e}}}$            & $1.286\pm0.038$ & $0.911\pm0.024$ & $7.085\pm0.077$ & $0.103\pm0.001$ & $0.086\pm0.004$ & $0.046\pm0.002$ \\
  19           \hfill  \ref{s.r.ss.galev.sss.corr}   & Uncorr. noise                                   & $1.277\pm0.037$ & $0.922\pm0.025$ & $7.100\pm0.077$ & $0.104\pm0.001$ & $0.086\pm0.004$ & $0.046\pm0.002$ \\
  \hline
  20           \hfill  \ref{s.r.ss.galev.sss.model}  & No censoring                                    & $1.277\pm0.039$ & $0.899\pm0.026$ & $7.109\pm0.077$ & $0.104\pm0.001$ & $0.091\pm0.004$ & $0.049\pm0.002$ \\
  21           \hfill  \ref{s.r.ss.galev.sss.model}  & LTS, \sc{clip}=3                                & $1.149\pm0.033$ & $0.896\pm0.025$ & $7.389\pm0.073$ & $0.100\pm0.004$ & $0.082\pm0.005$ & $0.047\pm0.003$ \\
  \hline
  22$\ddag$    \hfill  \ref{s.r.ss.mock.sss.mockfit} & $L \equiv M_\mathrm{vir} / \Upsilon_\star$      & $1.720\pm0.051$ & $1.198\pm0.033$ & $6.282\pm0.103$ & $0.137\pm0.002$ & $0.097\pm0.006$ & $0.042\pm0.003$ \\
  23$\ddag$    \hfill  \ref{s.r.ss.mock.sss.mockfit} & $L \equiv  M_\mathrm{vir} / \Upsilon_\star(\Sigma_\mathrm{vir})$ & $1.756\pm0.050$ & $1.196\pm0.033$ & $6.201\pm0.101$ & $0.138\pm0.002$ & $0.098\pm0.006$ & $0.042\pm0.003$ \\
  \hline
  \end{tabular}
  \end{center}
  \begin{flushleft}

  Columns: (1)~row identifier and reference in main text; (2)~brief description of the
  FP, with full details at reference in main text; (3--5)~best-fit
  FP coefficients and zero-point; (6--8)~FP scatter: the observed ($rms$) and
  intrinsic ($\sigma_{\log L}$) scatter in the direction of \logl and the
  intrinsic orthogonal scatter ($\sigma_\perp$); in addition the FP scatter has
  10\% systematic uncertainty, estimated by comparing different levels of
  outlier rejection.\\
  $^*$ The fiducial FP uses a censored 3-d Gaussian model, assuming correlated
  noise between \sige and \re and between
  \re and $L$; to reject outliers we mask the galaxies belonging to the
  1\textsuperscript{st} percentile of the posterior probability distribution and
  repeat the optimisation.\\
  $^\dagger$ These two tests use a
  restricted sample of 516 galaxies with both S{\'e}rsic and MGE photometry,
  selected with identical quality cuts.\\
  $^\ddag$ These two planes use mock luminosities derived from the virial mass
  estimator and from the stellar mass-to-light ratio \ups.
  \end{flushleft}
\end{table*}

\subsubsection{Outliers}\label{s.r.ss.galev.sss.outl}

We removed six galaxies that have low likelihood according to the fiducial
model (probability lower than the 1\textsuperscript{st} percentile of the
probability distribution, \ref{s.da.ss.outl}). Performing the optimisation while
including these six galaxies, the FP parameters $a$, $b$ and $c$ are unchanged:
their variation is at most $2\%$ (Table~\ref{t.r.bestfit}, row~5). The
same is true if we use a stricter rejection threshold and trim the
5\textsuperscript{th} percentile of the posterior distribution
(Table~\ref{t.r.bestfit}, row~6).

However, relaxing or tightening the rejection
criterion has a disproportionate effect on the $rms$ and intrinsic scatter,
which increase/decrease by 7--10\% (Table~\ref{t.r.bestfit}, columns 6--8).
Thus, even though the FP parameters are stable against outliers, we have to consider
an additional systematic uncertainty of 10\% on the tightness of the plane.
While this degeneracy between the strictness of the outlier rejection and the
$rms$ is undesirable, it does not affect the main conclusions of this work.
Furthermore, the intrinsic scatter of the FP is also degenerate
with the measurement uncertainties, which are difficult to quantify absolutely
(see \S~\ref{s.r.ss.galev.sss.corr}). For this reason, the systematic
uncertainty of 10\% on the FP intrinsic scatter is to be considered a lower
limit and is unlikely to decrease with increasing sample size.
We inspected the six outliers (crosses in \reffig{f.r.fpbench}),
and found that: one galaxy has a prominent neighbour and the measured \re is
overestimated by a factor of 2; one galaxy is a post-starburst galaxy and
lies at the edge of the colour-mass diagram; one galaxy has low surface
brightness and, even though it lies on the plane, it lies far from the FP
3-d Gaussian within the plane (Fig.~\ref{f.r.fpbench});  one galaxy has \sige
over-estimated by a factor of two (as determined by comparing to observations
in worse seeing conditions); and, finally, there are two galaxies where the
origin of the discrepancy is unclear.

Except for the outlier that has
low surface brightness, all the other outliers are rejected also by
the LTS algorithm (\S~\ref{s.r.ss.galev.sss.model}; for the LTS algorithm, there
is no preference for where galaxies lie \emph{within} the plane, hence the
outlier with low surface brightness is not penalised). We conclude by stressing
that the precise nature
of these six outliers is not germane to our discussion because, in our study
of the FP residuals, we are interested primarily in the FP parameters, which
we demonstrated to be insensitive to the rejection threshold adopted. We exclude
these six outliers from the study of the FP residuals.

\subsubsection{Adopted aperture}\label{s.r.ss.galev.sss.aper}

Any conventional choice of velocity dispersion does not seem to affect the FP. Our default
\sige is measured inside an elliptical aperture of area $\pi \mathrm{R_e}^2$
(\S~\ref{s.ds.ss.spec.sss.sigma}), but the FP parameters and scatter are
unchanged whether we use a smaller elliptical aperture of area $\pi (\mathrm{R_e}/4)^2$
(Table~\ref{t.r.bestfit}, row~7), a circular instead of elliptical
aperture of area $\pi \mathrm{R_e}^2$ (Table~\ref{t.r.bestfit}, row~8), or a
circular aperture of fixed apparent radius (1.5~arcsec radius;
Table~\ref{t.r.bestfit}, row~9). Notice that in these tests we used
exactly the same sample as the fiducial plane, even though using smaller
apertures in principle enables us to include more galaxies (i.e.\ the 24 large
galaxies with incomplete IFU coverage, see \S~\ref{s.r.ss.galev.sss.sample}).
Given the radial behaviour of aperture
$\sigma$ \citep{jorgensen+1996}, we expect an average increase of 0.024\,dex between our fiducial value of \sige and $\sigma_{\mathrm{e}/4}$.
To first order, this increase translates into a decrease in $a$ of a factor
$0.024 / \langle \log \sigma_\mathrm{e} \rangle \approx 0.014$, below our
significance  threshold (note that we used $\langle \log \sigma_\mathrm{e} \rangle = 2.18$~dex,
the median \logsige value over the sample). The measured change in $a$ is in line
with our expectations (the largest difference is only $0.7$ standard deviations
away from the fiducial value). In agreement with \citet{scott+2015}, we find
that measuring $\sigma$ inside circular apertures yields a slightly lower $rms$,
but with our sample size and systematic uncertainties this result is not
statistically significant.

\subsubsection{Adopted photometry}\label{s.r.ss.galev.sss.phot}

Replacing MGE circularised effective radii with major-axis effective radii
changes the FP parameters by 2\% (Table~\ref{t.r.bestfit}, row 10), below the
significance threshold of 3\%.
Even though major-axis effective radii are
systematically larger than circularised effective radii, for our ETG sample we
find that the average difference is modest ($\approx$10\% or 0.04~dex). This is
likely to change for a sample with a significantly different fraction of fast
rotators \citep{bernardi+2020}.

If we swap MGE for S{\'e}rsic photometry (including re-measuring \sige inside the
S{\'e}rsic half-light radius), we find significantly different FP
parameters. For a fair comparison, we ensured that the two FPs have exactly the
same galaxies: we repeat the MGE FP optimisation on the subset of galaxies that
have both S{\'e}rsic and MGE photometry (516 galaxies). For this restricted MGE-based
FP, we find the same parameters as for the fiducial
FP (the largest difference is 3\%; Table~\ref{t.r.bestfit}, row~11).
Compared to the restricted MGE FP, the
S{\'e}rsic-photometry FP has 7\% larger $a$ and smaller $b$ but the same value
of $c$ (Table~\ref{t.r.bestfit}, row~12). The observed $rms$ is 9\% larger,
which suggests we should increase the measurement uncertainties on the
S{\'e}rsic parameters; furthermore, for the S{\'e}rsic FP we assumed the same
correlated noise as for the MGE FP, whereas in reality S{\'e}rsic-based luminosity
and effective radii have stronger correlated noise than the MGE equivalents
\citepalias[e.g.][]{magoulas+2012}. Increasing the S{\'e}rsic measurement
uncertainties and their correlations does not change the $rms$, but reduces
the intrinsic scatter, which otherwise is 13\% larger than the MGE equivalent.
These results suggests that, for the data available here, the precision of the MGE
photometry is superior. Notice, however, that S{\'e}rsic photometry was derived
independently for field and cluster galaxies, and given that there is no overlap
between these two sets, it is not possible to calibrate the two measurements.
Repeating the comparison with a much larger sample, de~Vaucouleurs and MGE
photometry yield comparable results (F. D'Eugenio, in~prep.). In conclusion,
even though our tests are not definitive in assessing whether MGE or S{\'e}rsic
profiles yield the tightest FP, we infer that the adopted photometry introduces
a 7\% systematic uncertainty on the FP parameters.
Nonetheless, the main results of this paper are unchanged if we adopt S{\'e}rsic
photometry, including the ranking of residual correlations.

\subsubsection{Noise estimates and correlated noise}\label{s.r.ss.galev.sss.corr}

In the fiducial model, we determine the measurement uncertainties by comparing
MGE and S{\'e}rsic measurements, with the assumptions that
(i)~the observed $rms$ is mostly due to observational uncertainties and
(ii)~the MGE and S{\'e}rsic measurements have comparable uncertainties so each
contributes $1/\sqrt{2}$ to the observed $rms$. For \sige, we compare the scatter
between repeat observations. How does under/over-estimating the uncertainties
affect the FP parameters?

Increasing or decreasing the measurement uncertainties on \logsige by 50\% does
not change the FP parameters or scatter (changes are about one standard
deviation; Table~\ref{t.r.bestfit}, rows~13 and~16). The same is true if we
change the uncertainties on \logl (Table~\ref{t.r.bestfit}, rows~15 and~18). In
contrast, changing the uncertainty
on \logre has a measurable impact on the FP parameters: halving the uncertainty 
has no measurable effect, but increasing the uncertainties by 50\% changes the
values of $a$, $b$ and $c$ by up to three standard deviations
(Table~\ref{t.r.bestfit}, rows~14 and~17). Because the uncertainty
on the difference is $\sim\sqrt{2}$ larger than the standard deviation on the
fiducial parameters, these differences are not statistically significant.
However, further increasing the measurement uncertainties to $2\times$
larger than the fiducial value continues the trend and breaks the
significance threshold of three standard deviations. The different
behaviour of \logre compared to \logsige and \logl is due to the fact that
\logre has the largest measurement uncertainties.

This test also illustrates the degeneracy between measurement uncertainties and
the FP intrinsic scatter: increasing/decreasing the value of the uncertainties
does not change the observed $rms$, but decreases/increases both $\sigma_{\log L}$ and
$\sigma_\perp$ (Columns~6--8 in Table~\ref{t.r.bestfit}).

In our fiducial model, we assume correlated measurement errors between \re and
\sige, and between \re and $L$. Neglecting both correlations, considering
only one at a time, or doubling the value of the correlation does not change the
value of the best-fit parameters. The largest discrepancy with respect to the
fiducial FP occurs when neglecting all correlations: in this case the largest
differences are at the level of 1\%, below both statistical and sample
uncertainties. The effect of correlated noise is apparent however in the
intrinsic FP scatter; while the $rms$ is the same as reported for the fiducial
FP (Table~\ref{t.r.bestfit}, row~19), the orthogonal scatter is lower ($4\%$),
because neglecting correlated noise confounds the artificial tightening due to
correlated noise with the intrinsic tightness of the FP.

In conclusion, even though the absolute size and correlation of the measurement
uncertainties can affect the FP parameters and does affect the FP scatter, we
explicitly demonstrate that the analysis of the residual trends is unchanged within
the range of uncertainties explored here.

\subsubsection{Model choice}\label{s.r.ss.galev.sss.model}

\begin{figure}
  \includegraphics[type=pdf,ext=.pdf,read=.pdf,width=1.0\columnwidth]{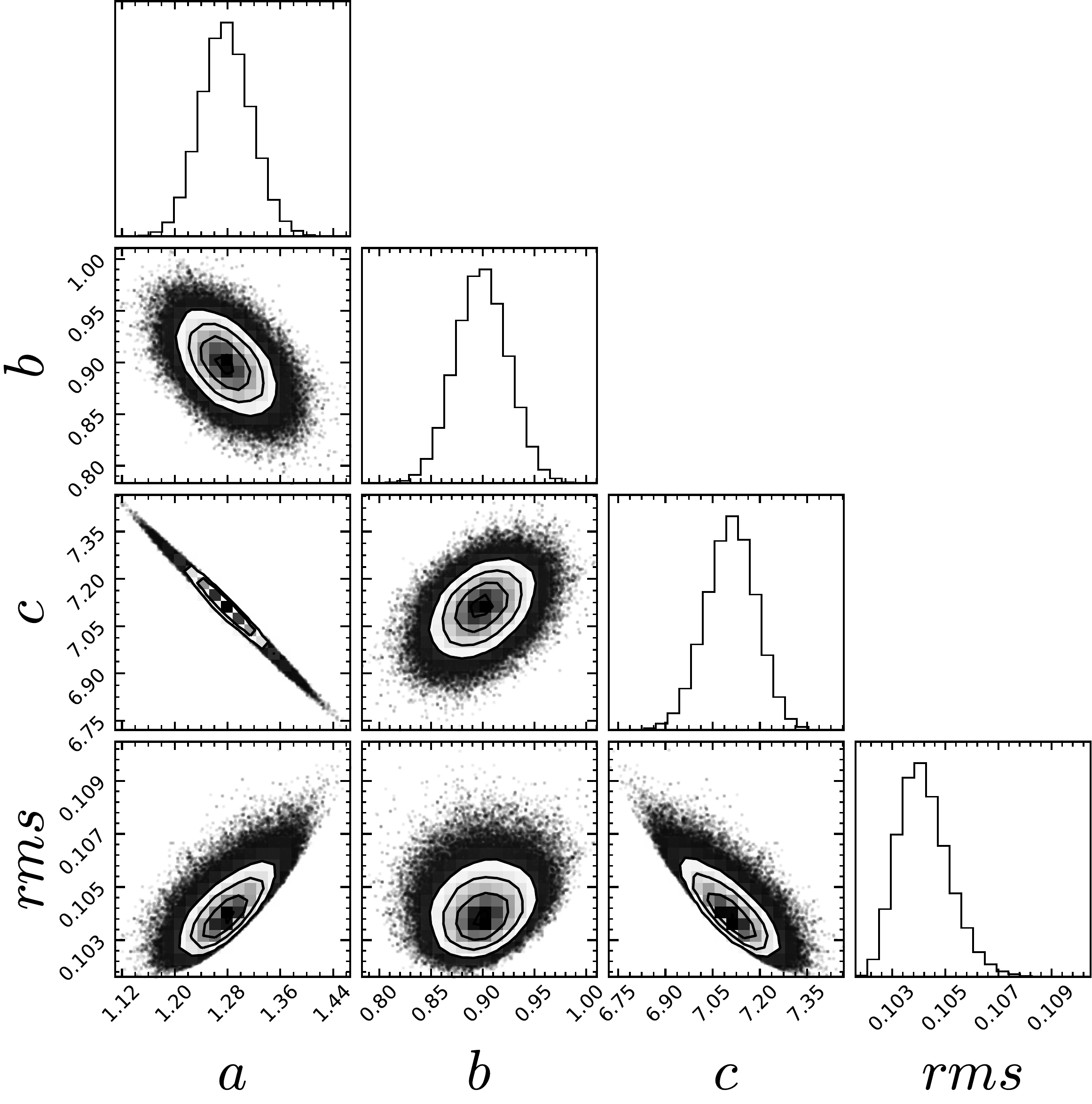}
  \caption{The posterior distribution for the fiducial FP, showing the 1-d and
  2-d distributions over $a$, $b$, $c$ and $rms$. This figure explains why
  the LTS algorithm has lower $rms$ and $a$ but larger $c$ compared to the
  fiducial FP (cf.\ rows~1 and~21 of Table~\ref{t.r.bestfit}). The LTS algorithm
  minimises the $rms$ but, because of the strong correlation between $rms$ and
  $a$ ($\rho=0.71$), the minimum possible $rms$ causes $a$ to be under-estimated.
  This bias, in turn, causes $c$ to be over-estimated, because of the strong
  anti-correlation between $a$ and $c$ ($\rho=-0.99$). 
  Notice that these considerations apply to a model with orthogonal scatter;
  assuming a probabilistic model with intrinsic scatter along \logl, the 3dG
  algorithm would be biased to steeper $a$ (\S~\ref{s.da.ss.comp}).
  }\label{f.r.abcrms}
\end{figure}

For a fair comparison between the 3dG and LTS algorithms, we repeat the 3dG fit
without censoring (Table~\ref{t.r.bestfit}, row~20). We compare this
FP to the LTS FP, and find the same value of $b$, but smaller $a$ and larger $c$
\citetext{four standard deviations; these results are qualitatively consistent with the
findings of \citealp{bernardi+2003}}.
As we have seen in \S~\ref{s.da.ss.rescorr}, this
large difference in $a$ and $c$ is explained as follows. First, there is a
strong correlation between $a$ and the observed $rms$; because direct fits
minimise the $rms$, they also tends to bias $a$ to lower values. Second, given that $a$
and $c$ are strongly anti-correlated, the decrease in $a$ must be compensated by
an increase in $c$ (Fig.~\ref{f.r.abcrms}).
In three dimensions, the normal to the
LTS FP $\hat{\bm{l}}_3$ is only $3.2\pm1.0\degree$ from the normal to the
fiducial FP; we report the components of $\hat{\bm{l}}_3$ along the eigenvectors
of the 3-d Gaussian model, as well as the enclosed angles (Table~\ref{t.r.eigen},
columns~4 \& 5).

We also find that the LTS FP, which ignores correlated noise, has intrinsic
scatter $\sigma_{\log L}$ 10\% smaller than the fiducial FP, but only 4\% smaller than the
intrinsic scatter inferred with the 3dG algorithm when ignoring
correlated noise (Table~\ref{t.r.bestfit}, row~19).

\subsubsection{Summary of method, model and sample uncertainties}

In summary, we find that the two changes that have the largest impact on the FP
parameters are the model choice (Table~\ref{t.r.bestfit}, row~21) and the method
used to measure the photometry (Table~\ref{t.r.bestfit}, row~12). Ignoring
correlated noise has a measurable impact on the inferred intrinsic scatter
(Table~\ref{t.r.bestfit}, row~19). With our sample size, all the other tests 
did not significantly change the FP parameters.
However, it is reasonable to assume that some of the reported changes are not
random, even though they are not statistically significant. The reason is that
some changes go in the direction expected from physical considerations
(e.g.\ using a smaller aperture to measure the kinematics yields on average higher
dispersion, and the resulting FP coefficient $a$ is therefore lower).
For this reason, (i)~changing more than one aspect at a time may well cause
significant changes to the FP parameters and (ii)~using a larger sample may
reveal that some of the highlighted changes are statistically significant.
Crucially, none of these tests affect, at least in a qualitative way, the main
results of our analysis: the relative significance of the residual correlations
is unchanged.

\subsection{Residuals with respect to the primary FP variables}\label{s.r.ss.nnl}

In the following, in order to uncover the origin of the FP scatter and the
age trend across the FP (Fig.~\ref{f.r.fpbench}), we study the residuals about
the FP as a function of various galaxy observables. The residuals are defined as
the difference \deltalogl between the observed \logl and the value of \logl
inferred from the fiducial FP
\begin{equation}
    \Delta \, \log L \equiv \log L - (a \, \log \sigma_\mathrm{e} + b \, \log R_\mathrm{e} + c)
\end{equation}
Our results are qualitatively unchanged if we use orthogonal
residuals, but we prefer the residuals in \logl for ease of comparison with
other works.

\begin{figure}
  \includegraphics[type=pdf,ext=.pdf,read=.pdf,width=1.0\columnwidth]{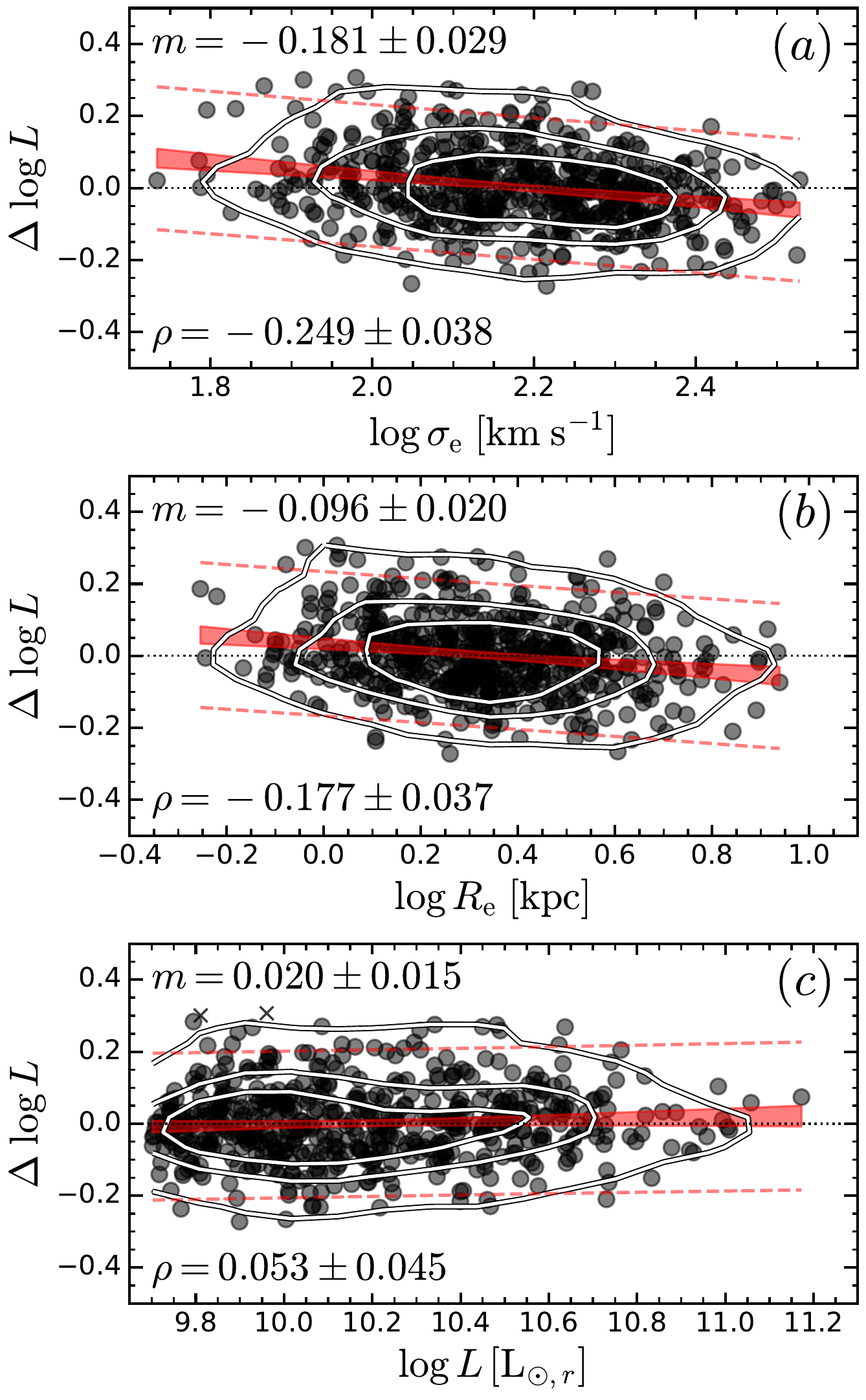}
  {\phantomsubcaption\label{f.galev.resid.nnl.a}
   \phantomsubcaption\label{f.galev.resid.nnl.b}
   \phantomsubcaption\label{f.galev.resid.nnl.c}}
  \caption{The residuals of the fiducial FP exhibit the expected correlations
  with the FP observables: \logsige (top row), \logre (middle row) and \logl
  (bottom row). Each circle represents a SAMI galaxy, the white contour lines
  enclose the 40\textsuperscript{th}, 68\textsuperscript{th} and
  96\textsuperscript{th} percentile of the data distribution. The red line
  traces the best-fit linear relation, the red regions are the 95\% confidence
  intervals, and the dashed red lines are the 95\% prediction intervals. The
  best-fit linear slope $m$ and the Spearman rank correlation coefficient are
  reported in the top left and bottom left corner of each panel.
  }\label{f.galev.resid.nnl}
\end{figure}

In \reffig{f.galev.resid.nnl} we show the residuals about the fiducial FP
(\deltalogl) as a function of the FP variables \logsige,
\logre and \logl. Circles represent individual galaxies and the solid white
contour lines enclose the 40\textsuperscript{th}, 68\textsuperscript{th}
and 96\textsuperscript{th} percentile of the galaxy distribution.
Naively, one would expect all three panels to show no correlation,
because the existence of a trend between \deltalogl and any of the three FP
variables suggests that the variable in question contains additional information
that could be used to further reduce the \deltalogl  residuals (the
target of many optimisation algorithms). However, this expectation is in
general incorrect, because the 3dG algorithm does not infer the model with the
\textit{least scatter}, but the model that is \textit{most likely} (or, more precisely, infers
the region of parameter space enclosing a given fraction of the posterior
probability; see \S~\ref{s.da.ss.rescorr}).
The fact that the true model can have (and in general does have) non-zero
correlation between the residuals and the independent variable is explicitly
shown in Table~\ref{t.ds.2dg}, column~3. Notice that in the three examples
with orthogonal intrinsic scatter, the 2dG algorithm recovers the true
correlations $\rho(\Delta, x)$ and $\rho(\Delta, y)$, whereas the best-fit LTS
model has $\rho(\Delta, x)$ closer to 0 than to the true value, which then
pushes $\rho(\Delta, y)$ too away from its true value. For this reason,
the presence or absence of residual correlations is not, by itself, a valid
method to assess the goodness of fit.

In order to quantify the correlations, we use {\sc lts\_linefit} to fit a
line to the data in each panel of \reffig{f.galev.resid.nnl}: the filled red 
regions are the 95\% confidence intervals and the dashed red lines are the 95\%
prediction intervals. The best-fit linear slope $m$ is reported in the top
left corner of each panel: in the absence of correlation, we expect $m$ to be
statistically consistent with zero.
As expected from 3dG and its underlying probability model, we find that the
residuals correlate with both \logsige and \logre: both the best-fit
slopes $m$ and the Spearman rank correlation coefficients $\rho$ are
different from zero (for \logsige, $m = -0.181 \pm 0.029$ and
$\rho = -0.249 \pm 0.038$, \reffig{f.galev.resid.nnl.a};
for \logre, $m = -0.096 \pm 0.020$ and $\rho = -0.177 \pm 0.037$,
\reffig{f.galev.resid.nnl.b}; see \reftab{t.r.resid}). In contrast, we find no
correlation between \deltalogl and \logl: both the best-fit slope
$m = 0.020 \pm 0.015$ and the correlation coefficient $\rho = 0.053 \pm 0.045$
are statistically consistent with zero (\reffig{f.galev.resid.nnl.c}). 
Given that even the \emph{true} model induces residual correlations between the
FP and its variables, the observed residual correlations are not, alone, an
indication of model mismatch. In addition, it has been suggested that the FP may
deviate from a log-linear relation \citep[e.g.][]{zaritsky+2006, wolf+2010},
which would also induce residual correlations. However, if non-linearity was
the cause of the observed residual trends, we would expect these residuals to
also show non-linearity, contrary to what we find in
Figs~\ref{f.galev.resid.nnl.a}--\subref{f.galev.resid.nnl.c}.

The presence and significance of residual trends with the FP
observables is critical to our analysis, because we set out to investigate the
trends between the FP residuals and galaxy observables. Given that most
galaxy observables correlate (or anti-correlate) with one or more of the FP
variables, we have to consider that the observed (anti)correlations might
reflect, at least in part, the correlations we have found so far.
For example, velocity dispersion correlates with stellar population age \citep[e.g.][]{
gallazzi+2005}, so that (at least in principle) a negative correlation
between \deltalogl and stellar population age could be due entirely to the observed negative
correlation between \deltalogl and \logsige. In the following, we will always
discuss whether an observed correlation would be enhanced or weakened by the
reported correlations between \deltalogl, \logsige and \logre.

\subsection{Structural trends}\label{s.r.ss.str}

\begin{figure*}
  \includegraphics[type=pdf,ext=.pdf,read=.pdf,width=1.0\textwidth]{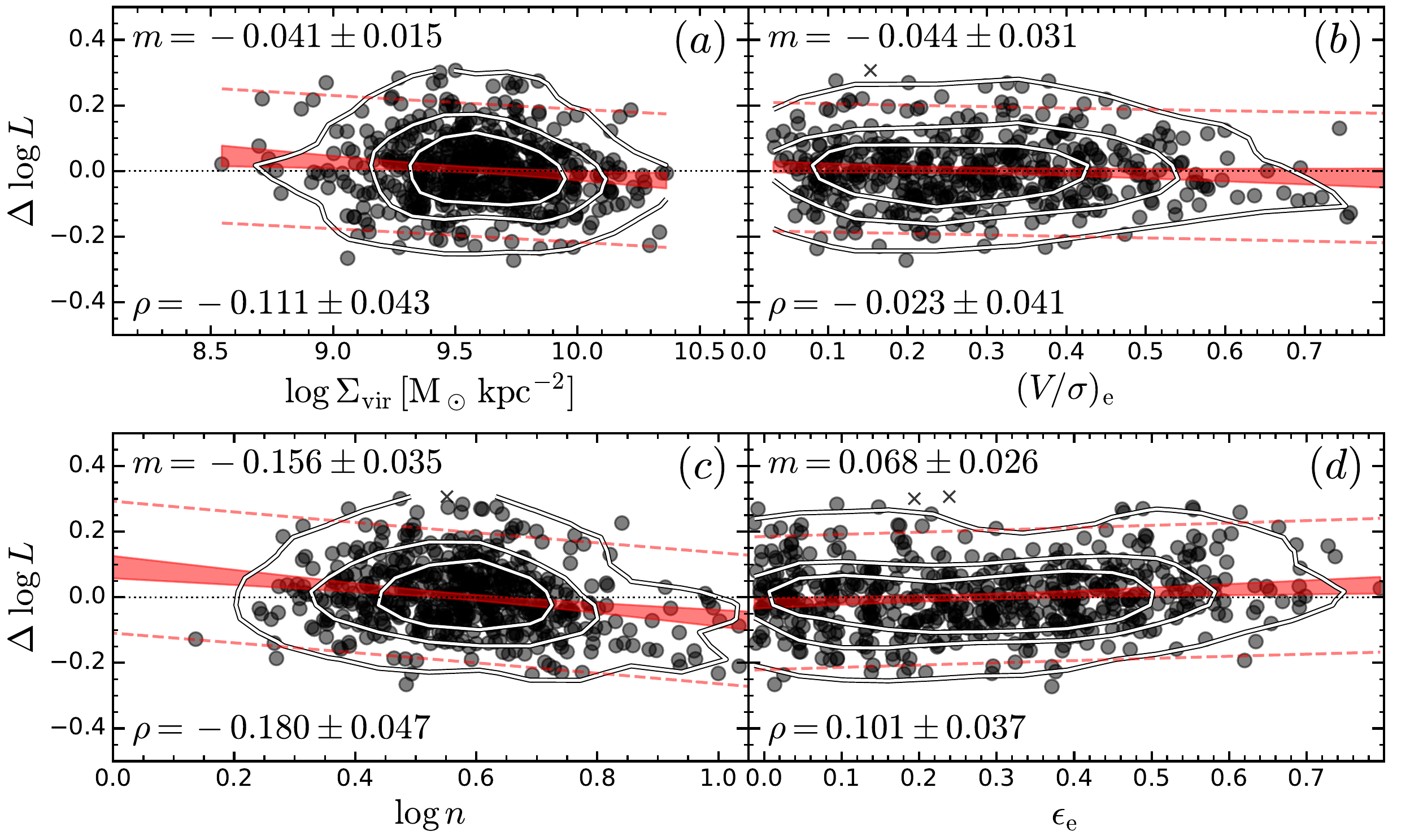}
  {\phantomsubcaption\label{f.galev.resid.str.a}
   \phantomsubcaption\label{f.galev.resid.str.b}
   \phantomsubcaption\label{f.galev.resid.str.c}
   \phantomsubcaption\label{f.galev.resid.str.d}}
  \caption{Residuals from the fiducial FP as a function of
  structural observables. There is a statistically-significant
  ($>$3 standard deviations) trend only with S{\'e}rsic index
  $n$ (panel~\subref{f.galev.resid.str.c}). We find no evidence of
  significant correlation with dynamical surface mass density
  \Sigvir (panel~\subref{f.galev.resid.str.a}), kinematic ratio
  $(V/\sigma)_\mathrm{e}$ (panel~\subref{f.galev.resid.str.b}), or
  galaxy shape $\epsilon_\mathrm{e}$ (panel~\subref{f.galev.resid.str.d}).
  Symbols are the same as in \reffig{f.galev.resid.nnl}.
  }\label{f.galev.resid.str}
\end{figure*}

In order to assess the effect of different dynamical properties on the FP
we study the presence and significance of trends between the FP residuals and
four structural variables: dynamical surface mass density ($\Sigma_\mathrm{vir}
\propto \sigma_\mathrm{e}^2 / R_\mathrm{e}$, \reffig{f.galev.resid.str.a}),
the ratio between streaming and random motions ($(V/\sigma)_\mathrm{e}$,
\reffig{f.galev.resid.str.b}), S\'ersic index ($n$, \reffig{f.galev.resid.str.c})
and ellipticity ($\epsilon_\mathrm{e}$, \reffig{f.galev.resid.str.d}).

We find a statistically-significant ($>$3 standard deviations) residual
trend for only one observable:
S{\'e}rsic index $n$, where the best-fit slope is $-0.156\pm0.035$ and the
correlation coefficient is $\rho = -0.180 \pm 0.047$, with a nominal
significance of $4.4$ standard deviations. The strength of this trend could
be over-estimated, because $n$ correlates with \logsige, which in turn
anticorrelates with \deltalogl.
Neither the dynamical density tracer \Sigvir
nor the kinematic ratio $(V/\sigma)_\mathrm{e}$ shows a significant
correlation, although we note the latter is a projected
quantity and the intrinsic (deprojected) value could have stronger correlation.
For surface mass density we find no correlation with the FP residuals even if
we swap \Sigvir with the stellar surface mass densities
$\Sigma_\star(R<R_\mathrm{e})$ and $\Sigma_\star(R<1\mathrm{kpc})$.
This lack of correlation may seem surprising, given the strong link between
$\Sigma_\star(R<1\mathrm{kpc})$ and quiescence \citep{cheung+2012, fang+2013,
woo+2015}, but this apparent discrepancy has a clear explanation 
(see \S~\ref{s.r.ss.resres}).

We also find no correlation with $\epsilon_\mathrm{e}$ (best-fit slope $0.068\pm0.026$ and
$\rho = 0.101\pm0.037$, $2.7$ standard deviations significance). However,
the strength and significance of this residual trend is
likely under-estimated because $\epsilon_\mathrm{e}$ anticorrelates with \logsige
\citep{vincent+ryden2005}. Moreover, among the FP variations shown in
Table~\ref{t.r.bestfit}, the fiducial FP (row~1) is the only one where the
significance of the residual correlation between \deltalogl and
$\epsilon_\mathrm{e}$ is below $3$ standard deviations, and the typical significance
is $\sim$4 standard deviations. Replacing
\re with $R_\mathrm{e}^\mathrm{maj}$ changes the \emph{sign} of the correlation
\citep[$\rho = -0.145$; cf.][]{bernardi+2020}.

Because $n$ is the structural parameter showing the most significant residual
correlation across the FP, we performed two additional tests. First, we
calculated the correlation between \deltalogl and $n$ for all the FP variations
in Table~\ref{t.r.bestfit}. We find the highest significance when using the
S\'{e}rsic FP (row 12, $\rho = -0.278$) and the least significance for the
$R_\mathrm{e}^\mathrm{maj}$ and LTS FPs (rows 10 and 21, $\rho = -0.061$ and
$\rho = -0.124$, respectively). In contrast, for stellar population age (the stellar population observable
with the most significant correlation), we find a minimum (in absolute value)
of $\rho = -0.335$ (for $R_\mathrm{e}^\mathrm{maj}$) and a
maximum of $\rho = -0.430$ (for the S\'{e}rsic FP). Second, we study the
correlation for two subsets in \logsige. For 330 galaxies with
$\sigma_\mathrm{e} < 170 \, \mathrm{km\,s^{-1}}$ we find no correlation ($\rho =
-0.08$, $P=0.1$; $n = 3.6\pm1.1$). In contrast, for as few as 71 galaxies with
$200 < \sigma_\mathrm{e} < 235 \, \mathrm{km\,s^{-1}}$ ($n = 4.3^{+1.7}_{-1.0}$),
we find a statistically significant correlation ($\rho = -0.265$, $P < 0.01$).

Taken together, these results suggest that: while the correlation between the FP
residuals and $n$ also exists for some limited ranges in \sige, it is driven
by galaxies with large S{\'e}rsic index ($n \gtrsim 4$). Given the distribution
of $n$ for our representative sample of ETG galaxies, structural differences
(non-homology) play a smaller role in the \emph{scatter} of the FP compared to
stellar population age (see \S~\ref{s.r.ss.ssp}). Alternatively, measurement uncertainties on
$n$ are so large that they damp the underlying physical trend. Notice that there
is no implication for the role of non-homology on the \emph{tilt} of the FP: as
the colour map in \reffig{f.r.fpsersic} shows, $n$ varies more along the plane than
across it. We study the effect of $n$ on the FP tilt in
\S~\ref{s.r.ss.mock.sss.misstilt}.

\subsection{Stellar population trends}\label{s.r.ss.ssp}

\begin{figure*}
  \includegraphics[type=pdf,ext=.pdf,read=.pdf,width=1.0\textwidth]{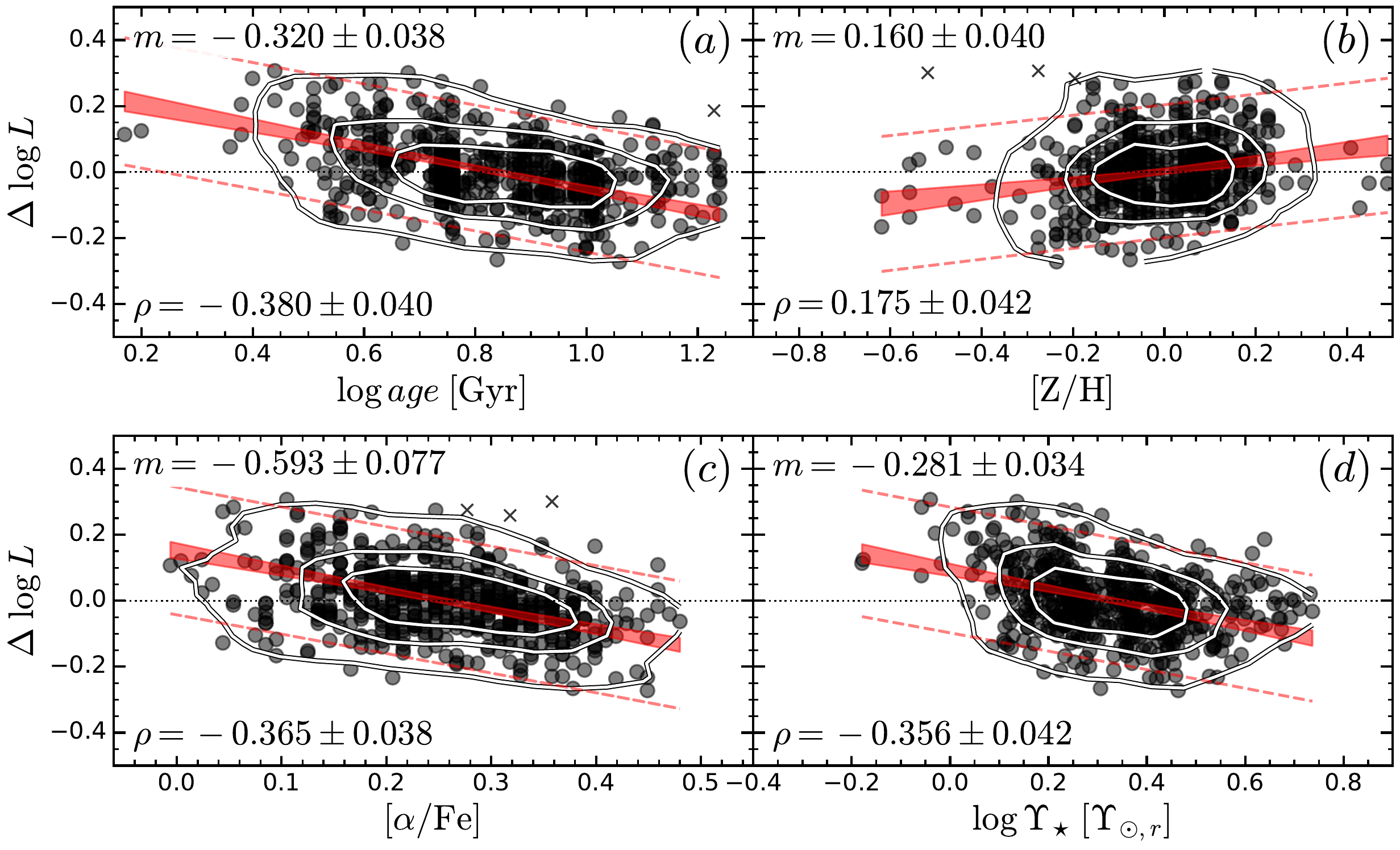}
  {\phantomsubcaption\label{f.galev.resid.ssp.a}
   \phantomsubcaption\label{f.galev.resid.ssp.b}
   \phantomsubcaption\label{f.galev.resid.ssp.c}
   \phantomsubcaption\label{f.galev.resid.ssp.d}}
  \caption{Correlations between the \deltalogl residuals of the fiducial FP and
  stellar-population parameters: (\subref{f.galev.resid.ssp.a})~\logage; 
  (\subref{f.galev.resid.ssp.b})~\ZH; (\subref{f.galev.resid.ssp.c})~\aFe; and
  (\subref{f.galev.resid.ssp.d})~\logups. The symbols are the same as in
  \reffig{f.galev.resid.nnl}. We find (anti)correlations for \logage, \aFe and
  \logups; among these three, the anticorrelations with \logage and \aFe have
  the highest statistical significance (more than seven standard deviations) and
  the highest rank correlation coefficient (in absolute value).
  }\label{f.galev.resid.ssp}
\end{figure*}

The residuals of the fiducial FP with respect to stellar-population observables
are illustrated in \reffig{f.galev.resid.ssp}. The symbols are the
same as in \reffig{f.galev.resid.nnl}; the only difference is the $x$ axis of
each panel. We show \deltalogl as a function of four observables related to
the simple stellar population (SSP) properties of our galaxies:
(\subref{f.galev.resid.ssp.a})~SSP age;
(\subref{f.galev.resid.ssp.b})~SSP metallicity, \ZH;
(\subref{f.galev.resid.ssp.c})~SSP $\alpha$-element enrichment, \aFe; and
(\subref{f.galev.resid.ssp.d})~$r-$band mass-to-light
ratio, $\Upsilon_\star$.

All the SSP observables show significant correlation. In order of increasing
significance, \deltalogl correlates with \ZH ($m = 0.160\pm0.040$ and $\rho =
0.175\pm0.042$) and anti-correlates with \aFe ($m = -0.593\pm0.077$ and $\rho =
-0.365\pm0.038$), with \ups ($m = -0.281\pm0.034$ and $\rho = -0.356\pm0.042$)
 and with age ($m = -0.320\pm0.038$ and $\rho = -0.380\pm0.040$).

SSP age from fixed apertures is already known to correlate with the FP residuals
\citep{forbes+1998, graves+faber2010, springob+2012}. As for $n$, we
check that the correlation with SSP age is not a secondary correlation due to
both \deltalogl and age (anti-)correlating with \sige, finding that, even at
fixed \sige, there are statistically significant trends. The median ages for
the 5\textsuperscript{th} and 95\textsuperscript{th} percentiles of the
residual distribution are 4.3 and 9.6\,Gyr, respectively.

However, removing young galaxies with
$age \lesssim 8$\,Gyr also removes the trend between age and \deltalogl.
This means that we find no correlation for galaxies with $\sigma_\mathrm{e}
\gtrsim 200 \, \mathrm{km\,s^{-1}}$, because of the correlation between age and
\sige. We also tested that SSP age retains the most significant correlation
with \deltalogl among all the alternative fits considered in Table~\ref{t.r.bestfit}.

\ups has the second-most significant correlation, followed by \aFe and \ZH.
Physically, we expect $\Upsilon_\star$ to be linked more directly to the FP
residuals than SSP age, but larger measurement uncertainties are likely to penalise
\ups compared to age.

The origin of the strong correlation with \aFe is unclear, because this
observable is connected only weakly to \ups, and certainly less so than \ZH.
To study the relation between \deltalogl, \aFe and age we proceed as follows.
When we consider a subset of ETGs with fixed \aFe ($0.2< [\mathrm{\alpha/Fe}]
<0.25$, $113$ targets), we find no correlation between \aFe and \deltalogl
($\rho = 0.11$, $P=0.2$) yet the correlation with age is still highly
significant ($\rho=-0.36$, $P=0.0001$). This suggests that age is the primary
driver of the correlation between \deltalogl and \aFe, because \aFe and age
are themselves strongly correlated ($\rho=0.50$, $P \ll 10^{-5}$). However, when
we take a slice of roughly constant age ($8 < age < 10 \; \mathrm{Gyr}$,
$117$ targets), even though we find no correlation between \deltalogl and age,
($\rho=-0.11$, $P=0.3$), \aFe still shows a tentative correlation: $\rho=-0.21$,
$P=0.02$. Although weak, this correlation relies on a much smaller number of
galaxies, and in the future it will be worth investigating this with a larger sample.
In \S~\ref{s.d.ss.resid} we suggest this trend might reflect age information that is captured
by \aFe but not by the absorption features used to estimate light-weighted SSP age
(for instance, because of large age uncertainties for old stellar populations).
Another possibility is that the observed correlation is due to the correlation
between \aFe and environment density \citep{liu+2016}, but our galaxies are more
massive than those considered by \citet{liu+2016}. Moreover, environment does not
appear to drive the correlation between \deltalogl and SSP observables; the
correlations of \aFe and age with \deltalogl are almost unchanged if we split
the sample by large-scale environment (i.e.\ cluster vs field/group) and, in
addition, the strength and significance of the correlation between \deltalogl
and environment density $\Sigma_5$ are lower than the correlations with both
$age$ and \aFe \citetext{$\rho=-0.18$; $\Sigma_5$ is the surface density of
galaxies inside the circle enclosing the five nearest $M_r<-18.5\,\mathrm{mag}$
neighbours within $\Delta \, v < 500 \, \mathrm{km\,s^{-1}}$, \citealp{brough+2017}}.

Finally, the correlation with \ZH also appears to be significant, but there
is no evidence of correlation at fixed age. Hence we deem the observed
correlation as an artefact of the strong degeneracy between age and metallicity.
This conclusion also agrees with the positive physical correlation between
\ZH and \ups at fixed age.

In conclusion, of the galaxy observables considered in Table~\ref{t.r.resid}, SSP
age always has the most significant correlation with \deltalogl. This is true
for all the FP variations considered in Table~\ref{t.r.bestfit}, so the fact that
SSP age is the best predictor of the FP scatter is independent (within reason) of
outlier rejection, under/over-estimation of uncertainties, adopted photometry,
chosen aperture, and algorithm/model choice.

\begin{table}
  \begin{center}
  \setlength{\tabcolsep}{2pt}
  \caption{Correlations between the residuals of the fiducial FP and the FP
  observables (rows~1--3), four structural parameters (rows~4--7) and four SSP
  parameters (rows~8--11). SSP age and mass-to-light ratio have the most
  significant correlations.
  }\label{t.r.resid}
  \begin{tabular}{ccccc}
  \hline
   X axis & $m\pm\sigma_m$ & $m/\sigma_m$ & $\rho$ & Fig. \\
   (1) & (2) & (3) & (4) & (5) \\
  \hline
  $\log \sigma_\mathrm{e}$                & $-0.181\pm0.029$\hphantom{-}& $-6.3$\hphantom{-}& $-0.249\pm0.038$ & \ref{f.galev.resid.nnl.a} \\
  $\log R_\mathrm{e}$                     & $-0.096\pm0.020$\hphantom{-}& $-4.8$\hphantom{-}& $-0.177\pm0.037$ & \ref{f.galev.resid.nnl.b} \\
  $\log L$                                & $\hphantom{-}0.020\pm0.015$\hphantom{-}& $\hphantom{-}1.3$\hphantom{-}& $\hphantom{-}0.053\pm0.045$ & \ref{f.galev.resid.nnl.c} \\
  \hline
  $\log \Sigma_\mathrm{vir}$              & $-0.041\pm0.015$\hphantom{-}& $-2.8$\hphantom{-}& $-0.111\pm0.043$ & \ref{f.galev.resid.str.a} \\
  $(V/\sigma)_\mathrm{e}$                    & $-0.044\pm0.031$\hphantom{-}& $-1.4$\hphantom{-}& $-0.023\pm0.041$ & \ref{f.galev.resid.str.b} \\
  $\log n$                                & $-0.156\pm0.035$\hphantom{-}& $-4.4$\hphantom{-}& $-0.180\pm0.047$ & \ref{f.galev.resid.str.c} \\
  $\epsilon_\mathrm{e}$                      & $\hphantom{-}0.068\pm0.026$\hphantom{-}& $\hphantom{-}2.7$\hphantom{-}& $\hphantom{-}0.101\pm0.037$ & \ref{f.galev.resid.str.d} \\
  \hline
  $\log age$                              & $-0.320\pm0.038$\hphantom{-}& $-8.4$\hphantom{-}& $-0.380\pm0.040$ & \ref{f.galev.resid.ssp.a} \\
  $[\mathrm{Z/H}]$                           & $\hphantom{-}0.160\pm0.040$\hphantom{-}& $\hphantom{-}4.0$\hphantom{-}& $\hphantom{-}0.175\pm0.042$ & \ref{f.galev.resid.ssp.b} \\
  $[\alpha/\mathrm{Fe}]$                     & $-0.593\pm0.077$\hphantom{-}& $-7.7$\hphantom{-}& $-0.365\pm0.038$ & \ref{f.galev.resid.ssp.c} \\
  $\log \Upsilon_\star$                      & $-0.281\pm0.034$\hphantom{-}& $-8.3$\hphantom{-}& $-0.356\pm0.042$ & \ref{f.galev.resid.ssp.d} \\
  \hline
  \end{tabular}
  \end{center} Columns: (1)~name of the observables being compared to the \deltalogl FP
  residuals; (2)~best-fit slope of the linear relation between the observable
  and \deltalogl; (3)~best-fit slope in units of the uncertainty; (4)~Spearman rank
  correlation coefficient, with bootstrapping uncertainties; (5)~figure showing
  \deltalogl vs observable.
\end{table}

\subsection{Structural trends of stellar mass-to-light ratio}\label{s.r.ss.ml}

\begin{figure*}
  \includegraphics[type=pdf,ext=.pdf,read=.pdf,width=1.0\textwidth]{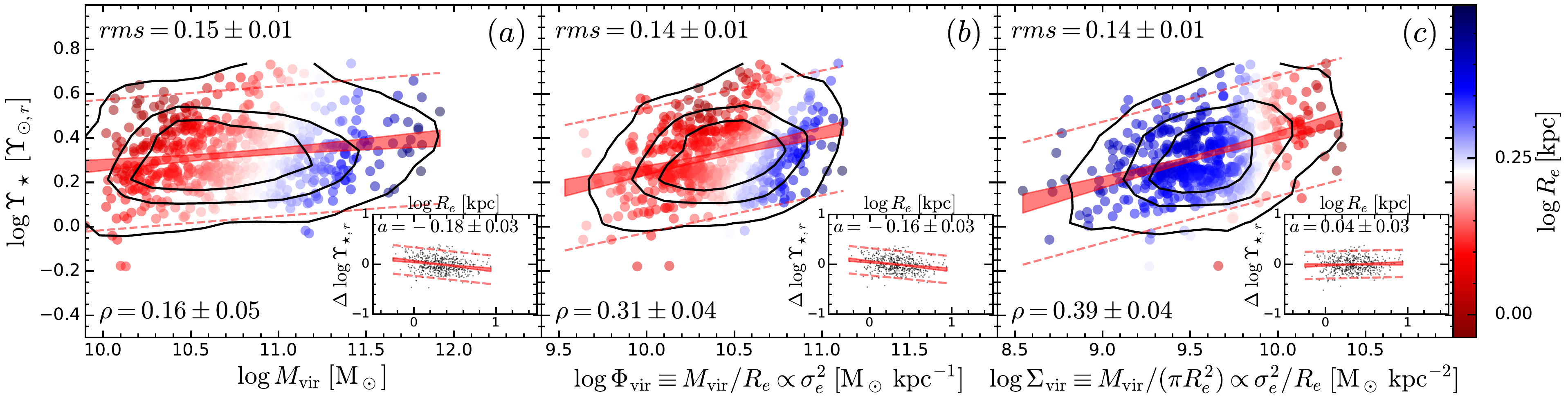}
  {\phantomsubcaption\label{f.r.ml.a}
   \phantomsubcaption\label{f.r.ml.b}
   \phantomsubcaption\label{f.r.ml.c}}
  \caption{The stellar mass-to-light ratio \logups is most naturally described as a
  function of surface mass density \Sigvir (panel~\subref{f.r.ml.c}).
  The relations of \ups with mass $M_\mathrm{vir}$ and gravitational potential
  $\Phi_\mathrm{vir}$ (panels~\subref{f.r.ml.a} and \subref{f.r.ml.b}) show
  more scatter and lower Spearman rank correlation coefficient $\rho$.
  The inset diagrams show the residuals of the best-fit relations as a function
  of effective radius; the relation between \logups and~\Sigvir is
  consistent with no residuals, unlike the other two relations.
  Low-\Sigvir galaxies have a range of sizes; on average, they are larger than
  the highest-\Sigvir galaxies.}\label{f.r.ml}
\end{figure*}

Having found that the most significant correlations of \deltalogl are with
SSP parameters, we investigate how these SSP parameters are related to the FP
observables \sige and \re. We relate these combinations to the virial mass
estimator $M_\mathrm{vir}$. Following \citetalias{cappellari+2006}, we define
$M_\mathrm{vir}$ by setting $\kappa = 5$ in equation~(\ref{eq.i.mvir}). This
assumption implies structural homology between galaxies, which
\citetalias{cappellari+2006} validate using dynamical modelling;
we also find weak evidence of structural non-homology
(\S~\ref{s.r.ss.mock.sss.misstilt}). We prefer to use the
expression with circularised \re instead of major-axis effective radii
\citepalias[e.g.][]{cappellari+2013a} because it yields less scatter both in the
FP and when compared to JAM-derived dynamical masses \citep{scott+2015}.
With this definition, we consider three combinations of \sige and \re:
$M_\mathrm{vir} \equiv 5 \sigma_\mathrm{e}^2 R_\mathrm{e} / G
\propto \sigma_\mathrm{e}^2 R_\mathrm{e}$,
$\Phi_\mathrm{vir} \equiv M_\mathrm{vir}/R_\mathrm{e} \propto \sigma_\mathrm{e}^2$ and
$\Sigma_\mathrm{vir} \equiv M_\mathrm{vir} / (\pi R_\mathrm{e}^2) \propto
\sigma_\mathrm{e}^2/R_\mathrm{e}$. \citetalias{barone+2018} and
\citet{barone+2020} have shown that the best
predictor\footnote{They use `best' in the same sense
as we do in this paper: lowest $rms$ and highest significance of the
correlation.} of SSP age is surface mass density \Sigvir, whereas
the best predictor of SSP \ZH is gravitational potential $\Phi_\mathrm{vir}$.
Here we focus on \ups, because it appears directly in the FP equation
(\ref{eq.i.fp}) through the expression of luminosity (equation \ref{eq.i.logl}).
In Fig.~\ref{f.r.ml}, we compare \ups to virial mass (panel~\subref{f.r.ml.a}),
gravitational potential (panel~\subref{f.r.ml.b}) and surface mass density
(panel~\subref{f.r.ml.c}).
The best-fit linear relations are shown by the solid red lines, with the 95\%
confidence intervals marked by the red shaded region, and the dashed red lines
enclosing the 95\% prediction intervals. The contours enclose the
40\textsuperscript{th}, 68\textsuperscript{th} and
96\textsuperscript{th} percentile of the galaxy distribution. The points are
colour-coded with LOESS-smoothed galaxy size, to highlight the presence or
absence of residual trends. It is apparent that the relations of \ups with both
$M_\mathrm{vir}$ and $\Phi_\mathrm{vir}$ present some residual trends with size,
because the colour hues are slanted with respect to the best-fit lines. In
contrast, galaxy size appears well mixed in the
\ups--\Sigvir plane, and the small leftover trend is
along the relation\footnote{Note that LOESS smoothing, by construction, hides
the mixing in the scatter plot.}. This intuition is quantified in three ways.
First, $\rho$ increases going from
the \ups--$M_\mathrm{vir}$ and \ups--$\Phi_\mathrm{vir}$
relations to the \ups--\Sigvir relation ($\rho = 0.16$,
0.30 and 0.38 respectively; see the bottom left corner of
Figs~\ref{f.r.ml.a}--\subref{f.r.ml.c}). Secondly, the observed
$rms$ decreases or stays constant from panels~\subref{f.r.ml.a} and~\subref{f.r.ml.b} to
panel~\subref{f.r.ml.c} ($rms = 0.15$, 0.14 and 0.14~dex
respectively; see the top left corner of Figs~\ref{f.r.ml.a}--\subref{f.r.ml.c}).
This decrease is small ($<$0.01~dex), but we have to consider that
the measurement uncertainties on \Sigvir are equal to the
measurement uncertainties on $M_\mathrm{vir}$, but $\sim$50\% larger than the
measurement uncertainties on $\Phi_\mathrm{vir}$ \citepalias[see the discussion in ][]{
barone+2018}. Finally, we study the residuals
$\Delta \log \Upsilon_\star$ of the three best-fit relations as a function of
galaxy size (inset panels of \reffig{f.r.ml}). We find that for the
\ups--$M_\mathrm{vir}$ and \ups--$\Phi_\mathrm{vir}$
relations the residuals $\Delta \log \Upsilon_\star$ have a statistically
significant correlation with \logre (best-fit slopes $r = -0.36 \pm 0.06$
and $r = -0.31 \pm 0.06$ respectively) whereas for the residuals of the
\ups--\Sigvir relation we find no such correlation ($r = 0.06 \pm
0.06$). These three results, taken together, suggest that the reason the
\ups--\Sigvir relation has the lowest scatter and highest
$\rho$ is because it takes into account the mass-to-light ratio
information encoded in the range of galaxy
sizes at fixed mass $M_\mathrm{vir}$ and at fixed potential $\Phi_\mathrm{vir}$.

In principle, this result could be due to correlated noise between \ups, \sige
and \re, but in practice this possibility is unlikely---the only
difference between the three panels of \reffig{f.r.ml} is \re, which is measured
from photometry and is therefore completely independent of both \ups and \sige, which are derived from spectroscopy. For the residual trends shown in the inset panels,
there is indeed a strong correlation between errors in $\Delta \log
\Upsilon_{\star}$ and \re, because measurement errors in \re propagate to \sige
\citep[via the aperture relation;][]{jorgensen+1996} or because \re appears
directly in the expression of $M_\mathrm{vir}$ and \Sigvir. The results displayed
take this correlation into account, but even when setting the correlation to zero our
results are qualitatively unchanged: only the residuals of the \ups--\Sigvir
relation have no correlation with galaxy size.
We further tested that our conclusions do not change if we repeat the above
analysis limiting the sample to elliptical galaxies, or if we use photometric
$M_\star$ instead of $M_\mathrm{vir}$, or if we swap circularised effective radii
for semi-major axis effective radii, or if we use {\sc galfit}-based effective
radii (either circularised or semi-major axis).

We conclude that, among the structural observables based on \sige and
\re, the best descriptor of \ups is surface mass density. This result is
consistent with: (i)~\ups being primarily
determined by age \citep{renzini1977, mould2020}
and (ii)~age
correlating more tightly with \Sigvir than either
$\Phi_\mathrm{vir}$ or $M_\mathrm{vir}$ \citetext{\citetalias{barone+2018},
\citealp{barone+2020}}. The best-fit relation between \ups and \Sigvir is
\begin{equation}\label{eq.r.upssigmadyn}
  \log \Upsilon_\star = (0.207\pm0.022) \log \Sigma_\mathrm{vir} - (1.654\pm0.210)
\end{equation}
with an observed \emph{rms} scatter of $0.14 \pm 0.01$~dex. We estimate the median
measurement uncertainties on both $\log \Sigma_\mathrm{vir}$ and \logups to be
0.05~dex, which yields a large
intrinsic scatter along \logups of $0.13 \pm 0.01$~dex. Using the most conservative
estimates for the uncertainties yields an intrinsic scatter of $0.10 \pm 0.01$~dex
(we used a median uncertainty of 0.11~dex and 0.10~dex
for $\log \Sigma_\mathrm{vir}$ and \logups, respectively).
If we model \logups as a function of both \logsige and \logre, we find
\begin{equation}\label{eq.r.upsplane}
\begin{split}
    \log \Upsilon_\star & = (0.445\pm0.041) \log \sigma_\mathrm{e} \\
        & - (0.187\pm0.032) \log R_\mathrm{e} - (0.584 \pm 0.089)
\end{split}
\end{equation}
in statistical agreement with the previous equation, once we express
\Sigvir as a function of \sige and \re. In particular, we find
that the ratio between the coefficients of \logre and \logsige is
$-0.42\pm0.08$, statistically consistent with the value $-0.5$ appropriate
if the correlation was with $\log \Sigma_\mathrm{vir}$.

We thus arrive at an apparent paradox: the residuals of the FP correlate
strongly with SSP \ups and the best predictor of \ups is surface mass density,
yet surface mass density itself shows no correlation with the FP residuals
(\S\ref{s.r.ss.ssp}, Fig.~\ref{f.galev.resid.str.a}). As we will see, the
solution to this apparent contradiction is that the intrinsic scatter of the FP
is partly due to the intrinsic scatter of the \ups--\Sigvir relation,
i.e.\ to the relatively broad range of \ups at fixed \sige and \re (0.14~dex),
resulting in a broad distribution of \ups at any position on the FP.

\subsection{The FP fully encapsulates the \texorpdfstring{\ups--\Sigvir}{Upsilon-Sigma} relation}\label{s.r.ss.resres}

\begin{figure}
  \includegraphics[type=pdf,ext=.pdf,read=.pdf,width=1.0\columnwidth]{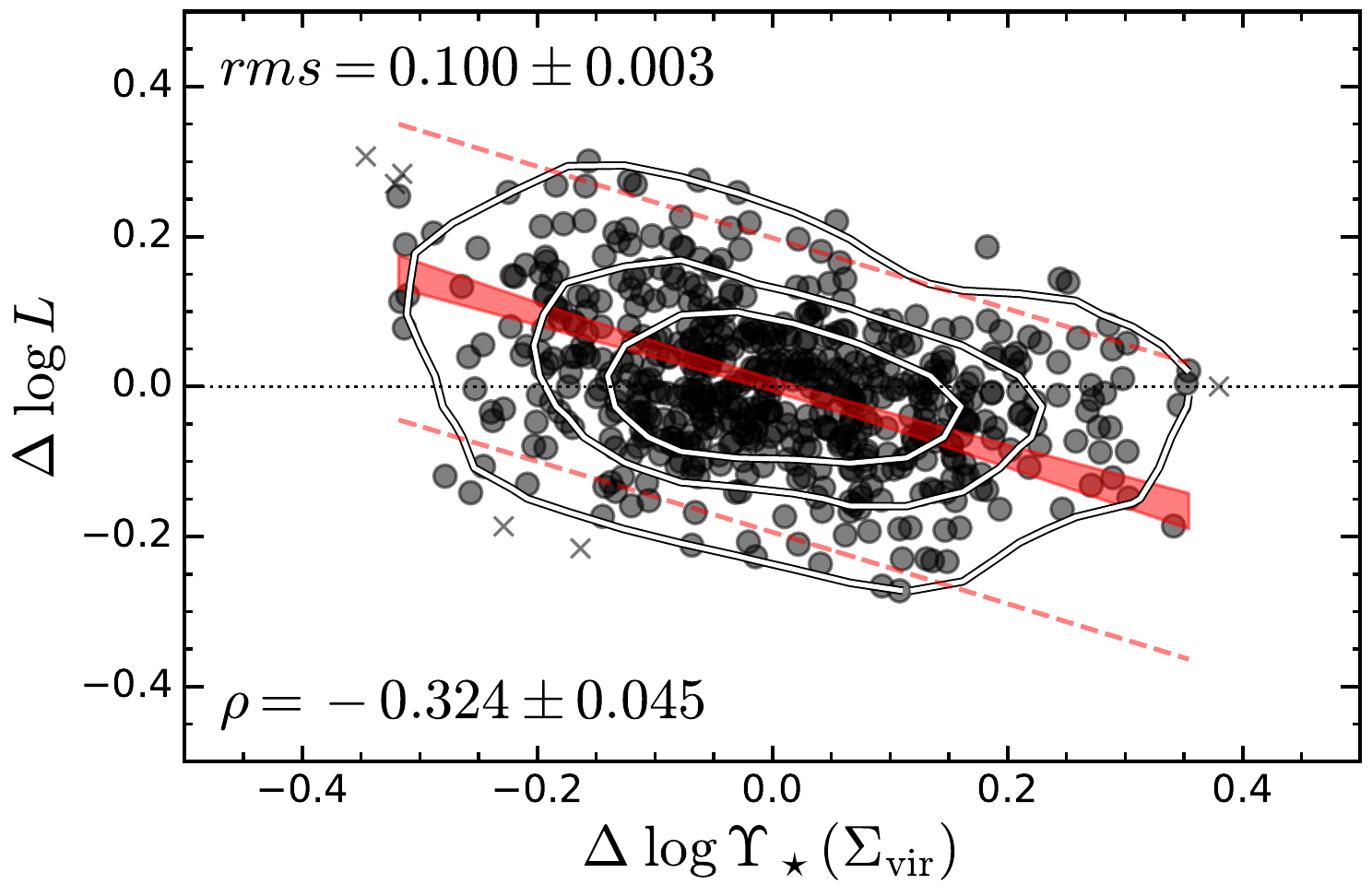}
  \caption{The residuals of the FP, \deltalogl, anticorrelate with the residuals of the \ups--\Sigvir relation, $\Delta\,\log\Upsilon_\star(\Sigma_\mathrm{vir})$:
  at any fixed position on the FP, galaxies that lie above the
  \ups--\Sigvir (or age--\Sigvir) relation have higher stellar mass-to-light
  ratio, are therefore less luminous compared to the average galaxy at that
  position, and so lie below the FP. The FP residuals do not correlate with \Sigvir
  (Fig.~\ref{f.galev.resid.str.a}) because the FP already encapsulates the
  mean \ups variation along the \ups--\Sigvir relation (Fig.~\ref{f.r.ml.c}).
  }\label{f.r.resres}
\end{figure}

To understand the origin of the correlation between the FP residuals
\deltalogl and \ups, we study the relation between \deltalogl and 
the residuals of the \ups--\Sigvir relation, $\Delta \, \log \Upsilon_\star (\Sigma_\mathrm{vir})$, as shown in
Fig.~\ref{f.r.resres}.

The negative correlation in Fig.~\ref{f.r.resres} means that, at fixed \sige and
\re (and so at any fixed position on the FP), galaxies lying above the
\ups--\Sigvir relation ($\Delta \, \log \Upsilon_\star (\Sigma_\mathrm{vir})
> 0$)
lie preferentially below the FP ($\Delta \, \log L < 0$).
The same is true if replacing \ups with SSP age. This result has a
straightforward interpretation: at any fixed position on the FP, galaxies
lying above the \ups--\Sigvir (or age--\Sigvir) relation have higher stellar
mass-to-light ratio and are thus less luminous compared to the average
galaxy at that position. The fact that the FP residuals do not correlate with
\Sigvir is because the FP already encapsulates the mean \ups variation along the
\ups--\Sigvir relation, so \Sigvir does not contain any additional information
that can reduce the FP scatter. The \ups--\Sigvir relation is embedded
in the FP through (part of) the FP tilt (the deviation of the best-fit
FP parameters from the virial values to the observed values).
It remains to be determined how much of the observed
tilt is explained by the \ups--\Sigvir relation.

Contrary to expectations, repeating the test of Fig.~\ref{f.r.resres} with 
the residuals of the best-fit \ups--$M_\mathrm{vir}$ and \ups--$\Phi_\mathrm{vir}$ 
relations gives results that are statistically consistent with Fig.~\ref{f.r.resres}. 
This lack of difference could be due to the large uncertainty in both 
$\Delta\,\log\Upsilon_\star(\Sigma_\mathrm{vir})$ and \deltalogl, but to test 
this hypothesis we need a larger ETG sample. Here we assume that, in accordance 
with the results of \S~\ref{s.r.ss.ml}, the \ups--\Sigvir is the most fundamental 
of the three relations.

\subsection{Mock Fundamental Planes}\label{s.r.ss.mock}

Having determined the dependence of \ups on the structural parameters \sige
and \re, we now investigate the effect of the \ups--\Sigvir
relation on the FP tilt and scatter. We do so by studying two mock FPs, to
test (i)~if using \ups inferred from the \ups--\Sigvir relation affects the
mock FP in the same way as using the measured \ups and (ii)~to quantify the
impact of systematic trends of \ups with \sige and \re on the FP tilt and
scatter.

We create two mock datasets by taking for each FP galaxy its measured \sige
and \re but replacing observed $L$ with a synthetic luminosity, calculated as
$L_\mathrm{synth} \equiv M_\mathrm{vir} / \Upsilon_\star$. The two mock datasets
differ only in how \ups, and so synthetic luminosity, is obtained.
With our definition of $M_\mathrm{vir}$, setting $\Upsilon_\star = 1$
yields the virial plane ($a = 2.000\pm0.017$ and $b = 1.000\pm0.012$). Because we
use the measured (or inferred) value of \ups, the mock planes measure how much
of the FP tilt relative to the virial plane is due to systematic variations of
\ups with \sige and \re, and how much of the FP scatter is due to the scatter in
\ups.

\begin{figure}
  \includegraphics[type=pdf,ext=.pdf,read=.pdf,width=1.0\columnwidth]{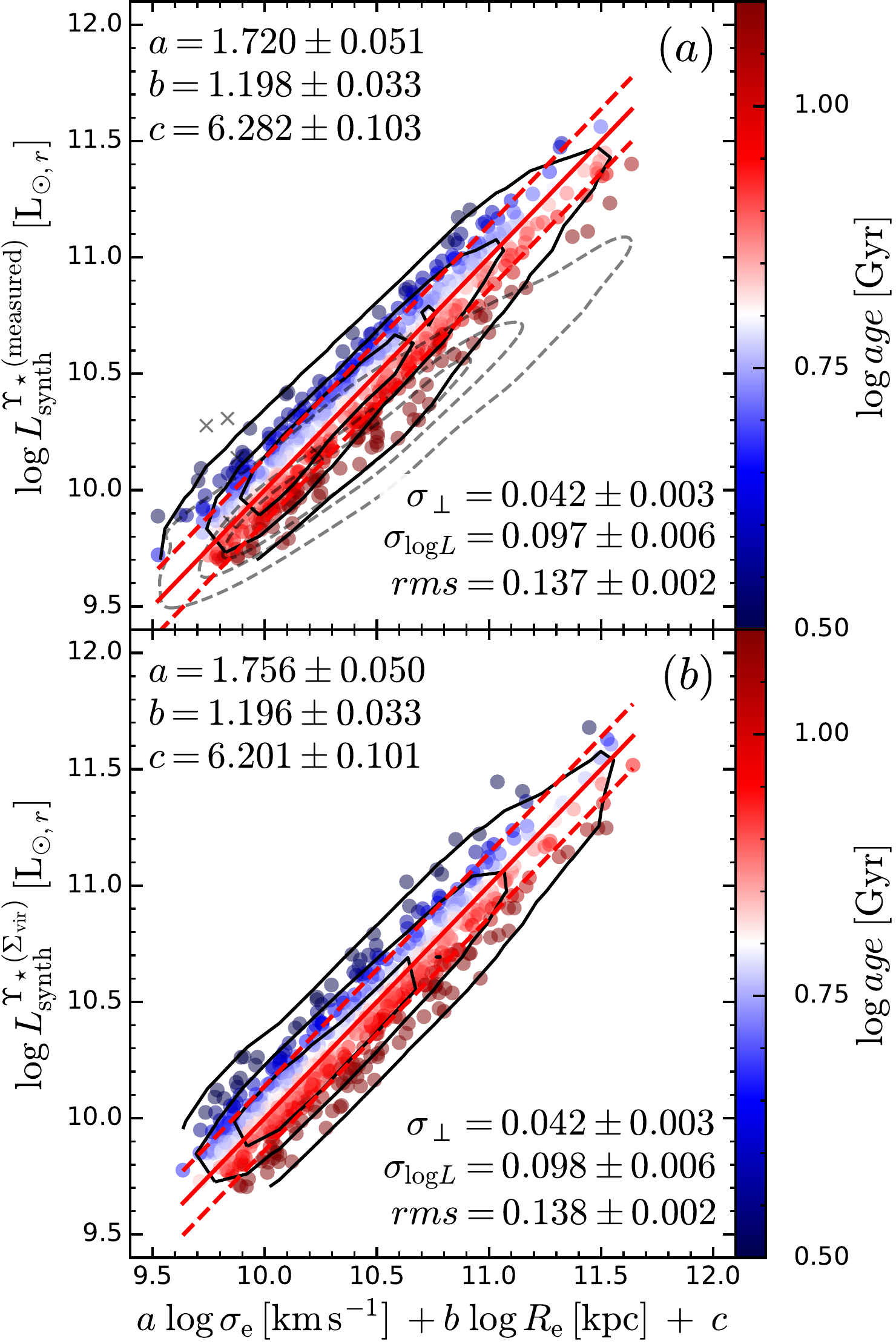}
  {\phantomsubcaption\label{f.r.mock.a}
   \phantomsubcaption\label{f.r.mock.b}}
  \caption{The mock FPs, showing that the systematic trends
  of SSP mass-to-light ratio across and through the plane are insufficient to
  explain all the tilt of the observed FP (compare the best-fit coefficients of
  the mock FPs with the FP coefficients in \reffig{f.r.fpbench}; the dashed
  contours show the projection of the fiducial FP, other symbols are the same as
  in \reffig{f.r.fpbench}).
  The synthetic luminosity is the virial mass divided by the SSP mass-to-light
  ratio, derived in turn either from the SSP age and metallicity measurements
  (panel~\subref{f.r.mock.a}) or from \Sigvir
  (panel~\subref{f.r.mock.b}) using equation~(\ref{eq.r.upssigmadyn}). In both
  panels the best-fit FP coefficients differ from both the virial 
  prediction and the fiducial FP, highlighting that the stellar-population
  relations are responsible for part, but not all, of the observed tilt.
  }\label{f.r.mock}
\end{figure}

For the first mock (Fig.~\ref{f.r.mock.a}), we take for each galaxy, alongside
\sige and \re, the measured \ups. We then derive the synthetic luminosity
$L^{\Upsilon_\star(\mathrm{measured})}_\mathrm{synth}$ by substituting these 
values in the expression for \logl (equation~\ref{eq.i.logl}); i.e.\ 
$L^{\Upsilon_\star(\mathrm{measured})}_\mathrm{synth} \equiv
M_\mathrm{vir}(\sigma_\mathrm{e}, R_\mathrm{e}) / \Upsilon_\star$.
Because we focus on how stellar-population
properties affect the FP, we set the structural factor $\kappa = 5$
\citepalias[following][]{cappellari+2006}, the stellar mass fraction
$f_\star = 1$, and assume a Chabrier IMF for every galaxy
($\Upsilon_\mathrm{IMF}/\Upsilon_\star = 1$); the impact of these assumptions
will be discussed in \S~\ref{s.r.ss.mock.sss.misstilt}. With these
definitions, the ratio $L^{\Upsilon_\star(\mathrm{measured})}_\mathrm{synth}/L$
leaves out the non-homology, dark-matter, and IMF terms of the FP (cf.\ equations
\ref{eq.i.logl} and \ref{eq.r.misstilt}).

For the second mock (Fig.~\ref{f.r.mock.b}), we take for each galaxy only two
measurements: \sige and \re. From these values, we calculate the dynamical
surface mass density \Sigvir, then use the empirical relation
(\ref{eq.r.upssigmadyn}) to infer the value of \logups; we also add Gaussian
random noise with standard deviation of $0.14$, equal to the $rms$ about the
best-fit \ups-\Sigvir relation.\footnote{For completeness, we also infer a mock
age, by using the empirical relation between the measured \logage and \Sigvir.
This inferred age is used to colour-code Fig.~\ref{f.r.mock.b} but has no role
in the analysis.}
For this second mock, the synthetic luminosity is then defined as
$L^{\Upsilon_\star(\Sigma_\mathrm{vir})}_\mathrm{synth} \equiv
M_\mathrm{vir}(\sigma_\mathrm{e}, R_\mathrm{e}) / \Upsilon_\star[\Sigma_\mathrm{vir}(
\sigma_\mathrm{e}, R_\mathrm{e})]$, where the mass-to-light ratio is given by the empirical
relation equation~(\ref{eq.r.upssigmadyn}).

Thus $L^{\Upsilon_\star(\mathrm{measured})}_\mathrm{synth}$ is a function of three
observables: \sige, \re and \ups, whereas
$L^{\Upsilon_\star(\Sigma_\mathrm{vir})}_\mathrm{synth}$ depends solely on
\sige and \re.

\subsubsection{Mock FP fit}\label{s.r.ss.mock.sss.mockfit}

The mock FPs are shown in \reffig{f.r.mock}, with the same symbols and colours
as the fiducial FP (\reffig{f.r.fpbench}). For
$L^{\Upsilon_\star(\Sigma_\mathrm{vir})}_\mathrm{synth}$ we do not use any
outlier rejection, because the data was generated from a Gaussian distribution
with no outliers. The best-fit parameters of the two mock FPs are in excellent
agreement with each other, implying that the empirical relation between \ups and
\Sigvir (\reffig{f.r.mock.b}) traces satisfactorily the systematic variations of
measured \ups with \sige and \re (\reffig{f.r.mock.a}). However, the mock FPs
have different scatter and tilt from the fiducial FP, and therefore trends in \ups
do not on their own explain the tilt and scatter of the FP. 

For the observed scatter, we find that the mock FPs have $rms = 0.137
\pm 0.002$\,dex (panel~\subref{f.r.mock.a}) and $0.138 \pm 0.002$\,
dex (panel~\subref{f.r.mock.b}), whereas the fiducial FP has $rms = 0.104
\pm 0.001$\,dex (\reffig{f.r.fpbench}). Despite appearances, these values are
consistent. First, the mock plane has significantly steeper slope, which amplifies
the ratio between the orthogonal scatter $\sigma_\perp$ and the scatter along
\logl; secondly, the measurement uncertainty in \logl is significantly smaller
than the uncertainty in \Lumsynth, which contains a
large contribution from the uncertainty in \logups 
\citep[typically 0.05--0.1\,dex;][]{gallazzi+bell2009}.
In fact, looking at the \textit{intrinsic orthogonal} scatter (which does not
suffer from amplification due to different tilts between the fiducial FP and the
mock plane), we find $\sigma_\perp = 0.042 \pm 0.003$\,dex, smaller than, but
comparable to, the scatter of the
fiducial FP ($\sigma_\perp = 0.048 \pm 0.002$\,dex). Subtracting
these numbers in quadrature, the `missing' scatter between the observed and
mock FP is $0.023 \pm 0.004$\,dex. This implies that, provided our
measurement uncertainties are correctly estimated, \emph{most ($\approx$75\%)
of the FP scatter is explained by stellar mass-to-light variations at fixed
\sige and \re}.

For the FP tilt, we find opposite results in the directions of \logsige and
\logre. We recall that the tilt of the fiducial FP with respect to the virial plane
is $0.706$ along the direction of \logsige and $0.088$ along the direction of
\logre (\S~\ref{s.r.ss.galev}). For \logsige, the best coefficients of the mock
FPs are $a = 1.720\pm0.051$ and $a = 1.756\pm0.050$, intermediate between the
virial coefficient $a = 2$ and the fiducial coefficient $a=1.294 \pm 0.039$.
Along \logsige, the systematic variation of \ups with \sige and \re tilts the
virial plane by $0.280$ (panel~\subref{f.r.mock.a}) or $0.244$
(panel~\subref{f.r.mock.b}); even though stellar \ups rotates the virial plane
in the right direction (i.e.\ closer to the observed FP), the magnitude of this
effect is insufficient to account for the observed tilt ($\approx$35--40\%).
In contrast, in the direction of \logre, we find
${b=1.198\pm0.033}$ and {$b=1.196\pm0.033$}, larger than both the fiducial
coefficient $b=0.912\pm0.025$ and the virial prediction $b = 1$; thus the
SSP relations on their own tilt the mock FP \textit{away} from the observed FP.

\begin{figure}
  \includegraphics[type=pdf,ext=.pdf,read=.pdf,width=1.0\columnwidth]{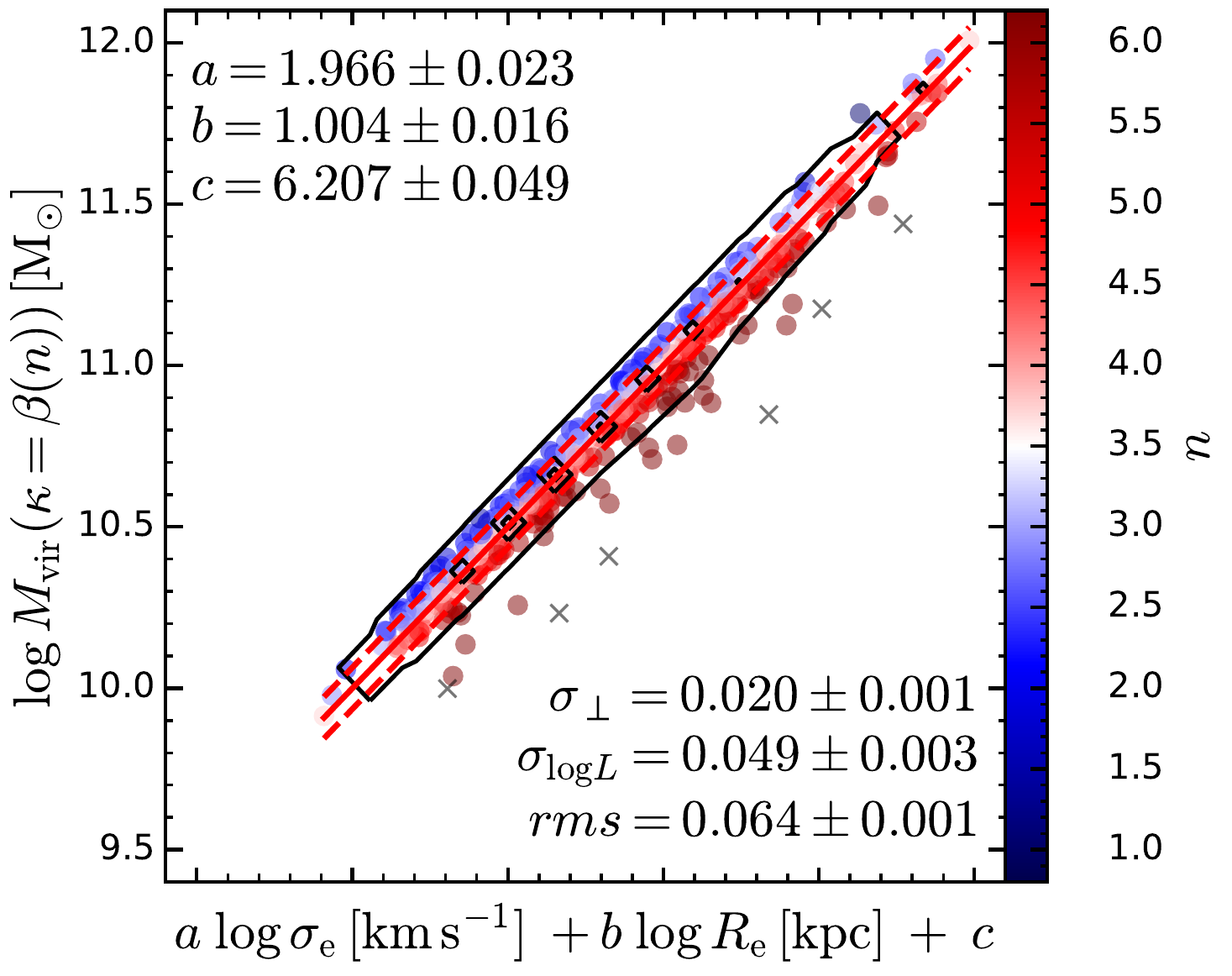}
  \caption{Edge-on view of the virial mass plane, showing the tilt and
  scatter due to the range of S{\'e}rsic index $n$ at fixed \sige and \re.
  Both the tilt and scatter are less than that caused by stellar-population
  variations (cf.\ Fig~\ref{f.r.mock.a}).
  }\label{f.r.misstilt}
\end{figure}

\subsubsection{`Missing' tilt and scatter}\label{s.r.ss.mock.sss.misstilt}

We now address the question about the `missing' tilt and scatter, i.e.\ the
difference between the fiducial and mock FPs. To do so, we consider
the ratio of the dynamical mass to the mass derived from the product of
the luminosity and the mass-to-light ratio
\begin{equation}\label{eq.r.misstilt}
    \frac{M_\mathrm{vir}(\kappa=5)}{L \, \Upsilon_\star} =
    \frac{5}{\kappa} \cdot \frac{\Upsilon_\mathrm{IMF}}{\Upsilon_\star} \cdot \frac{1}{f_\star} =
    \frac{L^{\Upsilon_\star(\mathrm{measured})}_\mathrm{synth}}{L}
\end{equation}
where the equality is derived from equations~\ref{eq.i.mvir} \&~\ref{eq.i.logl}
and $L^{\Upsilon_\star(\mathrm{measured})}_\mathrm{synth}$ is the quantity used
in the mock FP, Fig.~\ref{f.r.mock.a}.
Note that this mass ratio can also be viewed as the ratio of two mass-to-light 
ratios, one obtained from the observed virial mass and the observed luminosity 
and the other from the stellar population model; dividing 
$M_\mathrm{vir}/L$ by \ups removes the effect of measured SSP trends. 

The expression on the right of equation~(\ref{eq.r.misstilt}) can be
calculated from the observed values of \sige, \re, $L$ and \ups.
Modelling this quantity as a 3-d Gaussian as a function of \sige and \re, we find
the coefficients to be
\begin{equation}\label{eq.r.resid.tilt}
\begin{split}
    \log \frac{\Upsilon_{\mathrm{IMF}}/\Upsilon_{\star}}{\kappa f_\star} &
    = (0.561 \pm 0.076) \log \sigma_\mathrm{e}  \\
    + & (0.310 \pm 0.041) \log R_\mathrm{e} - (1.082 \pm 0.150)
\end{split}
\end{equation}
The ratio between the coefficients of \logre and \logsige is $b/a = 0.55\pm0.14$,
consistent with the value expected from a correlation with dynamical mass
\citep[in agreement with the findings of][]{graves+faber2010}. A correlation with
\sige only is unlikely ($P<10^{-4}$, with $b/a$ almost 4 standard
deviations from 0). Equation~\ref{eq.r.resid.tilt} also highlights
that \emph{both} \sige and \re play a role in the structural and/or IMF trends
\emph{along} the FP. Note, however, that the stellar mass fraction
varies significantly with radius, so an additional structural
difference could be due to the overall stellar mass fraction.

To isolate the contribution of structural non-homology to the FP tilt and
scatter, we study $M_\mathrm{vir}(\kappa=\beta(n))$, where
\begin{equation}\label{eq.r.kappan}
    \beta(n) = 8.87 - 0.831 \, n + 0.0241 \, n^2
\end{equation}
\citepalias[][their equation~20]{cappellari+2006}. We use
MGE \re with S{\'e}rsic-based $\kappa = \beta(n)$ precisely to isolate the
effect of non-homology, as captured by S{\'e}rsic index $n$: for
S{\'e}rsic mass (and light) profiles, using
$\beta(n)$ yields accurate dynamical masses based on
the S{\'e}rsic virial estimator, so using S{\'e}rsic \re in the definition of
$M_\mathrm{vir}(\kappa=\beta(n))$ would remove the effect of non-homology from
the FP tilt.

This model of the mass plane gives
\begin{equation}\label{eq.r.sertilt}
\begin{split}
    \log M_\mathrm{vir}(\kappa&=\beta(n)) = (1.966 \pm 0.023) \log \sigma_\mathrm{e} \\
    & + (1.004 \pm 0.016) \log R_\mathrm{e} + (6.207 \pm 0.049)
\end{split}
\end{equation}
with a scatter $rms = 0.064\pm0.001$~dex (see Fig.~\ref{f.r.misstilt}).
At face value, the
intrinsic scatter due to $n$-based non-homology is only $\sigma_\perp = 0.020
\pm 0.001$~dex, much smaller than we inferred for the SSP-induced scatter
($\sigma_\perp = 0.042$\,dex, Fig.~\ref{f.r.mock.a}), and contributing only 17\%
of the fiducial FP scatter ($\sigma_\perp = 0.048\pm0.002$\,dex,
Fig.~\ref{f.r.mock.a}). Intriguingly, this scatter is of the order of the
`missing' scatter after considering the contribution of stellar populations
($\sigma = 0.023 \pm 0.004$\, dex, \S~\ref{s.r.ss.mock.sss.mockfit}).
Clearly, this scatter is degenerate with
the measurement uncertainties on $\beta(n)$: if we over-estimated these
uncertainties, the contribution of $n$ to the FP scatter would be under-estimated.
The issue, however, is that even the observed $rms$ is not very large to begin with;
for the true intrinsic scatter in Fig.~\ref{f.r.misstilt} to be the same as we
infer for SSP variations, the observed $rms$ in Fig.~\ref{f.r.misstilt} must
be $\approx$0.1\,dex. To make matters
worse, we assumed uncertainties of order 6\% on $n$, but this value is likely
under-estimated for high-index galaxies (\S~\ref{s.ds.ss.ancillary}), i.e.\ for
precisely the same galaxies that dominate the scatter in
Fig.~\ref{eq.r.sertilt}. Therefore we conclude that non-homology, as measured by
S{\'e}rsic index, does not contribute significantly to the FP scatter, with a
conservative estimate of $\approx$ 20\%. Simply put: the variation of $n$ at
fixed \sige and \re is too little to cause significant scatter in the FP.

As for the tilt, we also find a very small deviation between the fit to the
model in Fig.~\ref{f.r.misstilt} and the virial plane.
Alternatively, if the measurement uncertainties on $n$ were severely
underestimated, the effect of systematic trends of $n$ with \sige and
\re could be damped, thereby masking the effect of non-homology in
Fig.~\ref{f.r.misstilt}. In this case, however, the $rms$ and the induced
scatter on the FP would be over-estimated. Note that there is a
degeneracy between the tilt and scatter: if our uncertainties were severely
under-estimated, then the effect of $n$ on the FP tilt would also be
under-estimated, but the effect on the FP scatter would be over-estimated.
One way to reconcile large scatter in both non-homology and
\ups is for $n$ and \ups to be correlated at fixed \sige and, possibly, \re.
Our data, however, offers no evidence of such a correlation.

Finally, we remark that, after considering both SSP variations and non-homology,
there is no space left for \emph{independent} scatter in $f_\star$ and
$\Upsilon_\mathrm{IMF}/\Upsilon_\star$, thus requiring that these two properties
(anti)correlate with \ups and/or $n$ and/or each other.

\section{Discussion}\label{s.d}

\subsection{The SAMI fiducial Fundamental Plane}\label{s.d.ss.benchfp}

In \S~\ref{s.r.ss.galev} we presented the fiducial FP for a volume- and
luminosity-limited sample ($z \leq 0.065$ and $L_r \geq 10^{9.7} \, \mathrm{L}_{r,\odot}$)
of early-type galaxies drawn from the SAMI Survey.
The best-fit values of the FP coefficients ($a=1.294\pm0.039$ and
$b=0.912\pm0.025$) are only in marginal agreement with the results from ATLAS$^{\rm 3D}$
\citepalias[$b = 1.249 \pm 0.044$ and $c=0.964\pm0.03$][top panel of their fig.12;
notice their different definition of the FP coefficients: our $(a, b, c)$
correspond to their $(b, c, a)$]{cappellari+2013a}.
We repeat our analysis with a volume- and mass-limited sample with
$M_\star \geq 6 \times 10^9 \, \mathrm{M_\odot}$ and $z < 0.05$
\citep[roughly equivalent to the
ATLAS$^{\rm 3D}$ selection criterion
$M_K < -21$\,mag;][]{cappellari+2011a}.
With this sample, and using the same algorithm as \citetalias{cappellari+2013a},
we find $a = 1.260 \pm 0.048$ and $b=0.931 \pm 0.032$, and an observed \emph{rms}
of $0.098 \pm 0.005$\,dex (ATLAS$^{\rm 3D}$ has 0.1\,dex).
Our fiducial FP is thus in excellent agreement with the ATLAS$^{\rm 3D}$ 
FP once we account for the different algorithm and sample selection.

When using \sige, the SAMI Pilot Survey found $\alpha = 0.79 \pm 0.07$ and
$\beta = 0.96 \pm 0.05$ \citep[][their Table~2, where they define
$\alpha \equiv a$ and $\beta \equiv b$]{scott+2015}.
These best-fit values differ from ours, but the
sample selection criteria also differ. We repeat our FP fit using
the same algorithm as \citet{scott+2015} and with an equivalent sample selection
(i.e.\ $\log L > 10^{10.2} \, \mathrm{L}_{\odot,r}$, only considering galaxies in
the clusters Abell~85, Abell~168 and Abell~2399), and find $a = 0.85 \pm 0.11$
and $b = 0.92 \pm 0.08$, in agreement with \citet{scott+2015}; our larger uncertainties are
estimated from bootstrapping the sample one hundred times.

From these comparisons, we conclude that the best-fit FP
coefficients depend not only on the model used
(cf.\ \S~\ref{s.r.ss.galev.sss.outl}), but
also on the properties of the sample considered.
These dependencies likely arise from inadequate models. For example, it is
well-known that the galaxy luminosity function is well-fit by a Schechter function,
and that the FP intrinsic scatter decreases with luminosity
\citep{hyde+bernardi2009}.
In addition, the plane model itself might only be an approximation: there is
evidence that the logarithm of the dynamical mass-to-light ratio is a quadratic
function of \logsige \citep{zaritsky+2006, wolf+2010}. By inserting this
quadratic relation in the virial equation, \citet{zaritsky+2006} obtain a
slanted parabolic cylinder (the `fundamental manifold of spheroids'). This
surface is well approximated by a plane, given our range in \sige and our
measurement uncertainties \citep{scott+2015}.
However, different samples give
different weights to their region on the manifold, and thus may yield
different plane approximations. Moreover, if we consider the FP as the projection of a higher dimensional
hyperplane involving age, then it is not surprising that different samples,
which have in general different age distributions, might give rise to
different projections.

It is worth clarifying that the trends we report between the FP residuals and
\sige and \re are not evidence for the non-linear nature of the galaxy manifold.
First, the residuals between a non-linear manifold and a linear model would
be non-linear, but we see no evidence of non-linearity in
Figs~\ref{f.galev.resid.nnl.a}--\subref{f.galev.resid.nnl.c}. Secondly, the trend
between \deltalogl and \sige is opposite to
expectations: if one recasts the expression of the galaxy manifold from
\citet{zaritsky+2006} in terms of \logl, the result is a concave function of
\logsige. Comparing this function to its linear approximation for the range of
\sige of our sample, we obtain a positive trend between the residuals and
\logsige, contrary to our findings. In order to study the non-linearity of the
FP, it would be best to extend the baseline in \sige, including a significant fraction
of quiescent dwarf galaxies \citetext{$\sigma \lesssim 30 \, \mathrm{km \,
s^{-1}}$, which might prove challenging without using adequate spectral
resolution, e.g.\ \citealp{barat+2019, barat+2020}, \citealp{scott+2020},
Eftekhari in~prep.}.

\subsection{Residual trends with structural and stellar-population parameters}\label{s.d.ss.resid}

In \S~\ref{s.r.ss.str} and \S~\ref{s.r.ss.ssp} we have studied the relation
between the residuals of the FP and various structural and stellar-population
observables. We concluded that SSP age is the strongest driver of the intrinsic
FP scatter, with a significance greater than eight standard deviations. This
result is independent of sample selection criteria: it persists (and becomes
stronger) if we drop our volume-limited requirement, if we include early-type
spirals, and if we select red galaxies regardless of their visual morphology.
It persists (and becomes weaker) if we select only elliptical galaxies, although the
sample size is smaller than the ETG sample.

The trend between \deltalogl and age also exists at fixed \sige, so it is
not a consequence of the correlation between age and \sige or the
anticorrelation between \sige and \deltalogl. Moreover, the trend persists and
remains the most significant even if we swap the 3dG algorithm for the LTS
algorithm, which gives no correlation between \deltalogl and \sige
(Appendix~\ref{a.s.lts}, Figs~\ref{f.a.fpbench} and \ref{f.a.galev.resid.ssp}).
The trend disappears only for the oldest galaxies, probably because of a
combination of physical and practical reasons: for old SSP ages, the effect of
age on \ups flattens out and the ages themselves have large uncertainties.

The existence of a trend between the residuals of the FP and SSP age has long
been known \citep{forbes+1998}. Here we demonstrate that this trend is the most
significant physical trend, regardless of a number of assumptions about the sample, outliers,
aperture size, photometry, uncertainties and optimisation. In addition, we
connect the trend between age and \deltalogl to the empirical relation between
surface mass density and age: among the observables based on galaxy mass and
size, surface mass density is known to be the best predictor of both SSP age
\citepalias{barone+2018} and light-weighted, full-spectral-fitting age
\citep{barone+2020}. Here we show the logical consequence, that surface
mass density is also the best predictor of SSP mass-to-light ratio \ups
(\S\ref{s.r.ss.ml}). Given that the FP residuals strongly
correlate with SSP age (and \ups) but do not correlate with surface mass
density, we conclude that: (i)~the FP fully captures the mean age (and \ups)
variation with $\Sigma_\mathrm{vir}$ and (ii)~the scatter about the
$\Sigma_\mathrm{vir}$--age relation is propagated to the FP. 
In particular, we argue that most ($\approx$75\%) of the FP scatter is due to
the broad distribution in SSP age at fixed $\Sigma_\mathrm{vir}$ (we find
an $rms$ of 0.17~dex; for \ups, the $rms$ in 0.14~dex; \reffig{f.r.ml.c}). The
large, size-dependent scatter
of age as a function of stellar mass is already in place at $z \approx 1$, and
is thought to arise from different evolutionary paths to quiescence leaving
different structural signatures on the structure of galaxies \citep{wu+2018,
deugenio+2020}.
Alternatively, the relation could be a consequence of surface mass density being
related to the cosmic epoch when the galaxy became quiescent (van der Wel et~al., in~prep.; Barone et~al, in~prep.).

\begin{figure}
  \includegraphics[type=pdf,ext=.pdf,read=.pdf,width=1.0\columnwidth]{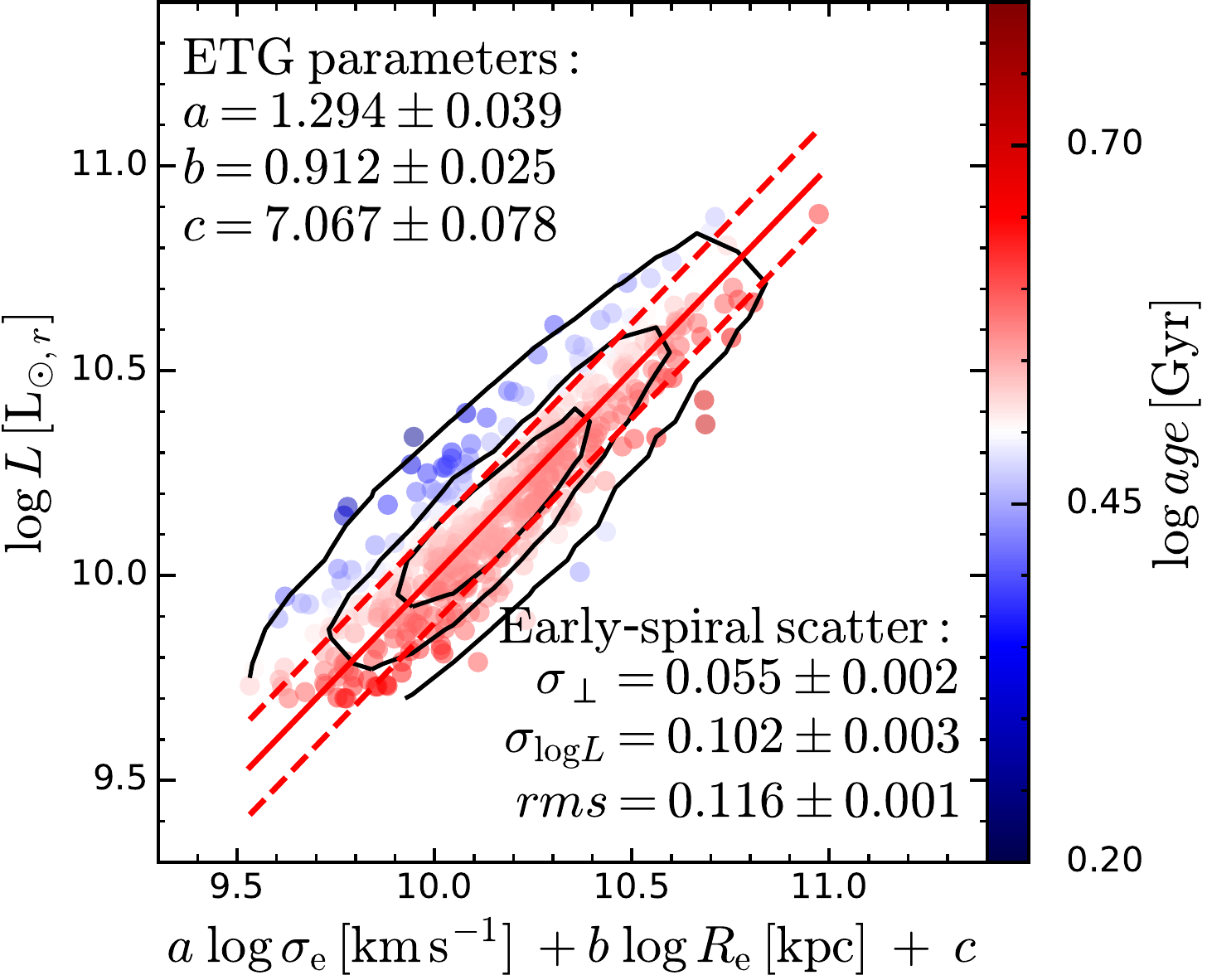}
  \caption{Location of the SAMI early spirals relative to the fiducial FP of
  ETGs. Circles represent early spirals ($1 < \mathrm{mtype} <= 2$,
  \S~\ref{s.ds.ss.ancillary}), colour-coded with LOESS-smoothed stellar age
  (the colours are mapped to the interval enclosing $\approx 68\%$ of the
  stellar age histogram).
  The red lines represent the fiducial FP of ETGs, with
  the best-fit parameters reported in the top left corner. The distribution
  of ETGs is represented by the solid black contours, enclosing the
  90\textsuperscript{th}, 67\textsuperscript{th} and 30\textsuperscript{th}
  percentiles of the ETG sample. The scatter reported in the bottom-right
  corner is the scatter of early spirals with respect to the fiducial FP.
  Early spirals fall very close to the fiducial FP (the median offset is
  just 0.03\,dex) and maintain the trend of increasing age across
  the FP.
  }\label{f.d.fpbench.esp}
\end{figure}

If the trend across the plane is due to stellar age, where do still younger
galaxies lie, relative to the fiducial FP? In \reffig{f.d.fpbench.esp} we
overlay 512 SAMI early spirals on the fiducial ETG FP. Each galaxy
is colour-coded according to its LOESS-smoothed age; the colours are different
from \reffig{f.r.fpbench}, as here they are mapped to the interval
$\mu(age)\pm\sigma(age)$, where $\mu$ and $\sigma$ are the median and
standard deviation of the early-spiral age distribution). Notice that, despite
their different properties (including ongoing star formation and, presumably,
larger dark matter fractions), early spirals lie remarkably close to the
fiducial FP. These galaxies have younger median ages compared to ETG sample
(3.5\,Gyr compared to 6.4\,Gyr), yet their offset from the
ETG FP is remarkably small (only $0.03 \, \mathrm{dex}$). This
surprising result
could be due to the competing effects of early spirals having lower median stellar
mass-to-light ratios but higher dark matter and/or dust fractions than ETGs, and
will be investigated in a future paper. However, it underscores the potential
for expanding to broader morphological samples the use of the FP and other
scaling relations as distance indicators \citep[e.g.\ ][]{barat+2019, barat+2020}.

We argued that the correlation between the FP residuals and SSP metallicity is
not physical, but rather an outcome of the age--metallicity degeneracy.
On the other hand, we found a significant correlation with SSP \aFe
(Fig.~\ref{f.galev.resid.ssp.c}), with evidence that this correlation also
holds at fixed SSP age. We checked that this correlation does not depend either
on global (cluster vs field/group) nor local environment (as measured by
the density of nearby galaxies), and so it is unlikely to be related to
the observed correlation between \aFe and cluster-centric distance \citep[
which is observed for galaxies that are $\sim$10 times less massive than
considered here,][]{liu+2016}. A possible explanation is offered by
downsizing, i.e.\ the oldest galaxies forming most rapidly \citep{cowie+1996,
thomas+2005, graves+2007}. Downsizing implies a positive correlation between age
and \aFe, as higher \aFe corresponds to shorter duration of star formation.
Because the age of the oldest stellar populations is difficult to measure, \aFe
may contain additional (or at least independent)
age information that is not captured by the indicators used to measure SSP age
itself. Another possibility is that \aFe could be related to the distribution 
of matter inside 1\re.

An important caveat is that all our stellar-population properties are
luminosity-weighted SSP equivalents, whereas real stellar populations have
a finite distribution of ages, metallicities and abundances, reflecting
complex star-formation histories.

Among the structural parameters, the FP residuals correlate most strongly with
S{\'e}rsic index $n$ and galaxy projected shape $\epsilon_\mathrm{e}$, but neither is as
significant as the trends with SSP properties.
For $\epsilon_\mathrm{e}$, the correlation probably results from the superposition of
three effects. The first effect is that high-$\epsilon_\mathrm{e}$ galaxies are oblate discs
observed close to edge-on and are therefore expected to be rotation-supported
and to have higher observed \sige than their face-on analogues. The second effect is that
if any dust  was present, it would increase the observed
mass-to-light ratio. These two effects would
make these galaxies under-luminous at their location on the FP, inducing a
negative trend between the FP residuals and $\epsilon_\mathrm{e}$, the opposite of what is seen. On the other hand, a third effect is that 
these disc-like ETGs have younger SSP ages than
spheroidal ETGs \citep[e.g.][]{vandesande+2018}, but the latter cannot project
to high $\epsilon_\mathrm{e}$, thus creating an anticorrelation between observed
$\epsilon_\mathrm{e}$ and age that makes flattened ETGs younger and therefore
more luminous than average at their location on the FP. The net effect of these
three effects in our sample is a weak
positive correlation with apparent shape; however, different samples, with a different
ratio of slow- to fast-rotators, may have different residual trends
\citep[e.g.][]{bernardi+2020}. This hypothesis can be tested using dynamical
models to infer the intrinsic shape.

Our results suggest that structural trends due to non-homology are not an
important driver of the FP scatter (Figs~\ref{f.r.fpsersic} and
\ref{f.galev.resid.str.c}), which is at variance with some other works
\citep{prugniel+simien1997, desmond+wechsler2017}, but in agreement with the
results from both dynamical modelling \citepalias{cappellari+2006} and strong
lensing \citep{bolton+2007, bolton+2008, koopmans+2009}, as well as from
stability considerations \citep[e.g.][]{nipoti+2002}.
It is important to recall, however, that although it seems unlikely, it is
still possible that large measurement uncertainties on the structural
observables prevent us from observing stronger trends with the FP residuals.
More importantly, given the strong dependence of the FP parameters on the
properties of the sample, we cannot exclude that structural trends would be
present with a different sample, for example by including a larger proportion
of massive galaxies than in our volume-limited sample. The fact that
non-homology has little impact on the FP scatter has no bearing to the FP tilt.

\subsection{Fundamental Plane tilt}\label{s.d.ss.tilt}

We have seen that the relation between the stellar-population \ups and
$\Sigma_\mathrm{vir}$ (equations~\ref{eq.r.upssigmadyn} and \ref{eq.r.upsplane})
accounts for approximately
$50\%$ of the FP tilt in \logsige. However, in \logre, stellar-population
effects tilt the FP \textit{away} from the observed plane. This result is a
direct consequence of the fact that SSP age and \ups follow \Sigvir, which is
a decreasing function of \re ($\Sigma_\mathrm{vir} \propto \sigma_\mathrm{e}^2 /
R_\mathrm{e}$). Given that non-homology (as captured by S{\'e}rsic index $n$)
seems to have little effect on the FP tilt (\S~\ref{s.r.ss.mock.sss.misstilt}),
after accounting for stellar-population effects the remaining FP tilt in
\logsige ($50\%$) and \logre ($250\%$) must be due to a combination of
systematic variations in the stellar-to-total mass fraction ($f_\star$) and/or
the IMF shape relative to our Chabrier assumption
(parameterised as $\Upsilon_{\mathrm{IMF}}/ \Upsilon_{\star}$).

There is considerable evidence in the literature that SSP variations account for
roughly half of the FP tilt \citep[e.g.][]{pahre+1995, pahre+1998,
prugniel+simien1997, gallazzi+2005, hyde+bernardi2009}. \citet{graves+faber2010} also reached this
conclusion, and added that the FP tilt is due to two contributions: SSP trends
tilt the FP only in the \logsige direction,
whereas dark matter/IMF trends tilt the FP along a direction that is proportional
to $\log M_\mathrm{vir}$. In contrast with their findings, we conclude that
both the measured SSP trends and the `missing contribution' tilt the FP along
both \logsige and \logre. The origin of this disagreement is difficult to
disentangle, because their result is based on stacking individual single-fibre
galaxy spectra and they use a different sample and optimisation, so a fair
comparison is particularly challenging.

To further constrain the terms $f_\star$ and
$\Upsilon_{\mathrm{IMF}}/ \Upsilon_{\star}$ in the FP equation (\ref{eq.i.logl})
requires dynamical models, a subject for future work.

\subsection{Fundamental Plane scatter}\label{s.d.ss.scatter}

The observed scatter about the fiducial FP is $0.104\pm0.001$\,dex.
With our estimate of the observational uncertainties, the intrinsic scatter is
$\sigma_\perp = 0.048 \pm 0.002$\,dex.
Given the observed correlation between the FP residuals and stellar population
age, we expect part of this scatter to contain information about the stellar
population age, an expectation that is confirmed by the strong correlation
between the FP residuals and the residuals of the \ups--\Sigvir
relation.

In \S~\ref{s.r.ss.mock.sss.mockfit} we used stellar population \ups to create a
mock FP, and find an intrinsic scatter $\sigma_\mathrm{z} = 0.042\pm0.003$
dex.
Thus, with our estimate of the observational uncertainties,
stellar population trends seem to account for most ($\approx$ 75\%) of the FP
scatter (in quadrature).

On the other hand, non-homology (as parameterised by S{\'e}rsic index $n$) seems
to account for $\approx$ 20\% of the FP scatter (\reffig{f.r.misstilt}); at
fixed \sige and \re, the range in $n$ is not broad enough to significantly
affect the FP.

Together, \ups and $n$ account for most of the FP scatter, leaving little space
for any scatter in the rest of the FP equation (\ref{eq.i.logl}), i.e.\ dark-matter
fraction $1-f_\star$ and IMF trends  $\Upsilon_{\mathrm{IMF}}/ \Upsilon_{\star}$.
This may point either to the under-estimation of the uncertainties on \ups and/or $n$
or to the existence of physical correlations between these galaxy properties
that reduce their impact on the FP scatter.

Dust could in principle reduce the FP scatter. At fixed \sige and \re, the
youngest galaxies have the lowest \ups and so the largest positive
deviations above the plane. If these galaxies also had the largest dust
fractions, dust attenuation would reduce the deviation and hence the scatter.
In practice this effect amounts to $rms = 0.01$--0.02\,dex
and is therefore negligible for the sample considered here
(Appendix~\ref{a.s.dust}). 

If SSP trends contribute a non-negligible fraction of the FP intrinsic scatter,
it should then be possible to use information encoded in SSP properties to
reduce this scatter, provided that the uncertainties in the SSP measurements
do not overwhelm the potential gains (as is presently the case with SSP age).

\section{Summary and conclusions}\label{s.c}

In this work, we used a volume- and mass-limited sample of
morphologically-selected early-type galaxies from the SAMI Galaxy Survey to
study the impact of systematic trends of stellar population parameters and
non-homology on the tilt and scatter of the Fundamental Plane (FP).
Our key advantage is the combination of high-quality aperture spectra with a
carefully selected sample and sophisticated analysis techniques, which yields 
both accurate and representative results.

Once the sample selection criteria are taken into account, the parameters of our
fiducial FP (\reftab{t.r.bestfit}) are consistent with previous works. We find
that:\vspace*{-6pt}

\begin{enumerate}
  \item The FP scatter is dominated by stellar population effects: among the
  structural properties considered, only S{\'e}rsic index anticorrelates with
  the FP residuals (four standard deviations, \S~\ref{s.r.ss.str} and 
  Table~\ref{t.r.resid}). In contrast, stellar age, $r-$band mass-to-light ratio
  and $\alpha$-element abundance all have statistically significant
  anticorrelations with the FP residuals (eight standard deviations,
  \S~\ref{s.r.ss.ssp} and Table~\ref{t.r.resid}).
  \item Our results are qualitatively unchanged if we alter a range of
  assumptions: sample selection (volume- and mass-limited vs no limits;
  ellipticals only vs early-types), outliers rejection, aperture size and shape,
  adopted photometry, measurement uncertainties and correlated noise, or model
  and algorithm used (3-d Gaussian vs least-trimmed squares, 
  Table~\ref{t.r.bestfit}).
  \item In agreement with previous works, we find that the relation between
  stellar mass-to-light ratio \ups and surface mass density \Sigvir is tighter
  than the relation between \ups and either virial mass or gravitational
  potential (\ups--\Sigvir relation, \S~\ref{s.r.ss.ml}).
  \item We find a strong anticorrelation between the residuals of the
  \ups--\Sigvir relation and the FP residuals. The strong correlation
  between the FP residuals and \ups (and stellar age) is due to the large
  intrinsic scatter in the \ups--\Sigvir (and age--\Sigvir) relations
  (\S~\ref{s.r.ss.resres}).
  \item For fixed IMF, stellar-population relations account for approximately 75\% of
  the FP scatter, whereas non-homology accounts for approximately 20\%
  (\S~\ref{s.r.ss.mock}).
  \item For fixed IMF, stellar-population relations 
  do not fully explain the FP tilt. In fact, in the direction of \logre, the
  stellar-population relations tilt the virial plane in the wrong direction with respect to
  the observed FP (\S~\ref{s.r.ss.mock}). Given the properties of our sample,
  non-homology appears to have a negligible effect on the FP tilt.
  \item The remaining FP tilt not explained by stellar populations or
  non-homology is presumably due to varying dark-matter fractions and
  systematic trends in IMF shape; this tilt is roughly proportional to virial mass.
\end{enumerate}

\section*{Acknowledgements}

We thank the referee for their constructive report, which improved the quality
of this manuscript. We also thank Michele Cappellari for his comments and suggestions
on the fitting algorithms and Elena Dalla Bont{\`a} for her suggestions.
FDE acknowledges funding through the H2020 ERC Consolidator Grant 683184, and
by the Australian Research Council Centre of Excellence for All-sky Astrophysics
(CAASTRO; grant CE110001020).
NS acknowledges support of an Australian Research Council Discovery Early Career
Research Award (project number DE190100375) funded by the Australian Government.
This research was supported by the Australian Research Council Centre of Excellence for All-sky Astrophysics (CAASTRO), through project number CE110001020, and by the Australian Research Council Centre of Excellence for All-sky Astrophysics in 3 Dimensions (ASTRO 3D), through project number CE170100013. 
RLD acknowledges travel and computer grants from Christ Church, Oxford and support from the Oxford Hintze Centre for Astrophysical Surveys which is funded by the Hintze Family Charitable Foundation.
RLD was also supported by the Science \& Technology Facilities Council grant numbers ST/H002456/1, ST/K00106X/1 and ST/J 002216/1.
JvdS acknowledges support of an Australian Research Council Discovery Early Career Research Award (project number DE200100461) funded by the Australian Government. 
SB acknowledges funding support from the Australian Research Council through a Future Fellowship (FT140101166).
JBH is supported by an ARC Laureate Fellowship that funds Jesse van de Sande and an ARC Federation Fellowship that funded the SAMI prototype. 
JJB acknowledges support of an Australian Research Council Future Fellowship (FT180100231).
MSO acknowledges the funding support from the Australian Research Council through a Future Fellowship (FT140100255).

The SAMI Galaxy Survey is based on observations made at the Anglo-Australian
Telescope. The Sydney-AAO Multi-object Integral-field spectrograph (SAMI) was
developed jointly by the University of Sydney and the Australian Astronomical
Observatory, and funded by ARC grants FF0776384 (Bland-Hawthorn) and
LE130100198. The SAMI input catalog is based on data taken from the Sloan
Digital Sky Survey, the GAMA Survey and the VST ATLAS Survey. The SAMI Galaxy
Survey is funded by the Australian Research Council Centre of Excellence for
All-sky Astrophysics (CAASTRO), through project number CE110001020, and other
participating institutions. The SAMI Galaxy Survey website is
http://sami-survey.org/ .

Funding for SDSS-III has been provided by the Alfred P. Sloan Foundation, the
Participating Institutions, the National Science Foundation, and the U.S.
Department of Energy Office of Science. The SDSS-III web site is
http://www.sdss3.org/ .

GAMA is a joint European-Australasian project based around a spectroscopic
campaign using the Anglo-Australian Telescope. The GAMA input catalogue is
based on data taken from the Sloan Digital Sky Survey and the UKIRT Infrared
Deep Sky Survey. Complementary imaging of the GAMA regions is being obtained
by a number of independent survey programmes including GALEX MIS, VST KiDS,
VISTA VIKING, WISE, Herschel-ATLAS, GMRT and ASKAP providing UV to radio
coverage. GAMA is funded by the STFC (UK), the ARC (Australia), the AAO, and
the participating institutions. The GAMA website is http://www.gama-survey.org/ .

Based on observations made with ESO Telescopes at the La Silla Paranal Observatory
under programme ID 179.A-2004.

This work made extensive use of the freely available
\href{http://www.debian.org}{Debian GNU/Linux} operative system. We used the
\href{http://www.python.org}{Python} programming language
\citep{vanrossum1995}, maintained and distributed by the Python Software
Foundation. We further acknowledge the use of
{\sc \href{https://pypy.org/project/scipy/}{numpy}} \citep{harris+2020},
{\sc \href{https://pypy.org/project/scipy/}{scipy}} \citep{jones+2001},
{\sc \href{https://pypy.org/project/matplotlib/}{matplotlib}} \citep{hunter2007},
{\sc \href{https://pypy.org/project/emcee/}{emcee}} \citep{foreman-mackey+2013},
{\sc \href{https://pypy.org/project/corner/}{corner}} \citep{foreman-mackey2016},
{\sc \href{https://pypy.org/project/astropy/}{astropy}} \citep{astropyco+2013},
{\sc \href{https://pypy.org/project/pathos/}{pathos}} \citep{mckerns+2011},
{\sc \href{https://pypi.org/project/mgefit/}{mgefit}} \citet{cappellari2002},
{\sc \href{https://pypi.org/project/ltsfit/}{ltsfit}} \citepalias{cappellari+2013a},
{\sc \href{https://pypi.org/project/loess/}{loess}} \citep{cappellari+2013b},
{\sc \href{https://pypi.org/project/dynesty/}{dynesty}} \citep{speagle2020}
and {\sc \href{https://www.astromatic.net/software/sextractor}{SExtractor}}
\citep{bertin+arnouts1996}.

During the preliminary analysis we have made extensive use of
{\sc \href{http://www.star.bris.ac.uk/~mbt/topcat/}{topcat}} \citep{taylor2005}.

\section*{Data availability}

The data used in this work is available in the public domain, through the
\href{https://docs.datacentral.org.au/sami}{SAMI Data Release 3} \citep{croom+2021}. Ancillary
data comes from the \href{http://gama-survey.org}{GAMA Data Release 3}
\citep{baldry+2018} and raw data is from
\href{https://classic.sdss.org/dr7/}{SDSS DR7} \citep{abazajian+2009},
\href{https://www.sdss3.org/dr9/}{SDSS DR9} \citep{ahn+2012} and
\href{http://casu.ast.cam.ac.uk/vstsp/imgquery/search}{VST}
\citep{shanks+2013, shanks+2015}.

Regularised MGE fits and the 3dG software can be obtained
\sendemail{francesco.deugenio@gmail.com}{SAMI FP: data request}{
contacting the corresponding author.}



\bibliographystyle{mnras}
\bibliography{samifp}

\begin{thebibliography}{}
\makeatletter
\relax
\def\mn@urlcharsother{\let\do\@makeother \do\$\do\&\do\#\do\^\do\_\do\%\do\~}
\def\mn@doi{\begingroup\mn@urlcharsother \@ifnextchar [ {\mn@doi@}
  {\mn@doi@[]}}
\def\mn@doi@[#1]#2{\def\@tempa{#1}\ifx\@tempa\@empty \href
  {http://dx.doi.org/#2} {doi:#2}\else \href {http://dx.doi.org/#2} {#1}\fi
  \endgroup}
\def\mn@eprint#1#2{\mn@eprint@#1:#2::\@nil}
\def\mn@eprint@arXiv#1{\href {http://arxiv.org/abs/#1} {{\tt arXiv:#1}}}
\def\mn@eprint@dblp#1{\href {http://dblp.uni-trier.de/rec/bibtex/#1.xml}
  {dblp:#1}}
\def\mn@eprint@#1:#2:#3:#4\@nil{\def\@tempa {#1}\def\@tempb {#2}\def\@tempc
  {#3}\ifx \@tempc \@empty \let \@tempc \@tempb \let \@tempb \@tempa \fi \ifx
  \@tempb \@empty \def\@tempb {arXiv}\fi \@ifundefined
  {mn@eprint@\@tempb}{\@tempb:\@tempc}{\expandafter \expandafter \csname
  mn@eprint@\@tempb\endcsname \expandafter{\@tempc}}}

\bibitem[\protect\citeauthoryear{{Abazajian} et~al.,}{{Abazajian}
  et~al.}{2009}]{abazajian+2009}
{Abazajian} K.~N.,  et~al., 2009, \mn@doi [\apjs]
  {10.1088/0067-0049/182/2/543}, \href
  {http://adsabs.harvard.edu/abs/2009ApJS..182..543A} {182, 543}

\bibitem[\protect\citeauthoryear{{Abolfathi} et~al.,}{{Abolfathi}
  et~al.}{2018}]{abolfathi+2018}
{Abolfathi} B.,  et~al., 2018, \mn@doi [\apjs] {10.3847/1538-4365/aa9e8a},
  \href {https://ui.adsabs.harvard.edu/abs/2018ApJS..235...42A} {235, 42}

\bibitem[\protect\citeauthoryear{{Ahn} et~al.,}{{Ahn} et~al.}{2012}]{ahn+2012}
{Ahn} C.~P.,  et~al., 2012, \mn@doi [\apjs] {10.1088/0067-0049/203/2/21}, \href
  {http://adsabs.harvard.edu/abs/2012ApJS..203...21A} {203, 21}

\bibitem[\protect\citeauthoryear{{Allen} et~al.,}{{Allen}
  et~al.}{2015}]{allen+2015}
{Allen} J.~T.,  et~al., 2015, \mn@doi [\mnras] {10.1093/mnras/stu2057}, \href
  {http://adsabs.harvard.edu/abs/2015MNRAS.446.1567A} {446, 1567}

\bibitem[\protect\citeauthoryear{{Astropy Collaboration} et~al.,}{{Astropy
  Collaboration} et~al.}{2013}]{astropyco+2013}
{Astropy Collaboration} et~al., 2013, \mn@doi [\aap]
  {10.1051/0004-6361/201322068}, \href
  {http://adsabs.harvard.edu/abs/2013A%26A...558A..33A} {558, A33}

\bibitem[\protect\citeauthoryear{{Baldry} et~al.,}{{Baldry}
  et~al.}{2012}]{baldry+2012}
{Baldry} I.~K.,  et~al., 2012, \mn@doi [\mnras]
  {10.1111/j.1365-2966.2012.20340.x}, \href
  {https://ui.adsabs.harvard.edu/abs/2012MNRAS.421..621B} {421, 621}

\bibitem[\protect\citeauthoryear{{Baldry} et~al.,}{{Baldry}
  et~al.}{2018}]{baldry+2018}
{Baldry} I.~K.,  et~al., 2018, \mn@doi [\mnras] {10.1093/mnras/stx3042}, \href
  {https://ui.adsabs.harvard.edu/abs/2018MNRAS.474.3875B} {474, 3875}

\bibitem[\protect\citeauthoryear{{Barat} et~al.,}{{Barat}
  et~al.}{2019}]{barat+2019}
{Barat} D.,  et~al., 2019, \mn@doi [\mnras] {10.1093/mnras/stz1439}, \href
  {https://ui.adsabs.harvard.edu/abs/2019MNRAS.487.2924B} {487, 2924}

\bibitem[\protect\citeauthoryear{{Barat}, {D'Eugenio}, {Colless}, {Sweet},
  {Groves}  \& {Cortese}}{{Barat} et~al.}{2020}]{barat+2020}
{Barat} D.,  {D'Eugenio} F.,  {Colless} M.,  {Sweet} S.~M.,  {Groves} B.,
  {Cortese} L.,  2020, \mn@doi [\mnras] {10.1093/mnras/staa2716}, \href
  {https://ui.adsabs.harvard.edu/abs/2020MNRAS.tmp.2556B} {}

\bibitem[\protect\citeauthoryear{{Barone} et~al.,}{{Barone}
  et~al.}{2018}]{barone+2018}
{Barone} T.~M.,  et~al., 2018, \mn@doi [\apj] {10.3847/1538-4357/aaaf6e}, \href
  {http://adsabs.harvard.edu/abs/2018ApJ...856...64B} {856, 64}

\bibitem[\protect\citeauthoryear{{Barone}, {D'Eugenio}, {Colless}  \&
  {Scott}}{{Barone} et~al.}{2020}]{barone+2020}
{Barone} T.~M.,  {D'Eugenio} F.,  {Colless} M.,   {Scott} N.,  2020, \mn@doi
  [\apj] {10.3847/1538-4357/ab9951}, \href
  {https://ui.adsabs.harvard.edu/abs/2020ApJ...898...62B} {898, 62}

\bibitem[\protect\citeauthoryear{{Bender}, {Burstein}  \& {Faber}}{{Bender}
  et~al.}{1992}]{bender+1992}
{Bender} R.,  {Burstein} D.,   {Faber} S.~M.,  1992, \mn@doi [\apj]
  {10.1086/171940}, \href {http://adsabs.harvard.edu/abs/1992ApJ...399..462B}
  {399, 462}

\bibitem[\protect\citeauthoryear{{Bernardi} et~al.,}{{Bernardi}
  et~al.}{2003}]{bernardi+2003}
{Bernardi} M.,  et~al., 2003, \mn@doi [\aj] {10.1086/367794}, \href
  {https://ui.adsabs.harvard.edu/abs/2003AJ....125.1866B} {125, 1866}

\bibitem[\protect\citeauthoryear{{Bernardi}, {Dom{\'\i}nguez S{\'a}nchez},
  {Margalef-Bentabol}, {Nikakhtar}  \& {Sheth}}{{Bernardi}
  et~al.}{2020}]{bernardi+2020}
{Bernardi} M.,  {Dom{\'\i}nguez S{\'a}nchez} H.,  {Margalef-Bentabol} B.,
  {Nikakhtar} F.,   {Sheth} R.~K.,  2020, \mn@doi [\mnras]
  {10.1093/mnras/staa1064}, \href
  {https://ui.adsabs.harvard.edu/abs/2020MNRAS.494.5148B} {494, 5148}

\bibitem[\protect\citeauthoryear{{Bertin}}{{Bertin}}{2011}]{bertin2011}
{Bertin} E.,  2011, in {Evans} I.~N.,  {Accomazzi} A.,  {Mink} D.~J.,   {Rots}
  A.~H.,  eds,  Astronomical Society of the Pacific Conference Series Vol. 442,
  Astronomical Data Analysis Software and Systems XX. p.~435

\bibitem[\protect\citeauthoryear{{Bertin} \& {Arnouts}}{{Bertin} \&
  {Arnouts}}{1996}]{bertin+arnouts1996}
{Bertin} E.,  {Arnouts} S.,  1996, \mn@doi [\aaps] {10.1051/aas:1996164}, \href
  {http://adsabs.harvard.edu/abs/1996A%26AS..117..393B} {117, 393}

\bibitem[\protect\citeauthoryear{{Bertin}, {Ciotti}  \& {Del
  Principe}}{{Bertin} et~al.}{2002}]{bertin+2002}
{Bertin} G.,  {Ciotti} L.,   {Del Principe} M.,  2002, \mn@doi [\aap]
  {10.1051/0004-6361:20020248}, \href
  {https://ui.adsabs.harvard.edu/abs/2002A&A...386..149B} {386, 149}

\bibitem[\protect\citeauthoryear{Beutler et~al.,}{Beutler
  et~al.}{2011}]{beutler+2011}
Beutler F.,  et~al., 2011, \mn@doi [Monthly Notices of the Royal Astronomical
  Society] {10.1111/j.1365-2966.2011.19250.x}, 416, 3017

\bibitem[\protect\citeauthoryear{{Bezanson}, {Franx}  \& {van
  Dokkum}}{{Bezanson} et~al.}{2015}]{bezanson+2015}
{Bezanson} R.,  {Franx} M.,   {van Dokkum} P.~G.,  2015, \mn@doi [\apj]
  {10.1088/0004-637X/799/2/148}, \href
  {https://ui.adsabs.harvard.edu/abs/2015ApJ...799..148B} {799, 148}

\bibitem[\protect\citeauthoryear{{Bland-Hawthorn} et~al.,}{{Bland-Hawthorn}
  et~al.}{2011}]{bland-hawthorn+2011}
{Bland-Hawthorn} J.,  et~al., 2011, \mn@doi [Optics Express]
  {10.1364/OE.19.002649}, \href
  {http://adsabs.harvard.edu/abs/2011OExpr..19.2649B} {19, 2649}

\bibitem[\protect\citeauthoryear{{Blanton} \& {Roweis}}{{Blanton} \&
  {Roweis}}{2007}]{blanton+roweis2007}
{Blanton} M.~R.,  {Roweis} S.,  2007, \mn@doi [\aj] {10.1086/510127}, \href
  {http://adsabs.harvard.edu/abs/2007AJ....133..734B} {133, 734}

\bibitem[\protect\citeauthoryear{{Bolton}, {Burles}, {Treu}, {Koopmans}  \&
  {Moustakas}}{{Bolton} et~al.}{2007}]{bolton+2007}
{Bolton} A.~S.,  {Burles} S.,  {Treu} T.,  {Koopmans} L.~V.~E.,   {Moustakas}
  L.~A.,  2007, \mn@doi [\apjl] {10.1086/521357}, \href
  {http://adsabs.harvard.edu/abs/2007ApJ...665L.105B} {665, L105}

\bibitem[\protect\citeauthoryear{{Bolton}, {Treu}, {Koopmans}, {Gavazzi},
  {Moustakas}, {Burles}, {Schlegel}  \& {Wayth}}{{Bolton}
  et~al.}{2008}]{bolton+2008}
{Bolton} A.~S.,  {Treu} T.,  {Koopmans} L.~V.~E.,  {Gavazzi} R.,  {Moustakas}
  L.~A.,  {Burles} S.,  {Schlegel} D.~J.,   {Wayth} R.,  2008, \mn@doi [\apj]
  {10.1086/589989}, \href {http://adsabs.harvard.edu/abs/2008ApJ...684..248B}
  {684, 248}

\bibitem[\protect\citeauthoryear{{Brough} et~al.,}{{Brough}
  et~al.}{2017}]{brough+2017}
{Brough} S.,  et~al., 2017, \mn@doi [\apj] {10.3847/1538-4357/aa7a11}, \href
  {https://ui.adsabs.harvard.edu/abs/2017ApJ...844...59B} {844, 59}

\bibitem[\protect\citeauthoryear{{Bryant}, {Bland-Hawthorn}, {Fogarty},
  {Lawrence}  \& {Croom}}{{Bryant} et~al.}{2014}]{bryant+2014}
{Bryant} J.~J.,  {Bland-Hawthorn} J.,  {Fogarty} L.~M.~R.,  {Lawrence} J.~S.,
  {Croom} S.~M.,  2014, \mn@doi [\mnras] {10.1093/mnras/stt2254}, \href
  {http://adsabs.harvard.edu/abs/2014MNRAS.438..869B} {438, 869}

\bibitem[\protect\citeauthoryear{{Bryant} et~al.,}{{Bryant}
  et~al.}{2015}]{bryant+2015}
{Bryant} J.~J.,  et~al., 2015, \mn@doi [\mnras] {10.1093/mnras/stu2635}, \href
  {http://adsabs.harvard.edu/abs/2015MNRAS.447.2857B} {447, 2857}

\bibitem[\protect\citeauthoryear{{Bundy} et~al.,}{{Bundy}
  et~al.}{2015}]{bundy+2015}
{Bundy} K.,  et~al., 2015, \mn@doi [\apj] {10.1088/0004-637X/798/1/7}, \href
  {http://adsabs.harvard.edu/abs/2015ApJ...798....7B} {798, 7}

\bibitem[\protect\citeauthoryear{{Cappellari}}{{Cappellari}}{2002}]{cappellari2002}
{Cappellari} M.,  2002, \mn@doi [\mnras] {10.1046/j.1365-8711.2002.05412.x},
  \href {http://adsabs.harvard.edu/abs/2002MNRAS.333..400C} {333, 400}

\bibitem[\protect\citeauthoryear{{Cappellari}}{{Cappellari}}{2016}]{cappellari2016}
{Cappellari} M.,  2016, \mn@doi [\araa] {10.1146/annurev-astro-082214-122432},
  \href {http://adsabs.harvard.edu/abs/2016ARA%26A..54..597C} {54, 597}

\bibitem[\protect\citeauthoryear{{Cappellari}}{{Cappellari}}{2017}]{cappellari2017}
{Cappellari} M.,  2017, \mn@doi [\mnras] {10.1093/mnras/stw3020}, \href
  {https://ui.adsabs.harvard.edu/abs/2017MNRAS.466..798C} {466, 798}

\bibitem[\protect\citeauthoryear{{Cappellari} \& {Emsellem}}{{Cappellari} \&
  {Emsellem}}{2004}]{cappellari+emsellem2004}
{Cappellari} M.,  {Emsellem} E.,  2004, \mn@doi [\pasp] {10.1086/381875}, \href
  {http://adsabs.harvard.edu/abs/2004PASP..116..138C} {116, 138}

\bibitem[\protect\citeauthoryear{{Cappellari} et~al.,}{{Cappellari}
  et~al.}{2006}]{cappellari+2006}
{Cappellari} M.,  et~al., 2006, \mn@doi [\mnras]
  {10.1111/j.1365-2966.2005.09981.x}, \href
  {http://adsabs.harvard.edu/abs/2006MNRAS.366.1126C} {366, 1126}

\bibitem[\protect\citeauthoryear{{Cappellari} et~al.,}{{Cappellari}
  et~al.}{2011}]{cappellari+2011a}
{Cappellari} M.,  et~al., 2011, \mn@doi [\mnras]
  {10.1111/j.1365-2966.2010.18174.x}, \href
  {http://adsabs.harvard.edu/abs/2011MNRAS.413..813C} {413, 813}

\bibitem[\protect\citeauthoryear{{Cappellari} et~al.,}{{Cappellari}
  et~al.}{2013a}]{cappellari+2013a}
{Cappellari} M.,  et~al., 2013a, \mn@doi [\mnras] {10.1093/mnras/stt562}, \href
  {http://adsabs.harvard.edu/abs/2013MNRAS.432.1709C} {432, 1709}

\bibitem[\protect\citeauthoryear{{Cappellari} et~al.,}{{Cappellari}
  et~al.}{2013b}]{cappellari+2013b}
{Cappellari} M.,  et~al., 2013b, \mn@doi [\mnras] {10.1093/mnras/stt644}, \href
  {http://adsabs.harvard.edu/abs/2013MNRAS.432.1862C} {432, 1862}

\bibitem[\protect\citeauthoryear{{Carollo}, {Danziger}  \& {Buson}}{{Carollo}
  et~al.}{1993}]{carollo+1993}
{Carollo} C.~M.,  {Danziger} I.~J.,   {Buson} L.,  1993, \mn@doi [\mnras]
  {10.1093/mnras/265.3.553}, \href
  {https://ui.adsabs.harvard.edu/abs/1993MNRAS.265..553C} {265, 553}

\bibitem[\protect\citeauthoryear{{Chabrier}}{{Chabrier}}{2003}]{chabrier2003}
{Chabrier} G.,  2003, \mn@doi [\pasp] {10.1086/376392}, \href
  {http://adsabs.harvard.edu/abs/2003PASP..115..763C} {115, 763}

\bibitem[\protect\citeauthoryear{{Cheung} et~al.,}{{Cheung}
  et~al.}{2012}]{cheung+2012}
{Cheung} E.,  et~al., 2012, \mn@doi [\apj] {10.1088/0004-637X/760/2/131}, \href
  {https://ui.adsabs.harvard.edu/abs/2012ApJ...760..131C} {760, 131}

\bibitem[\protect\citeauthoryear{{Chiu}, {Ko}  \& {Shu}}{{Chiu}
  et~al.}{2017}]{chiu+2017}
{Chiu} M.-C.,  {Ko} C.-M.,   {Shu} C.,  2017, \mn@doi [\prd]
  {10.1103/PhysRevD.95.063020}, \href
  {https://ui.adsabs.harvard.edu/abs/2017PhRvD..95f3020C} {95, 063020}

\bibitem[\protect\citeauthoryear{{Ciotti}, {Lanzoni}  \& {Renzini}}{{Ciotti}
  et~al.}{1996}]{ciotti+1996}
{Ciotti} L.,  {Lanzoni} B.,   {Renzini} A.,  1996, \mn@doi [\mnras]
  {10.1093/mnras/282.1.1}, \href
  {https://ui.adsabs.harvard.edu/abs/1996MNRAS.282....1C} {282, 1}

\bibitem[\protect\citeauthoryear{{Colless}, {Saglia}, {Burstein}, {Davies},
  {McMahan}  \& {Wegner}}{{Colless} et~al.}{2001}]{colless+2001}
{Colless} M.,  {Saglia} R.~P.,  {Burstein} D.,  {Davies} R.~L.,  {McMahan}
  R.~K.,   {Wegner} G.,  2001, \mn@doi [\mnras]
  {10.1046/j.1365-8711.2001.04044.x}, \href
  {http://adsabs.harvard.edu/abs/2001MNRAS.321..277C} {321, 277}

\bibitem[\protect\citeauthoryear{{Conroy} \& {van Dokkum}}{{Conroy} \& {van
  Dokkum}}{2012}]{conroy+vandokkum2012}
{Conroy} C.,  {van Dokkum} P.~G.,  2012, \mn@doi [\apj]
  {10.1088/0004-637X/760/1/71}, \href
  {http://adsabs.harvard.edu/abs/2012ApJ...760...71C} {760, 71}

\bibitem[\protect\citeauthoryear{{Cortese} et~al.,}{{Cortese}
  et~al.}{2016}]{cortese+2016}
{Cortese} L.,  et~al., 2016, \mn@doi [\mnras] {10.1093/mnras/stw1891}, \href
  {http://adsabs.harvard.edu/abs/2016MNRAS.463..170C} {463, 170}

\bibitem[\protect\citeauthoryear{{Cowie}, {Songaila}, {Hu}  \& {Cohen}}{{Cowie}
  et~al.}{1996}]{cowie+1996}
{Cowie} L.~L.,  {Songaila} A.,  {Hu} E.~M.,   {Cohen} J.~G.,  1996, \mn@doi
  [\aj] {10.1086/118058}, \href
  {https://ui.adsabs.harvard.edu/abs/1996AJ....112..839C} {112, 839}

\bibitem[\protect\citeauthoryear{{Croom} et~al.,}{{Croom}
  et~al.}{2012}]{croom+2012}
{Croom} S.~M.,  et~al., 2012, \mn@doi [\mnras]
  {10.1111/j.1365-2966.2011.20365.x}, \href
  {http://adsabs.harvard.edu/abs/2012MNRAS.421..872C} {421, 872}

\bibitem[\protect\citeauthoryear{{Croom} et~al.,}{{Croom}
  et~al.}{2021}]{croom+2021}
{Croom} S.~M.,  et~al., 2021, \mn@doi [\mnras]
  {10.1111/j.1365-2966.2011.20365.x}, \href
  {http://adsabs.harvard.edu/abs/2012MNRAS.421..872C} {421, 872}

\bibitem[\protect\citeauthoryear{{D'Eugenio} et~al.,}{{D'Eugenio}
  et~al.}{2020}]{deugenio+2020}
{D'Eugenio} F.,  et~al., 2020, \mn@doi [\mnras] {10.1093/mnras/staa1937}, \href
  {https://ui.adsabs.harvard.edu/abs/2020MNRAS.497..389D} {497, 389}

\bibitem[\protect\citeauthoryear{{Dalla Bont{\`a}}, {Davies}, {Houghton},
  {D'Eugenio}  \& {M{\'e}ndez-Abreu}}{{Dalla Bont{\`a}}
  et~al.}{2018}]{dallabonta+2018}
{Dalla Bont{\`a}} E.,  {Davies} R.~L.,  {Houghton} R.~C.~W.,  {D'Eugenio} F.,
  {M{\'e}ndez-Abreu} J.,  2018, \mn@doi [\mnras] {10.1093/mnras/stx2477}, \href
  {https://ui.adsabs.harvard.edu/abs/2018MNRAS.474..339D} {474, 339}

\bibitem[\protect\citeauthoryear{DeGroot}{DeGroot}{1970}]{degroot1970}
DeGroot M.,  1970, Optimal statistical decisions.
McGraw-Hill, New York, NY [u.a]

\bibitem[\protect\citeauthoryear{{Desmond} \& {Wechsler}}{{Desmond} \&
  {Wechsler}}{2017}]{desmond+wechsler2017}
{Desmond} H.,  {Wechsler} R.~H.,  2017, \mn@doi [\mnras]
  {10.1093/mnras/stw2804}, \href
  {http://adsabs.harvard.edu/abs/2017MNRAS.465..820D} {465, 820}

\bibitem[\protect\citeauthoryear{{Djorgovski} \& {Davis}}{{Djorgovski} \&
  {Davis}}{1987}]{djorgovski+davis1987}
{Djorgovski} S.,  {Davis} M.,  1987, \mn@doi [\apj] {10.1086/164948}, \href
  {http://adsabs.harvard.edu/abs/1987ApJ...313...59D} {313, 59}

\bibitem[\protect\citeauthoryear{{Dressler}, {Lynden-Bell}, {Burstein},
  {Davies}, {Faber}, {Terlevich}  \& {Wegner}}{{Dressler}
  et~al.}{1987}]{dressler+1987}
{Dressler} A.,  {Lynden-Bell} D.,  {Burstein} D.,  {Davies} R.~L.,  {Faber}
  S.~M.,  {Terlevich} R.,   {Wegner} G.,  1987, \mn@doi [\apj]
  {10.1086/164947}, \href {http://adsabs.harvard.edu/abs/1987ApJ...313...42D}
  {313, 42}

\bibitem[\protect\citeauthoryear{{Driver} et~al.,}{{Driver}
  et~al.}{2011}]{driver+2011}
{Driver} S.~P.,  et~al., 2011, \mn@doi [\mnras]
  {10.1111/j.1365-2966.2010.18188.x}, \href
  {http://adsabs.harvard.edu/abs/2011MNRAS.413..971D} {413, 971}

\bibitem[\protect\citeauthoryear{{Emsellem}, {Monnet}  \& {Bacon}}{{Emsellem}
  et~al.}{1994}]{emsellem+1994}
{Emsellem} E.,  {Monnet} G.,   {Bacon} R.,  1994, \aap, \href
  {http://adsabs.harvard.edu/abs/1994A%26A...285..723E} {285, 723}

\bibitem[\protect\citeauthoryear{{Emsellem} et~al.,}{{Emsellem}
  et~al.}{2011}]{emsellem+2011}
{Emsellem} E.,  et~al., 2011, \mn@doi [\mnras]
  {10.1111/j.1365-2966.2011.18496.x}, \href
  {http://adsabs.harvard.edu/abs/2011MNRAS.414..888E} {414, 888}

\bibitem[\protect\citeauthoryear{{Falc{\'o}n-Barroso},
  {S{\'a}nchez-Bl{\'a}zquez}, {Vazdekis}, {Ricciardelli}, {Cardiel}, {Cenarro},
  {Gorgas}  \& {Peletier}}{{Falc{\'o}n-Barroso}
  et~al.}{2011}]{falcon-barroso+2011}
{Falc{\'o}n-Barroso} J.,  {S{\'a}nchez-Bl{\'a}zquez} P.,  {Vazdekis} A.,
  {Ricciardelli} E.,  {Cardiel} N.,  {Cenarro} A.~J.,  {Gorgas} J.,
  {Peletier} R.~F.,  2011, \mn@doi [\aap] {10.1051/0004-6361/201116842}, \href
  {http://adsabs.harvard.edu/abs/2011A%26A...532A..95F} {532, A95}

\bibitem[\protect\citeauthoryear{{Fang}, {Faber}, {Koo}  \& {Dekel}}{{Fang}
  et~al.}{2013}]{fang+2013}
{Fang} J.~J.,  {Faber} S.~M.,  {Koo} D.~C.,   {Dekel} A.,  2013, \mn@doi [\apj]
  {10.1088/0004-637X/776/1/63}, \href
  {https://ui.adsabs.harvard.edu/abs/2013ApJ...776...63F} {776, 63}

\bibitem[\protect\citeauthoryear{{Feroz}, {Hobson}  \& {Bridges}}{{Feroz}
  et~al.}{2009}]{feroz+2009}
{Feroz} F.,  {Hobson} M.~P.,   {Bridges} M.,  2009, \mn@doi [\mnras]
  {10.1111/j.1365-2966.2009.14548.x}, \href
  {https://ui.adsabs.harvard.edu/abs/2009MNRAS.398.1601F} {398, 1601}

\bibitem[\protect\citeauthoryear{{Forbes}, {Ponman}  \& {Brown}}{{Forbes}
  et~al.}{1998}]{forbes+1998}
{Forbes} D.~A.,  {Ponman} T.~J.,   {Brown} R.~J.~N.,  1998, \mn@doi [\apjl]
  {10.1086/311715}, \href {http://adsabs.harvard.edu/abs/1998ApJ...508L..43F}
  {508, L43}

\bibitem[\protect\citeauthoryear{Foreman-Mackey}{Foreman-Mackey}{2016}]{foreman-mackey2016}
Foreman-Mackey D.,  2016, \mn@doi [The Journal of Open Source Software]
  {10.21105/joss.00024}, 1, 24

\bibitem[\protect\citeauthoryear{{Foreman-Mackey}, {Hogg}, {Lang}  \&
  {Goodman}}{{Foreman-Mackey} et~al.}{2013}]{foreman-mackey+2013}
{Foreman-Mackey} D.,  {Hogg} D.~W.,  {Lang} D.,   {Goodman} J.,  2013, \mn@doi
  [\pasp] {10.1086/670067}, \href
  {http://adsabs.harvard.edu/abs/2013PASP..125..306F} {125, 306}

\bibitem[\protect\citeauthoryear{{Gallazzi} \& {Bell}}{{Gallazzi} \&
  {Bell}}{2009}]{gallazzi+bell2009}
{Gallazzi} A.,  {Bell} E.~F.,  2009, \mn@doi [\apjs]
  {10.1088/0067-0049/185/2/253}, \href
  {http://adsabs.harvard.edu/abs/2009ApJS..185..253G} {185, 253}

\bibitem[\protect\citeauthoryear{{Gallazzi}, {Charlot}, {Brinchmann}, {White}
  \& {Tremonti}}{{Gallazzi} et~al.}{2005}]{gallazzi+2005}
{Gallazzi} A.,  {Charlot} S.,  {Brinchmann} J.,  {White} S.~D.~M.,   {Tremonti}
  C.~A.,  2005, \mn@doi [\mnras] {10.1111/j.1365-2966.2005.09321.x}, \href
  {http://adsabs.harvard.edu/abs/2005MNRAS.362...41G} {362, 41}

\bibitem[\protect\citeauthoryear{{Gallazzi}, {Charlot}, {Brinchmann}  \&
  {White}}{{Gallazzi} et~al.}{2006}]{gallazzi+2006}
{Gallazzi} A.,  {Charlot} S.,  {Brinchmann} J.,   {White} S. D.~M.,  2006,
  \mn@doi [\mnras] {10.1111/j.1365-2966.2006.10548.x}, \href
  {https://ui.adsabs.harvard.edu/abs/2006MNRAS.370.1106G} {370, 1106}

\bibitem[\protect\citeauthoryear{{Ganda} et~al.,}{{Ganda}
  et~al.}{2007}]{ganda+2007}
{Ganda} K.,  et~al., 2007, \mn@doi [\mnras] {10.1111/j.1365-2966.2007.12121.x},
  \href {http://adsabs.harvard.edu/abs/2007MNRAS.380..506G} {380, 506}

\bibitem[\protect\citeauthoryear{{Goodman} \& {Weare}}{{Goodman} \&
  {Weare}}{2010}]{goodman+weare2010}
{Goodman} J.,  {Weare} J.,  2010, \mn@doi [Communications in Applied
  Mathematics and Computational Science] {10.2140/camcos.2010.5.65}, \href
  {https://ui.adsabs.harvard.edu/abs/2010CAMCS...5...65G} {5, 65}

\bibitem[\protect\citeauthoryear{{Graham} \& {Colless}}{{Graham} \&
  {Colless}}{1997}]{graham+colless1997}
{Graham} A.,  {Colless} M.,  1997, \mn@doi [\mnras] {10.1093/mnras/287.1.221},
  \href {https://ui.adsabs.harvard.edu/abs/1997MNRAS.287..221G} {287, 221}

\bibitem[\protect\citeauthoryear{{Graves} \& {Faber}}{{Graves} \&
  {Faber}}{2010}]{graves+faber2010}
{Graves} G.~J.,  {Faber} S.~M.,  2010, \mn@doi [\apj]
  {10.1088/0004-637X/717/2/803}, \href
  {http://adsabs.harvard.edu/abs/2010ApJ...717..803G} {717, 803}

\bibitem[\protect\citeauthoryear{{Graves}, {Faber}, {Schiavon}  \&
  {Yan}}{{Graves} et~al.}{2007}]{graves+2007}
{Graves} G.~J.,  {Faber} S.~M.,  {Schiavon} R.~P.,   {Yan} R.,  2007, \mn@doi
  [\apj] {10.1086/522325}, \href
  {https://ui.adsabs.harvard.edu/abs/2007ApJ...671..243G} {671, 243}

\bibitem[\protect\citeauthoryear{{Graves}, {Faber}  \& {Schiavon}}{{Graves}
  et~al.}{2009}]{graves+2009}
{Graves} G.~J.,  {Faber} S.~M.,   {Schiavon} R.~P.,  2009, \mn@doi [\apj]
  {10.1088/0004-637X/698/2/1590}, \href
  {https://ui.adsabs.harvard.edu/abs/2009ApJ...698.1590G} {698, 1590}

\bibitem[\protect\citeauthoryear{{Green} et~al.,}{{Green}
  et~al.}{2018}]{green+2018}
{Green} A.~W.,  et~al., 2018, \mn@doi [\mnras] {10.1093/mnras/stx3135}, \href
  {http://adsabs.harvard.edu/abs/2018MNRAS.475..716G} {475, 716}

\bibitem[\protect\citeauthoryear{{Harris} et~al.,}{{Harris}
  et~al.}{2020}]{harris+2020}
{Harris} C.~R.,  et~al., 2020, \mn@doi [\nat] {10.1038/s41586-020-2649-2},
  \href {https://ui.adsabs.harvard.edu/abs/2020Natur.585..357H} {585, 357}

\bibitem[\protect\citeauthoryear{{Higson}, {Handley}, {Hobson}  \&
  {Lasenby}}{{Higson} et~al.}{2019}]{higson+2019}
{Higson} E.,  {Handley} W.,  {Hobson} M.,   {Lasenby} A.,  2019, \mn@doi
  [Statistics and Computing] {10.1007/s11222-018-9844-0}, \href
  {https://ui.adsabs.harvard.edu/abs/2019S&C....29..891H} {29, 891}

\bibitem[\protect\citeauthoryear{{Hill} et~al.,}{{Hill}
  et~al.}{2011}]{hill+2011}
{Hill} D.~T.,  et~al., 2011, \mn@doi [\mnras]
  {10.1111/j.1365-2966.2010.17950.x}, \href
  {https://ui.adsabs.harvard.edu/abs/2011MNRAS.412..765H} {412, 765}

\bibitem[\protect\citeauthoryear{{Holden}, {van der Wel}, {Kelson}, {Franx}  \&
  {Illingworth}}{{Holden} et~al.}{2010}]{holden+2010}
{Holden} B.~P.,  {van der Wel} A.,  {Kelson} D.~D.,  {Franx} M.,
  {Illingworth} G.~D.,  2010, \mn@doi [\apj] {10.1088/0004-637X/724/1/714},
  \href {https://ui.adsabs.harvard.edu/abs/2010ApJ...724..714H} {724, 714}

\bibitem[\protect\citeauthoryear{{Hudson}, {Smith}, {Lucey}, {Schlegel}  \&
  {Davies}}{{Hudson} et~al.}{1999}]{hudson+1999}
{Hudson} M.~J.,  {Smith} R.~J.,  {Lucey} J.~R.,  {Schlegel} D.~J.,   {Davies}
  R.~L.,  1999, \mn@doi [\apjl] {10.1086/311883}, \href
  {https://ui.adsabs.harvard.edu/abs/1999ApJ...512L..79H} {512, L79}

\bibitem[\protect\citeauthoryear{{Hunter}}{{Hunter}}{2007}]{hunter2007}
{Hunter} J.~D.,  2007, \mn@doi [Computing in Science and Engineering]
  {10.1109/MCSE.2007.55}, \href
  {https://ui.adsabs.harvard.edu/abs/2007CSE.....9...90H} {9, 90}

\bibitem[\protect\citeauthoryear{{Hyde} \& {Bernardi}}{{Hyde} \&
  {Bernardi}}{2009}]{hyde+bernardi2009}
{Hyde} J.~B.,  {Bernardi} M.,  2009, \mn@doi [\mnras]
  {10.1111/j.1365-2966.2009.14783.x}, \href
  {http://adsabs.harvard.edu/abs/2009MNRAS.396.1171H} {396, 1171}

\bibitem[\protect\citeauthoryear{{Johnson} et~al.,}{{Johnson}
  et~al.}{2014}]{johnson+2014}
{Johnson} A.,  et~al., 2014, \mn@doi [\mnras] {10.1093/mnras/stu1615}, \href
  {http://adsabs.harvard.edu/abs/2014MNRAS.444.3926J} {444, 3926}

\bibitem[\protect\citeauthoryear{Jones, Oliphant, Peterson  et~al.}{Jones
  et~al.}{2001}]{jones+2001}
Jones E.,  Oliphant T.,  Peterson P.,   et~al., 2001, {SciPy}: Open source
  scientific tools for {Python}, \url {http://www.scipy.org/}

\bibitem[\protect\citeauthoryear{{J{\o}rgensen}, {Franx}  \&
  {Kjaergaard}}{{J{\o}rgensen} et~al.}{1996}]{jorgensen+1996}
{J{\o}rgensen} I.,  {Franx} M.,   {Kjaergaard} P.,  1996, \mn@doi [\mnras]
  {10.1093/mnras/280.1.167}, \href
  {http://adsabs.harvard.edu/abs/1996MNRAS.280..167J} {280, 167}

\bibitem[\protect\citeauthoryear{{Kelly}}{{Kelly}}{2007}]{kelly2007}
{Kelly} B.~C.,  2007, \mn@doi [\apj] {10.1086/519947}, \href
  {https://ui.adsabs.harvard.edu/abs/2007ApJ...665.1489K} {665, 1489}

\bibitem[\protect\citeauthoryear{{Kelvin} et~al.,}{{Kelvin}
  et~al.}{2012}]{kelvin+2012}
{Kelvin} L.~S.,  et~al., 2012, \mn@doi [\mnras]
  {10.1111/j.1365-2966.2012.20355.x}, \href
  {http://adsabs.harvard.edu/abs/2012MNRAS.421.1007K} {421, 1007}

\bibitem[\protect\citeauthoryear{{Kelvin} et~al.,}{{Kelvin}
  et~al.}{2014}]{kelvin+2014a}
{Kelvin} L.~S.,  et~al., 2014, \mn@doi [\mnras] {10.1093/mnras/stt2391}, \href
  {http://adsabs.harvard.edu/abs/2014MNRAS.439.1245K} {439, 1245}

\bibitem[\protect\citeauthoryear{{Kobayashi}}{{Kobayashi}}{2005}]{kobayashi2005}
{Kobayashi} C.,  2005, \mn@doi [\mnras] {10.1111/j.1365-2966.2005.09248.x},
  \href {https://ui.adsabs.harvard.edu/abs/2005MNRAS.361.1216K} {361, 1216}

\bibitem[\protect\citeauthoryear{{Koopmans} et~al.,}{{Koopmans}
  et~al.}{2009}]{koopmans+2009}
{Koopmans} L.~V.~E.,  et~al., 2009, \mn@doi [\apjl]
  {10.1088/0004-637X/703/1/L51}, \href
  {http://adsabs.harvard.edu/abs/2009ApJ...703L..51K} {703, L51}

\bibitem[\protect\citeauthoryear{{Krajnovi{\'c}} et~al.,}{{Krajnovi{\'c}}
  et~al.}{2011}]{krajnovic+2011}
{Krajnovi{\'c}} D.,  et~al., 2011, \mn@doi [\mnras]
  {10.1111/j.1365-2966.2011.18560.x}, \href
  {https://ui.adsabs.harvard.edu/abs/2011MNRAS.414.2923K} {414, 2923}

\bibitem[\protect\citeauthoryear{{Kroupa}}{{Kroupa}}{2001}]{kroupa2001}
{Kroupa} P.,  2001, \mn@doi [\mnras] {10.1046/j.1365-8711.2001.04022.x}, \href
  {http://adsabs.harvard.edu/abs/2001MNRAS.322..231K} {322, 231}

\bibitem[\protect\citeauthoryear{{Liu} et~al.,}{{Liu} et~al.}{2016}]{liu+2016}
{Liu} Y.,  et~al., 2016, \mn@doi [\apj] {10.3847/0004-637X/818/2/179}, \href
  {https://ui.adsabs.harvard.edu/abs/2016ApJ...818..179L} {818, 179}

\bibitem[\protect\citeauthoryear{{Loveday} et~al.,}{{Loveday}
  et~al.}{2012}]{loveday+2012}
{Loveday} J.,  et~al., 2012, \mn@doi [\mnras]
  {10.1111/j.1365-2966.2011.20111.x}, \href
  {http://adsabs.harvard.edu/abs/2012MNRAS.420.1239L} {420, 1239}

\bibitem[\protect\citeauthoryear{{Magoulas} et~al.,}{{Magoulas}
  et~al.}{2012}]{magoulas+2012}
{Magoulas} C.,  et~al., 2012, \mn@doi [\mnras]
  {10.1111/j.1365-2966.2012.21421.x}, \href
  {http://adsabs.harvard.edu/abs/2012MNRAS.427..245M} {427, 245}

\bibitem[\protect\citeauthoryear{{Maraston}}{{Maraston}}{2005}]{maraston2005}
{Maraston} C.,  2005, \mn@doi [\mnras] {10.1111/j.1365-2966.2005.09270.x},
  \href {http://adsabs.harvard.edu/abs/2005MNRAS.362..799M} {362, 799}

\bibitem[\protect\citeauthoryear{{McDermid} et~al.,}{{McDermid}
  et~al.}{2015}]{mcdermid+2015}
{McDermid} R.~M.,  et~al., 2015, \mn@doi [\mnras] {10.1093/mnras/stv105}, \href
  {http://adsabs.harvard.edu/abs/2015MNRAS.448.3484M} {448, 3484}

\bibitem[\protect\citeauthoryear{{McKerns}, {Strand}, {Sullivan}, {Fang}  \&
  {Aivazis}}{{McKerns} et~al.}{2011}]{mckerns+2011}
{McKerns} M.~M.,  {Strand} L.,  {Sullivan} T.,  {Fang} A.,   {Aivazis}
  M.~A.~G.,  2011, in Proc. 10th Python in Sci. Conf.. Texas, p.~76 (\mn@eprint
  {arXiv} {1202.1056})

\bibitem[\protect\citeauthoryear{{Mehlert}, {Thomas}, {Saglia}, {Bender}  \&
  {Wegner}}{{Mehlert} et~al.}{2003}]{mehlert+2003}
{Mehlert} D.,  {Thomas} D.,  {Saglia} R.~P.,  {Bender} R.,   {Wegner} G.,
  2003, \mn@doi [\aap] {10.1051/0004-6361:20030886}, \href
  {https://ui.adsabs.harvard.edu/abs/2003A&A...407..423M} {407, 423}

\bibitem[\protect\citeauthoryear{{Metropolis}, {Rosenbluth}, {Rosenbluth},
  {Teller}  \& {Teller}}{{Metropolis} et~al.}{1953}]{metropolis+1953}
{Metropolis} N.,  {Rosenbluth} A.~W.,  {Rosenbluth} M.~N.,  {Teller} A.~H.,
  {Teller} E.,  1953, \mn@doi [\jcp] {10.1063/1.1699114}, \href
  {https://ui.adsabs.harvard.edu/abs/1953JChPh..21.1087M} {21, 1087}

\bibitem[\protect\citeauthoryear{{Mould}}{{Mould}}{2020}]{mould2020}
{Mould} J.,  2020, \mn@doi [Frontiers in Astronomy and Space Sciences]
  {10.3389/fspas.2020.00021}, \href
  {https://ui.adsabs.harvard.edu/abs/2020FrASS...7...21M} {7, 21}

\bibitem[\protect\citeauthoryear{{Nelan}, {Smith}, {Hudson}, {Wegner}, {Lucey},
  {Moore}, {Quinney}  \& {Suntzeff}}{{Nelan} et~al.}{2005}]{nelan+2005}
{Nelan} J.~E.,  {Smith} R.~J.,  {Hudson} M.~J.,  {Wegner} G.~A.,  {Lucey}
  J.~R.,  {Moore} S. A.~W.,  {Quinney} S.~J.,   {Suntzeff} N.~B.,  2005,
  \mn@doi [\apj] {10.1086/431962}, \href
  {https://ui.adsabs.harvard.edu/abs/2005ApJ...632..137N} {632, 137}

\bibitem[\protect\citeauthoryear{{Nipoti}, {Londrillo}  \& {Ciotti}}{{Nipoti}
  et~al.}{2002}]{nipoti+2002}
{Nipoti} C.,  {Londrillo} P.,   {Ciotti} L.,  2002, \mn@doi [\mnras]
  {10.1046/j.1365-8711.2002.05356.x}, \href
  {http://adsabs.harvard.edu/abs/2002MNRAS.332..901N} {332, 901}

\bibitem[\protect\citeauthoryear{{Oke} \& {Gunn}}{{Oke} \&
  {Gunn}}{1983}]{oke+gunn1983}
{Oke} J.~B.,  {Gunn} J.~E.,  1983, \mn@doi [\apj] {10.1086/160817}, \href
  {https://ui.adsabs.harvard.edu/abs/1983ApJ...266..713O} {266, 713}

\bibitem[\protect\citeauthoryear{{Oldham}, {Houghton}  \& {Davies}}{{Oldham}
  et~al.}{2017}]{oldham+2017}
{Oldham} L.~J.,  {Houghton} R.~C.~W.,   {Davies} R.~L.,  2017, \mn@doi [\mnras]
  {10.1093/mnras/stw2791}, \href
  {https://ui.adsabs.harvard.edu/abs/2017MNRAS.465.2101O} {465, 2101}

\bibitem[\protect\citeauthoryear{{Owers} et~al.,}{{Owers}
  et~al.}{2017}]{owers+2017}
{Owers} M.~S.,  et~al., 2017, \mn@doi [\mnras] {10.1093/mnras/stx562}, \href
  {http://adsabs.harvard.edu/abs/2017MNRAS.468.1824O} {468, 1824}

\bibitem[\protect\citeauthoryear{{Owers} et~al.,}{{Owers}
  et~al.}{2019}]{owers+2019}
{Owers} M.~S.,  et~al., 2019, \mn@doi [\apj] {10.3847/1538-4357/ab0201}, \href
  {https://ui.adsabs.harvard.edu/abs/2019ApJ...873...52O} {873, 52}

\bibitem[\protect\citeauthoryear{{Pahre}, {Djorgovski}  \& {de
  Carvalho}}{{Pahre} et~al.}{1995}]{pahre+1995}
{Pahre} M.~A.,  {Djorgovski} S.~G.,   {de Carvalho} R.~R.,  1995, \mn@doi
  [\apjl] {10.1086/309740}, \href
  {http://adsabs.harvard.edu/abs/1995ApJ...453L..17P} {453, L17}

\bibitem[\protect\citeauthoryear{{Pahre}, {Djorgovski}  \& {de
  Carvalho}}{{Pahre} et~al.}{1998}]{pahre+1998}
{Pahre} M.~A.,  {Djorgovski} S.~G.,   {de Carvalho} R.~R.,  1998, \mn@doi [\aj]
  {10.1086/300544}, \href {http://adsabs.harvard.edu/abs/1998AJ....116.1591P}
  {116, 1591}

\bibitem[\protect\citeauthoryear{{Peng}, {Ho}, {Impey}  \& {Rix}}{{Peng}
  et~al.}{2002}]{peng+2002}
{Peng} C.~Y.,  {Ho} L.~C.,  {Impey} C.~D.,   {Rix} H.-W.,  2002, \mn@doi [\aj]
  {10.1086/340952}, \href {http://adsabs.harvard.edu/abs/2002AJ....124..266P}
  {124, 266}

\bibitem[\protect\citeauthoryear{{Planck Collaboration} et~al.,}{{Planck
  Collaboration} et~al.}{2016}]{planck+2016}
{Planck Collaboration} et~al., 2016, \mn@doi [\aap]
  {10.1051/0004-6361/201525830}, \href
  {https://ui.adsabs.harvard.edu/\#abs/2016A&A...594A..13P} {594, A13}

\bibitem[\protect\citeauthoryear{{Press}, {Teukolsky}, {Vetterling}  \&
  {Flannery}}{{Press} et~al.}{2007}]{press+2007}
{Press} W.~H.,  {Teukolsky} S.~A.,  {Vetterling} W.~T.,   {Flannery} B.~P.,
  2007, {Numerical recipes: The art of scientific computing}, 3rd edn.
Cambridge Univ. Press, Cambridge, UK

\bibitem[\protect\citeauthoryear{{Prichard} et~al.,}{{Prichard}
  et~al.}{2017}]{prichard+2017}
{Prichard} L.~J.,  et~al., 2017, \mn@doi [\apj] {10.3847/1538-4357/aa96a6},
  \href {https://ui.adsabs.harvard.edu/abs/2017ApJ...850..203P} {850, 203}

\bibitem[\protect\citeauthoryear{{Prugniel} \& {Simien}}{{Prugniel} \&
  {Simien}}{1997}]{prugniel+simien1997}
{Prugniel} P.,  {Simien} F.,  1997, \aap, \href
  {http://adsabs.harvard.edu/abs/1997A%26A...321..111P} {321, 111}

\bibitem[\protect\citeauthoryear{{Renzini}}{{Renzini}}{1977}]{renzini1977}
{Renzini} A.,  1977, in {Bouvier} P.,  {Maeder} A.,  eds, Saas-Fee Advanced
  Course 7: Advanced Stages in Stellar Evolution. p.~151

\bibitem[\protect\citeauthoryear{{Riess} et~al.,}{{Riess}
  et~al.}{2016}]{riess+2016}
{Riess} A.~G.,  et~al., 2016, \mn@doi [\apj] {10.3847/0004-637X/826/1/56},
  \href {https://ui.adsabs.harvard.edu/\#abs/2016ApJ...826...56R} {826, 56}

\bibitem[\protect\citeauthoryear{{Robotham} \& {Obreschkow}}{{Robotham} \&
  {Obreschkow}}{2015}]{robotham+obreschkow2015}
{Robotham} A.~S.~G.,  {Obreschkow} D.,  2015, \mn@doi [\pasa]
  {10.1017/pasa.2015.33}, \href
  {https://ui.adsabs.harvard.edu/abs/2015PASA...32...33R} {32, e033}

\bibitem[\protect\citeauthoryear{Rousseeuw \& Driessen}{Rousseeuw \&
  Driessen}{2006}]{rousseeuw+driessen2006}
Rousseeuw P.~J.,  Driessen K.,  2006, \mn@doi [Data Min. Knowl. Discov.]
  {10.1007/s10618-005-0024-4}, 12, 29

\bibitem[\protect\citeauthoryear{{Saglia} et~al.,}{{Saglia}
  et~al.}{2010}]{saglia+2010}
{Saglia} R.~P.,  et~al., 2010, \mn@doi [\aap] {10.1051/0004-6361/201014703},
  \href {https://ui.adsabs.harvard.edu/abs/2010A&A...524A...6S} {524, A6}

\bibitem[\protect\citeauthoryear{{Saglia} et~al.,}{{Saglia}
  et~al.}{2016}]{saglia+2016}
{Saglia} R.~P.,  et~al., 2016, \mn@doi [\aap] {10.1051/0004-6361/201014703e},
  \href {https://ui.adsabs.harvard.edu/abs/2016A&A...596C...1S} {596, C1}

\bibitem[\protect\citeauthoryear{{Said}, {Colless}, {Magoulas}, {Lucey}  \&
  {Hudson}}{{Said} et~al.}{2020}]{said+2020}
{Said} K.,  {Colless} M.,  {Magoulas} C.,  {Lucey} J.~R.,   {Hudson} M.~J.,
  2020, \mn@doi [\mnras] {10.1093/mnras/staa2032}, \href
  {https://ui.adsabs.harvard.edu/abs/2020MNRAS.tmp.2141S} {}

\bibitem[\protect\citeauthoryear{{S{\'a}nchez-Bl{\'a}zquez}
  et~al.,}{{S{\'a}nchez-Bl{\'a}zquez} et~al.}{2006}]{sanchez-blazquez+2006}
{S{\'a}nchez-Bl{\'a}zquez} P.,  et~al., 2006, \mn@doi [\mnras]
  {10.1111/j.1365-2966.2006.10699.x}, \href
  {http://adsabs.harvard.edu/abs/2006MNRAS.371..703S} {371, 703}

\bibitem[\protect\citeauthoryear{{S{\'a}nchez-Bl{\'a}zquez}, {Forbes},
  {Strader}, {Brodie}  \& {Proctor}}{{S{\'a}nchez-Bl{\'a}zquez}
  et~al.}{2007}]{sanchez-blazquez+2007}
{S{\'a}nchez-Bl{\'a}zquez} P.,  {Forbes} D.~A.,  {Strader} J.,  {Brodie} J.,
  {Proctor} R.,  2007, \mn@doi [\mnras] {10.1111/j.1365-2966.2007.11647.x},
  \href {https://ui.adsabs.harvard.edu/abs/2007MNRAS.377..759S} {377, 759}

\bibitem[\protect\citeauthoryear{{S{\'a}nchez} et~al.,}{{S{\'a}nchez}
  et~al.}{2012}]{sanchez+2012}
{S{\'a}nchez} S.~F.,  et~al., 2012, \mn@doi [\aap]
  {10.1051/0004-6361/201117353}, \href
  {http://adsabs.harvard.edu/abs/2012A%26A...538A...8S} {538, A8}

\bibitem[\protect\citeauthoryear{{Saracco}, {Gargiulo}, {La Barbera},
  {Annunziatella}  \& {Marchesini}}{{Saracco} et~al.}{2020}]{saracco+2020}
{Saracco} P.,  {Gargiulo} A.,  {La Barbera} F.,  {Annunziatella} M.,
  {Marchesini} D.,  2020, \mn@doi [\mnras] {10.1093/mnras/stz3109}, \href
  {https://ui.adsabs.harvard.edu/abs/2020MNRAS.491.1777S} {491, 1777}

\bibitem[\protect\citeauthoryear{{Schiavon}}{{Schiavon}}{2007}]{schiavon2007}
{Schiavon} R.~P.,  2007, \mn@doi [\apjs] {10.1086/511753}, \href
  {http://adsabs.harvard.edu/abs/2007ApJS..171..146S} {171, 146}

\bibitem[\protect\citeauthoryear{{Scott} et~al.,}{{Scott}
  et~al.}{2009}]{scott+2009}
{Scott} N.,  et~al., 2009, \mn@doi [\mnras] {10.1111/j.1365-2966.2009.15275.x},
  \href {http://adsabs.harvard.edu/abs/2009MNRAS.398.1835S} {398, 1835}

\bibitem[\protect\citeauthoryear{{Scott} et~al.,}{{Scott}
  et~al.}{2015}]{scott+2015}
{Scott} N.,  et~al., 2015, \mn@doi [\mnras] {10.1093/mnras/stv1127}, \href
  {http://adsabs.harvard.edu/abs/2015MNRAS.451.2723S} {451, 2723}

\bibitem[\protect\citeauthoryear{{Scott} et~al.,}{{Scott}
  et~al.}{2017}]{scott+2017}
{Scott} N.,  et~al., 2017, \mn@doi [\mnras] {10.1093/mnras/stx2166}, \href
  {http://adsabs.harvard.edu/abs/2017MNRAS.472.2833S} {472, 2833}

\bibitem[\protect\citeauthoryear{{Scott} et~al.,}{{Scott}
  et~al.}{2018}]{scott+2018}
{Scott} N.,  et~al., 2018, \mn@doi [\mnras] {10.1093/mnras/sty2355}, \href
  {http://adsabs.harvard.edu/abs/2018MNRAS.481.2299S} {481, 2299}

\bibitem[\protect\citeauthoryear{{Scott} et~al.,}{{Scott}
  et~al.}{2020}]{scott+2020}
{Scott} N.,  et~al., 2020, \mn@doi [\mnras] {10.1093/mnras/staa2042}, \href
  {https://ui.adsabs.harvard.edu/abs/2020MNRAS.497.1571S} {497, 1571}

\bibitem[\protect\citeauthoryear{{Scrimgeour} et~al.,}{{Scrimgeour}
  et~al.}{2016}]{scrimgeour+2016}
{Scrimgeour} M.~I.,  et~al., 2016, \mn@doi [\mnras] {10.1093/mnras/stv2146},
  \href {https://ui.adsabs.harvard.edu/abs/2016MNRAS.455..386S} {455, 386}

\bibitem[\protect\citeauthoryear{{Shanks} et~al.,}{{Shanks}
  et~al.}{2013}]{shanks+2013}
{Shanks} T.,  et~al., 2013, The Messenger, \href
  {http://adsabs.harvard.edu/abs/2013Msngr.154...38S} {154, 38}

\bibitem[\protect\citeauthoryear{{Shanks} et~al.,}{{Shanks}
  et~al.}{2015}]{shanks+2015}
{Shanks} T.,  et~al., 2015, \mn@doi [\mnras] {10.1093/mnras/stv1130}, \href
  {http://adsabs.harvard.edu/abs/2015MNRAS.451.4238S} {451, 4238}

\bibitem[\protect\citeauthoryear{{Sharp} et~al.,}{{Sharp}
  et~al.}{2006}]{sharp+2006}
{Sharp} R.,  et~al., 2006, in Society of Photo-Optical Instrumentation
  Engineers (SPIE) Conference Series. p. 62690G (\mn@eprint {}
  {astro-ph/0606137}), \mn@doi{10.1117/12.671022}

\bibitem[\protect\citeauthoryear{{Sharp} et~al.,}{{Sharp}
  et~al.}{2015}]{sharp+2015}
{Sharp} R.,  et~al., 2015, \mn@doi [\mnras] {10.1093/mnras/stu2055}, \href
  {http://adsabs.harvard.edu/abs/2015MNRAS.446.1551S} {446, 1551}

\bibitem[\protect\citeauthoryear{{Skilling}}{{Skilling}}{2004}]{skilling2004}
{Skilling} J.,  2004, in {Fischer} R.,  {Preuss} R.,   {Toussaint} U.~V.,  eds,
   American Institute of Physics Conference Series Vol. 735, Bayesian Inference
  and Maximum Entropy Methods in Science and Engineering: 24th International
  Workshop on Bayesian Inference and Maximum Entropy Methods in Science and
  Engineering. pp 395--405, \mn@doi{10.1063/1.1835238}

\bibitem[\protect\citeauthoryear{{Skilling}}{{Skilling}}{2006}]{skilling2006}
{Skilling} J.,  2006, \mn@doi [Bayesian Anal.] {10.1214/06-BA127}, 1, 833

\bibitem[\protect\citeauthoryear{{Speagle}}{{Speagle}}{2020}]{speagle2020}
{Speagle} J.~S.,  2020, \mn@doi [\mnras] {10.1093/mnras/staa278}, \href
  {https://ui.adsabs.harvard.edu/abs/2020MNRAS.493.3132S} {493, 3132}

\bibitem[\protect\citeauthoryear{{Speagle}, {Steinhardt}, {Capak}  \&
  {Silverman}}{{Speagle} et~al.}{2014}]{speagle+2014}
{Speagle} J.~S.,  {Steinhardt} C.~L.,  {Capak} P.~L.,   {Silverman} J.~D.,
  2014, \mn@doi [\apjs] {10.1088/0067-0049/214/2/15}, \href
  {http://adsabs.harvard.edu/abs/2014ApJS..214...15S} {214, 15}

\bibitem[\protect\citeauthoryear{{Springob} et~al.,}{{Springob}
  et~al.}{2012}]{springob+2012}
{Springob} C.~M.,  et~al., 2012, \mn@doi [\mnras]
  {10.1111/j.1365-2966.2011.19900.x}, \href
  {http://adsabs.harvard.edu/abs/2012MNRAS.420.2773S} {420, 2773}

\bibitem[\protect\citeauthoryear{{Springob} et~al.,}{{Springob}
  et~al.}{2014}]{springob+2014}
{Springob} C.~M.,  et~al., 2014, \mn@doi [\mnras] {10.1093/mnras/stu1743},
  \href {https://ui.adsabs.harvard.edu/abs/2014MNRAS.445.2677S} {445, 2677}

\bibitem[\protect\citeauthoryear{{Taylor}}{{Taylor}}{2005}]{taylor2005}
{Taylor} M.~B.,  2005, in {Shopbell} P.,  {Britton} M.,   {Ebert} R.,  eds,
  Astronomical Society of the Pacific Conference Series Vol. 347, Astronomical
  Data Analysis Software and Systems XIV. p.~29

\bibitem[\protect\citeauthoryear{{Taylor} et~al.,}{{Taylor}
  et~al.}{2011}]{taylor+2011}
{Taylor} E.~N.,  et~al., 2011, \mn@doi [\mnras]
  {10.1111/j.1365-2966.2011.19536.x}, \href
  {http://adsabs.harvard.edu/abs/2011MNRAS.418.1587T} {418, 1587}

\bibitem[\protect\citeauthoryear{{Thob} et~al.,}{{Thob}
  et~al.}{2019}]{thob+2019}
{Thob} A.~C.~R.,  et~al., 2019, \mn@doi [\mnras] {10.1093/mnras/stz448}, \href
  {http://adsabs.harvard.edu/abs/2019MNRAS.485..972T} {485, 972}

\bibitem[\protect\citeauthoryear{{Thomas}, {Maraston}, {Bender}  \& {Mendes de
  Oliveira}}{{Thomas} et~al.}{2005}]{thomas+2005}
{Thomas} D.,  {Maraston} C.,  {Bender} R.,   {Mendes de Oliveira} C.,  2005,
  \mn@doi [\apj] {10.1086/426932}, \href
  {https://ui.adsabs.harvard.edu/abs/2005ApJ...621..673T} {621, 673}

\bibitem[\protect\citeauthoryear{{Thomas}, {Maraston}, {Schawinski}, {Sarzi}
  \& {Silk}}{{Thomas} et~al.}{2010}]{thomas+2010}
{Thomas} D.,  {Maraston} C.,  {Schawinski} K.,  {Sarzi} M.,   {Silk} J.,  2010,
  \mn@doi [\mnras] {10.1111/j.1365-2966.2010.16427.x}, \href
  {http://adsabs.harvard.edu/abs/2010MNRAS.404.1775T} {404, 1775}

\bibitem[\protect\citeauthoryear{{Tonry}, {Blakeslee}, {Ajhar}  \&
  {Dressler}}{{Tonry} et~al.}{2000}]{tonry+2000}
{Tonry} J.~L.,  {Blakeslee} J.~P.,  {Ajhar} E.~A.,   {Dressler} A.,  2000,
  \mn@doi [\apj] {10.1086/308409}, \href
  {https://ui.adsabs.harvard.edu/abs/2000ApJ...530..625T} {530, 625}

\bibitem[\protect\citeauthoryear{{Trager}, {Worthey}, {Faber}, {Burstein}  \&
  {Gonz{\'a}lez}}{{Trager} et~al.}{1998}]{trager+1998}
{Trager} S.~C.,  {Worthey} G.,  {Faber} S.~M.,  {Burstein} D.,   {Gonz{\'a}lez}
  J.~J.,  1998, \mn@doi [\apjs] {10.1086/313099}, \href
  {http://adsabs.harvard.edu/abs/1998ApJS..116....1T} {116, 1}

\bibitem[\protect\citeauthoryear{Vincent \& Ryden}{Vincent \&
  Ryden}{2005}]{vincent+ryden2005}
Vincent R.~A.,  Ryden B.~S.,  2005, \mn@doi [The Astrophysical Journal]
  {10.1086/428765}, 623, 137

\bibitem[\protect\citeauthoryear{{Wolf}, {Martinez}, {Bullock}, {Kaplinghat},
  {Geha}, {Mu{\~n}oz}, {Simon}  \& {Avedo}}{{Wolf} et~al.}{2010}]{wolf+2010}
{Wolf} J.,  {Martinez} G.~D.,  {Bullock} J.~S.,  {Kaplinghat} M.,  {Geha} M.,
  {Mu{\~n}oz} R.~R.,  {Simon} J.~D.,   {Avedo} F.~F.,  2010, \mn@doi [\mnras]
  {10.1111/j.1365-2966.2010.16753.x}, \href
  {https://ui.adsabs.harvard.edu/abs/2010MNRAS.406.1220W} {406, 1220}

\bibitem[\protect\citeauthoryear{{Woo}, {Dekel}, {Faber}  \& {Koo}}{{Woo}
  et~al.}{2015}]{woo+2015}
{Woo} J.,  {Dekel} A.,  {Faber} S.~M.,   {Koo} D.~C.,  2015, \mn@doi [\mnras]
  {10.1093/mnras/stu2755}, \href
  {https://ui.adsabs.harvard.edu/abs/2015MNRAS.448..237W} {448, 237}

\bibitem[\protect\citeauthoryear{{Worthey}, {Faber}, {Gonzalez}  \&
  {Burstein}}{{Worthey} et~al.}{1994}]{worthey+1994}
{Worthey} G.,  {Faber} S.~M.,  {Gonzalez} J.~J.,   {Burstein} D.,  1994,
  \mn@doi [\apjs] {10.1086/192087}, \href
  {http://adsabs.harvard.edu/abs/1994ApJS...94..687W} {94, 687}

\bibitem[\protect\citeauthoryear{{Wright} et~al.,}{{Wright}
  et~al.}{2016}]{wright+2016}
{Wright} A.~H.,  et~al., 2016, \mn@doi [\mnras] {10.1093/mnras/stw832}, \href
  {https://ui.adsabs.harvard.edu/abs/2016MNRAS.460..765W} {460, 765}

\bibitem[\protect\citeauthoryear{{Wu} et~al.,}{{Wu} et~al.}{2018}]{wu+2018}
{Wu} P.-F.,  et~al., 2018, \mn@doi [\apj] {10.3847/1538-4357/aae822}, \href
  {http://adsabs.harvard.edu/abs/2018ApJ...868...37W} {868, 37}

\bibitem[\protect\citeauthoryear{{Wuyts}, {van Dokkum}, {Kelson}, {Franx}  \&
  {Illingworth}}{{Wuyts} et~al.}{2004}]{wuyts+2004}
{Wuyts} S.,  {van Dokkum} P.~G.,  {Kelson} D.~D.,  {Franx} M.,   {Illingworth}
  G.~D.,  2004, \mn@doi [\apj] {10.1086/381746}, \href
  {https://ui.adsabs.harvard.edu/abs/2004ApJ...605..677W} {605, 677}

\bibitem[\protect\citeauthoryear{{Zaritsky}, {Gonzalez}  \&
  {Zabludoff}}{{Zaritsky} et~al.}{2006}]{zaritsky+2006}
{Zaritsky} D.,  {Gonzalez} A.~H.,   {Zabludoff} A.~I.,  2006, \mn@doi [\apj]
  {10.1086/498672}, \href {http://adsabs.harvard.edu/abs/2006ApJ...638..725Z}
  {638, 725}

\bibitem[\protect\citeauthoryear{{Zibetti}, {Gallazzi}, {Hirschmann},
  {Consolandi}, {Falc{\'o}n-Barroso}, {van de Ven}  \& {Lyubenova}}{{Zibetti}
  et~al.}{2020}]{zibetti+2020}
{Zibetti} S.,  {Gallazzi} A.~R.,  {Hirschmann} M.,  {Consolandi} G.,
  {Falc{\'o}n-Barroso} J.,  {van de Ven} G.,   {Lyubenova} M.,  2020, \mn@doi
  [\mnras] {10.1093/mnras/stz3205}, \href
  {https://ui.adsabs.harvard.edu/abs/2020MNRAS.491.3562Z} {491, 3562}

\bibitem[\protect\citeauthoryear{{da Cunha}, {Charlot}  \& {Elbaz}}{{da Cunha}
  et~al.}{2008}]{dacunha+2008}
{da Cunha} E.,  {Charlot} S.,   {Elbaz} D.,  2008, \mn@doi [\mnras]
  {10.1111/j.1365-2966.2008.13535.x}, \href
  {http://adsabs.harvard.edu/abs/2008MNRAS.388.1595D} {388, 1595}

\bibitem[\protect\citeauthoryear{{de Graaff} et~al.,}{{de Graaff}
  et~al.}{2020}]{degraaff+2020}
{de Graaff} A.,  et~al., 2020, \mn@doi [\apjl] {10.3847/2041-8213/abc428},
  \href {https://ui.adsabs.harvard.edu/abs/2020ApJ...903L..30D} {903, L30}

\bibitem[\protect\citeauthoryear{{de Zeeuw} et~al.,}{{de Zeeuw}
  et~al.}{2002}]{dezeeuw+2002}
{de Zeeuw} P.~T.,  et~al., 2002, \mn@doi [\mnras]
  {10.1046/j.1365-8711.2002.05059.x}, \href
  {http://adsabs.harvard.edu/abs/2002MNRAS.329..513D} {329, 513}

\bibitem[\protect\citeauthoryear{{van Dokkum} \& {Stanford}}{{van Dokkum} \&
  {Stanford}}{2003}]{vandokkum+stanford2003}
{van Dokkum} P.~G.,  {Stanford} S.~A.,  2003, \mn@doi [\apj] {10.1086/345989},
  \href {https://ui.adsabs.harvard.edu/abs/2003ApJ...585...78V} {585, 78}

\bibitem[\protect\citeauthoryear{{van Dokkum} \& {van der Marel}}{{van Dokkum}
  \& {van der Marel}}{2007}]{vandokkum+vandermarel2007}
{van Dokkum} P.~G.,  {van der Marel} R.~P.,  2007, \mn@doi [\apj]
  {10.1086/509633}, \href
  {https://ui.adsabs.harvard.edu/abs/2007ApJ...655...30V} {655, 30}

\bibitem[\protect\citeauthoryear{{van Rossum}}{{van
  Rossum}}{1995}]{vanrossum1995}
{van Rossum} G.,  1995, CWI Technical Report, CS-R9526

\bibitem[\protect\citeauthoryear{{van de Sande}, {Kriek}, {Franx}, {Bezanson}
  \& {van Dokkum}}{{van de Sande} et~al.}{2014}]{vandesande+2014}
{van de Sande} J.,  {Kriek} M.,  {Franx} M.,  {Bezanson} R.,   {van Dokkum}
  P.~G.,  2014, \mn@doi [\apjl] {10.1088/2041-8205/793/2/L31}, \href
  {https://ui.adsabs.harvard.edu/abs/2014ApJ...793L..31V} {793, L31}

\bibitem[\protect\citeauthoryear{{van de Sande} et~al.,}{{van de Sande}
  et~al.}{2017a}]{vandesande+2017b}
{van de Sande} J.,  et~al., 2017a, \mn@doi [\mnras] {10.1093/mnras/stx1751},
  \href {http://adsabs.harvard.edu/abs/2017MNRAS.472.1272V} {472, 1272}

\bibitem[\protect\citeauthoryear{{van de Sande} et~al.,}{{van de Sande}
  et~al.}{2017b}]{vandesande+2017a}
{van de Sande} J.,  et~al., 2017b, \mn@doi [\apj]
  {10.3847/1538-4357/835/1/104}, \href
  {http://adsabs.harvard.edu/abs/2017ApJ...835..104V} {835, 104}

\bibitem[\protect\citeauthoryear{{van de Sande} et~al.,}{{van de Sande}
  et~al.}{2018}]{vandesande+2018}
{van de Sande} J.,  et~al., 2018, \mn@doi [Nature Astronomy]
  {10.1038/s41550-018-0436-x}, \href
  {http://adsabs.harvard.edu/abs/2018NatAs...2..483V} {2, 483}

\bibitem[\protect\citeauthoryear{{van de Sande} et~al.,}{{van de Sande}
  et~al.}{2020}]{vandesande+2021}
{van de Sande} J.,  et~al., 2020, arXiv e-prints, \href
  {https://ui.adsabs.harvard.edu/abs/2020arXiv201108199V} {p. arXiv:2011.08199}

\bibitem[\protect\citeauthoryear{{van der Wel}, {Franx}, {van Dokkum}  \&
  {Rix}}{{van der Wel} et~al.}{2004}]{vanderwel+2004}
{van der Wel} A.,  {Franx} M.,  {van Dokkum} P.~G.,   {Rix} H.~W.,  2004,
  \mn@doi [\apjl] {10.1086/381887}, \href
  {https://ui.adsabs.harvard.edu/abs/2004ApJ...601L...5V} {601, L5}

\makeatother
\end{thebibliography}



\appendix

\section{Normalising a censored 3-d Gaussian}\label{a.s.fi}

To calculate the normalisation factor $f_i$ defined in
equation~(\ref{eq.app.fi}), we introduce the matrix $\bm{A}_i$ given by
\begin{equation}
  \left(\bm{R}^T \bm{\Sigma} \bm{R} + \bm{E}_i\right)^{-1} \equiv \bm{A}_i^{-1} \equiv
  \begin{bmatrix}
  a_{11} & a_{12} & a_{13} \\
  a_{21} & a_{22} & a_{23} \\
  a_{13} & a_{23} & a_{33} \\
   \end{bmatrix}
\end{equation}
where, for brevity, we dropped the dependence on $i$ from the matrix elements.
With this notation, the equation for $f_i$ becomes
\begin{equation}
    \dfrac{1}{f_i(L_\mathrm{min})} =
    \int_{-\infty}^{\infty} dx_1
    \int_{-\infty}^{\infty} dx_2
    \int_{\log L_\mathrm{min} - \mu3}^{\infty} dx_3
    \mathcal{N}_{\bm{0},\bm{A}_i}(\bm{x})
\end{equation}
The argument of the exponential function in the Gaussian function above can be
decomposed into the sum of three squares as
\begin{equation}
  \begin{split}
     -\dfrac{1}{2} \langle & \bm{x}, \bm{A}_i^{-1} \bm{x} \rangle = 
     -\dfrac{1}{2} \left(\xi_1^2 + \xi_2^2 + \xi_3^2 \right) \\
     \xi_1 & \equiv \sqrt{a_{11}} x_1 + \dfrac{a_{12} x_2 + a_{13} x_3}{\sqrt{a_{11}}} \\
     \xi_2 & \equiv  \displaystyle{\sqrt{a_{22} - \dfrac{a_{12}^2}{a_{11}}}} x_2
  + \dfrac{a_{23} - \dfrac{a_{12} a_{13}}{a_{11}}}{\sqrt{a_{22} - \dfrac{a_{12}^2}{a_{11}}}} x_3\\
     \xi_3 & \equiv \sqrt{ a_{33} \hspace{-0.2em} - \dfrac{a_{13}^2}{a_{11}}
  \hspace{-0.2em} - \dfrac{\left(a_{23} - \dfrac{a_{12} a_{13}}{a_{11}}\right)^2}{a_{22} - \dfrac{a_{12}^2}{a_{11}}}} x_3 \equiv \sqrt{\mathcal{D}_i} x_3
  \end{split}
\end{equation}
We can therefore change variables in the multiple integral and, since the
adopted change of variables has a Jacobian determinant equal to
$\sqrt{\det{\bm{A}_i}}$, we get
\begin{equation}
    \dfrac{1}{f_i(L_\mathrm{min})} =
    \int_{-\infty}^{\infty} d\xi_1
    \int_{-\infty}^{\infty} d\xi_2
    \int_{\sqrt{\mathcal{D}_i}(\log L_\mathrm{min} - \mu3)}^{\infty} d\xi_3 
    \mathcal{N}_{\bm{0},\bm{I}}(\bm{\xi})
\end{equation}
This readily evaluates to
\begin{equation}
    \dfrac{1}{f_i(L_\mathrm{min})} =
    \dfrac{1}{2} \left[1 - \mathrm{erf}\left(\sqrt{\dfrac{\mathcal{D}_i}{2}}\left(\log L_\mathrm{min}-\mu_3\right)\right)\right]
\end{equation}
and $\mathcal{D}_i$ can be simplified to
\begin{equation}
    \mathcal{D}_i = \dfrac{\det{\bm{A}_i^{-1}}}{a_{11} a_{22} - a_{12}^2}
\end{equation}
It may also be useful to recall that, in the more general case where $x_3$ is
censored at both extremes $\alpha$ and $\beta$, then
\begin{equation}
    \dfrac{1}{f_i(L_\mathrm{min})} =
    \dfrac{1}{2} \left[\mathrm{erf}\left(\sqrt{\dfrac{\mathcal{D}_i}{2}}\left(\beta-\mu_3\right)\right) - 
    \mathrm{erf}\left(\sqrt{\dfrac{\mathcal{D}_i}{2}}\left(\alpha-\mu_3\right)\right)\right]
\end{equation}
Where more than one variable is censored, the integrating factor can be computed
by Cholesky decomposition \citepalias[see ][their equation~A3]{magoulas+2012}.

\section{
Orthogonal vs direct formulation of the least-squares problem with correlated, heteroscedastic measurement uncertainties
}\label{a.s.lq}

In this section, we provide a formula to include correlated, heteroscedastic 
measurement uncertainties in the expression for the direct fit to 2-d and
3-d linear relations. While they may or may not be biased for galaxy-evolution
studies, there is no questioning the importance of direct-fit methods to the FP
use as a distance indicator. Because these methods minimise the uncertainty on
the dependent variable from which distances are inferred, they also minimise the
uncertainty on the FP-derived distances \citep[e.g.][]{bernardi+2003, said+2020}.
Despite this importance, we are not aware of any study providing the correct
expression to include correlated uncertainties. For comparison, we also give
the corresponding expressions for the orthogonal fit, even though these are
already available in the literature \citep{robotham+obreschkow2015}.

Generally, in the FP literature, direct fits optimise the $\chi^2$ defined by
\begin{equation}
  \chi^2_\mathrm{direct} \equiv \sum_{i=0}^{N-1} \frac{(z_i - m \, x_i - b \, y_i - c)^2}{\sigma_\mathrm{int}^2}
\end{equation}
where $a$, $b$ and $c$ are the FP coefficients, $\sigma_\mathrm{int}$ is the
intrinsic scatter along the $z$ axis, and $(x_i,y_i,z_i)$, $i \in {0, \dots, N}$,
are the data. Orthogonal fits minimise
\begin{equation}
  \chi^2_\perp = \sum_{i=0}^{N-1} \frac{(z_i - a \, x_i - b \, y_i - c)^2}{\sigma_\perp^2 (a^2 + b^2 + 1)}
\end{equation}
where $\sigma_\perp$ is the FP intrinsic scatter, orthogonal to the FP. When
adding uncorrelated measurement uncertainties ($\sigma_{x,i}, \sigma_{y,i},
\sigma_{z,i})$, the expression for $\chi^2_\perp$ is generally
modified to be
\begin{equation}\label{eq.app.orth}
  \chi^2_\perp = \sum_{i=0}^{N-1} \frac{(z_i - a \, x_i - b \, y_i - c)^2}{\sigma_\perp^2 + \sigma_{z,i}^2 + a^2 \sigma_{x,i}^2 + b^2 \sigma_{y,i}^2}
\end{equation}
However, we argue that this expression is properly appropriate to
$\chi^2_\mathrm{direct}$, with the substitution $\sigma_\perp \rightarrow
\sigma_\mathrm{int}$. This claim can be verified by considering
the perfect symmetry between $\sigma_{z,i}$ (along the $z$ axis) and
$\sigma_\perp$ (that should be orthogonal to the plane); clearly, if
equation (\ref{eq.app.orth}) is to be the expression for
$\chi^2_\perp$, these two terms cannot add in an unweighted
quadrature: either $\sigma_\perp$ must be projected along the $z-$axis or
$\sigma_{z,i}$ must be projected orthogonal to the plane before they can
be added in quadrature. Moreover, taking the
limit ($\sigma_{x,i}, \sigma_{y,i}, \sigma_{z,i}) \rightarrow \mathbf{0}$,
equation (\ref{eq.app.orth}) implies $\chi^2_\perp \rightarrow
\chi^2_\mathrm{direct}$, which is incorrect (again, with the substitution
$\sigma_\perp \rightarrow \sigma_\mathrm{int}$).

Hence we propose to modify equation (\ref{eq.app.orth}) as follows
\begin{equation}
  \chi^2_\perp \equiv \sum_{i=0}^{N-1} \frac{(z_i - a \, x_i - b \, y_i - c)^2}{
    (a^2 + b^2 + 1) \sigma_\perp^2 + \langle (a, b, -1), \mathbf{E}_i (a, b, -1) \rangle}
\end{equation}
which manifestly tends to the correct expression for
$\chi^2_\perp$ when measurement uncertainties are negligible. In vector
notation, defining the unit normal to the plane as $\mathbf{\hat{n}} \equiv
(a, b, -1) / \sqrt{a^2 + b^2 + 1}$, $d \equiv c/\sqrt{a^2 + b^2 + 1}$ as the
orthogonal distance between the plane and the origin, and $\mathbf{E}_i$ as the
covariance matrix of measurement uncertainties, the expression becomes
\begin{equation}
  \chi^2_\perp \equiv \sum_{i=0}^{N-1} \frac{(\langle \mathbf{\hat{n}}, \mathbf{x}_i \rangle - d)^2}{
    \sigma_\perp^2 + \langle \mathbf{\hat{n}}, \mathbf{E}_i \mathbf{\hat{n}} \rangle}
\end{equation}
which is rotationally symmetric. Similarly, the expression for the direct fit is
\begin{equation}
  \chi^2_\mathrm{direct} \equiv \sum_{i=0}^{N-1} \frac{(z_i - a \, x_i - b \, y_i - c)^2}{
    \sigma_\mathrm{int}^2 + \langle (a, b, -1), \mathbf{E}_i (a, b, -1) \rangle}
\end{equation}
which, as it should be, cannot be written in rotationally-symmetric form, but
has complete symmetry between $\sigma_\mathrm{int}$ and $\sigma_{z,i} =
\sqrt{[\mathbf{E}_i]_{z,z}}$.
In principle, this expression can be derived from equation~(29) of \citet{kelly2007}, using the
same approach as they use to derive the corresponding 2-d expression (their
equation~24).

These expressions can be easily derived in the 2-d case, by considering the
probability distribution of a 2-d Gaussian with uncorrelated scatter along two
independent directions $\mathbf{v_1}$ and $\mathbf{v_2}$ (direct fit) or along
two orthogonal directions (orthogonal fit). If we write the Gaussian correlation
matrix in the reference frame $[\mathbf{v_1}, \mathbf{v_2}]$ as
\begin{equation}
  \mathbf{\Sigma} \equiv
  \begin{bmatrix}
  \sigma_1^2      & 0          \\
  0               & \sigma_2^2 \\
   \end{bmatrix}
\end{equation}
then the expressions for the 2-d analogues of the $\chi^2$ can be obtained
by maximising the likelihood function in the limit $\sigma_1 \rightarrow \infty$.
If we write the transformation between the coordinates $(v_1, v_2)$ and $(x, y)$
as $\mathbf{v} = \mathbf{T} \mathbf{x}$, the probability of the data given
the model can be written as
\begin{equation}
    p(\bm{x}_i|\mathrm{model}) = \prod_{i=0}^{N-1} \mathcal{N}_{\mu, \mathbf{T}^{-1} \mathbf{\Sigma} (\mathbf{T}^T)^{-1} + \mathbf{E}_i}(\mathbf{x}_i)
\end{equation}
where $\mathbf{\mu}$ is the centroid and
\begin{equation}
  \mathbf{T} \equiv
  \begin{bmatrix}
  1 / \cos \vartheta                & 0 \\
  - \sin \vartheta / \cos \vartheta & 1 \\
   \end{bmatrix}
\end{equation}
if the scatter $\sigma_2$ is along the $z$ axis (direct fit) whereas
\begin{equation}
  \mathbf{T} \equiv
  \begin{bmatrix}
    \cos \vartheta & \sin \vartheta \\
  - \sin \vartheta & \cos \vartheta \\
   \end{bmatrix}
\end{equation}
if the scatter $\sigma_2$ is orthogonal to the line (orthogonal fit). The
expression for $p(\bm{x}_i|\mathrm{model})$ can be written explicitly. For
each data point $i$, the argument of the Gaussian exponentials are (apart from a
factor $-1/2$)
\begin{equation}
  \chi^2_{i,\mathrm{direct}} \equiv \frac{(\sin \vartheta \, (x_i - \mu_x) - \cos \vartheta \, (y_i - \mu_y))^2}{\splitfrac{
    \cos \vartheta^2 \sigma_\mathrm{int}^2 + \sin \vartheta^2 \sigma_{x,i}^2 + \cos \vartheta^2 \sigma_{y,i}^2}{- 2 \rho_i \sin \vartheta \cos \vartheta \sigma_{x,i} \sigma_{y,i}}}
\end{equation}
and
\begin{equation}
  \chi^2_{i,\perp} \equiv
  \frac{(\sin \vartheta \, (x_i - \mu_x) - \cos \vartheta \, (y_i - \mu_y))^2}{\splitfrac{
    \sigma_\perp^2 + \sin \vartheta^2 \sigma_{x,i}^2 + \cos \vartheta^2 \sigma_{y,i}^2}{- 2 \rho_i \sin \vartheta \cos \vartheta \sigma_{x,i} \sigma_{y,i}}}
\end{equation}
Notice that in the limit $\sigma_1 \rightarrow \infty$ the probability
distribution $p(\bm{x}_i|\mathrm{model})$ is no longer integrable. Using this
distribution as a probability requires some truncation along $x$ at a `large'
distance from the centroid $\mu$. With this caveat, maximising the
`probability' of the `infinitely-extended' Gaussian is akin to minimising the
sum of the squares $\chi^2_i$ (we neglect the normalisation factor). So the
expression for $\chi^2_\mathrm{direct}$ in 2-d can be written either as
\begin{equation}
  \chi^2_\mathrm{direct} = \sum_{i=0}^{N-1} \frac{(\sin \vartheta \, (x_i - \mu_x) - \cos \vartheta \, (y_i - \mu_y))^2}{\splitfrac{
    \cos \vartheta^2 \sigma_\mathrm{int}^2 + \sin \vartheta^2 \sigma_{x,i}^2 + \cos \vartheta^2 \sigma_{y,i}^2}{- 2 \rho_i \sin \vartheta \cos \vartheta \sigma_{x,i} \sigma_{y,i}}}
\end{equation}
which reduces the bias for steep slopes, or in the more familiar form
\begin{equation}
  \chi^2_\mathrm{direct} = \sum_{i=0}^{N-1} \frac{(a x_i + b - y_i)^2}{
    \sigma_\mathrm{int}^2 + a^2 \sigma_{x,i}^2 + \sigma_{y,i}^2 - 2 \rho_i a \sigma_{x,i} \sigma_{y,i}}
\end{equation}
where $a = \tan \vartheta$ and and $b = \mu_y - a \mu_x$ are the slope and
zero-point of the line. The expressions for the orthogonal 2-d fit are
\begin{equation}
  \chi^2_\perp = \sum_{i=0}^{N-1} \frac{(\sin \vartheta \, (x_i - \mu_x) - \cos \vartheta \, (y_i - \mu_y))^2}{
    \splitfrac{\sigma_\perp^2 + \sin \vartheta^2 \sigma_{x,i}^2 + \cos \vartheta^2 \sigma_{y,i}^2}{- 2 \rho_i \sin \vartheta \cos \vartheta \sigma_{x,i} \sigma_{y,i}}}
\end{equation}
and
\begin{equation}
  \chi^2_\perp = \sum_{i=0}^{N-1} \frac{(a x_i + b - y_i)^2}{
    (1 + a^2) \sigma_\perp^2 + a^2 \sigma_{x,i}^2 + \sigma_{y,i}^2 - 2 \rho_i a \sigma_{x,i} \sigma_{y,i}}
\end{equation}
Clearly, some constraints on $\sigma_\mathrm{intr}$ and $\sigma_\perp$ are
required before minimising these expressions.

\section{The {\sc lts\_planefit} analysis}\label{a.s.lts}

\begin{figure}
  \includegraphics[type=pdf,ext=.pdf,read=.pdf,width=1.0\columnwidth]{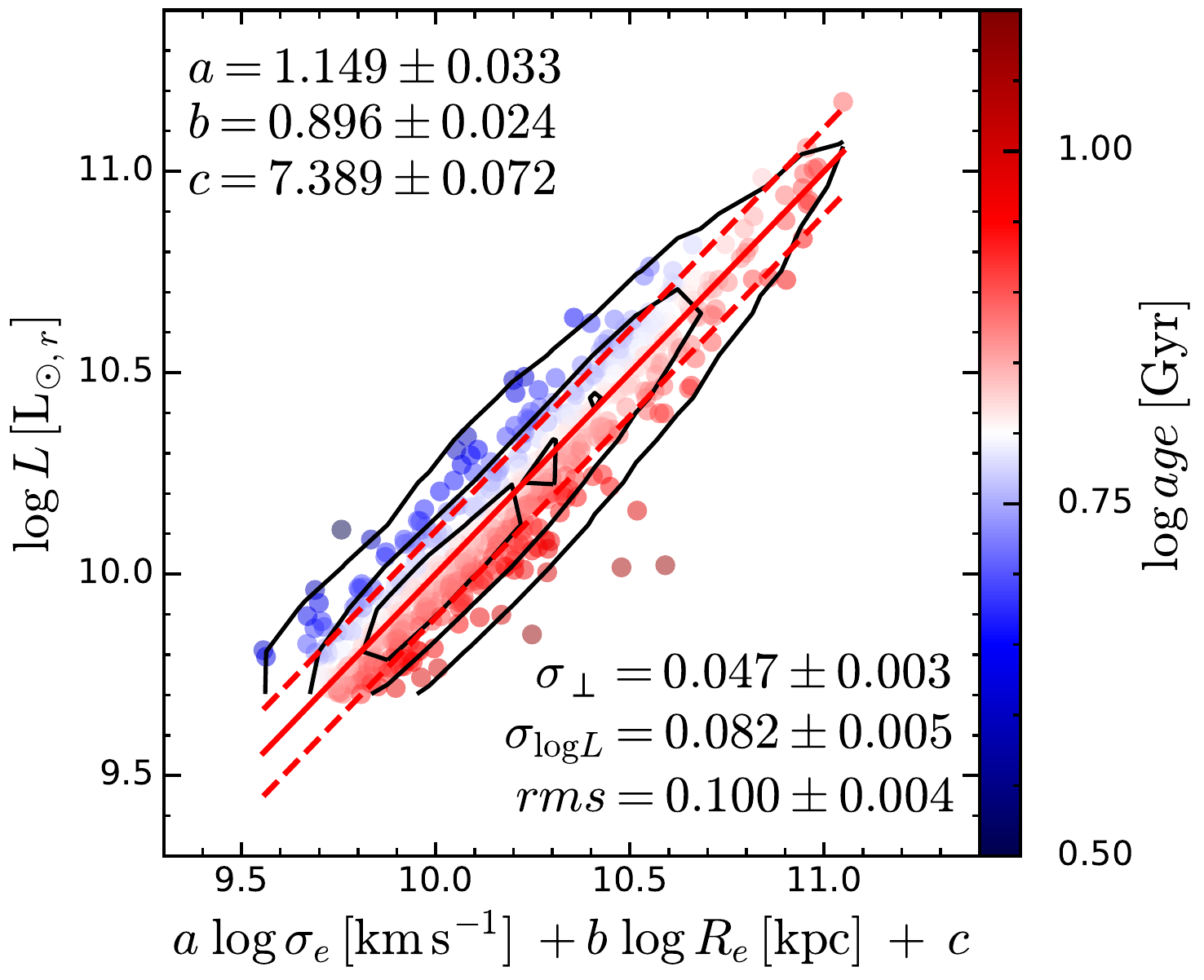}
  \caption{The LTS FP for the SAMI ETGs, showing a clear age gradient
  across the plane. Each circle represents a SAMI galaxy, colour-coded by the
  (LOESS-smoothed) SSP age. The best-fit FP is traced by
  the solid red line; the dashed red lines encompass $\pm rms$.
  The black contours enclose the 90\textsuperscript{th},
  67\textsuperscript{th} and 30\textsuperscript{th} percentiles of the
  distribution. 
  There is a clear age gradient across the FP: at fixed \sige and \re, old
  galaxies (red hues) are under-luminous and tend to lie below the
  best-fit plane.
  }\label{f.a.fpbench}
\end{figure}

\begin{figure}
  \includegraphics[type=pdf,ext=.pdf,read=.pdf,width=1.0\columnwidth]{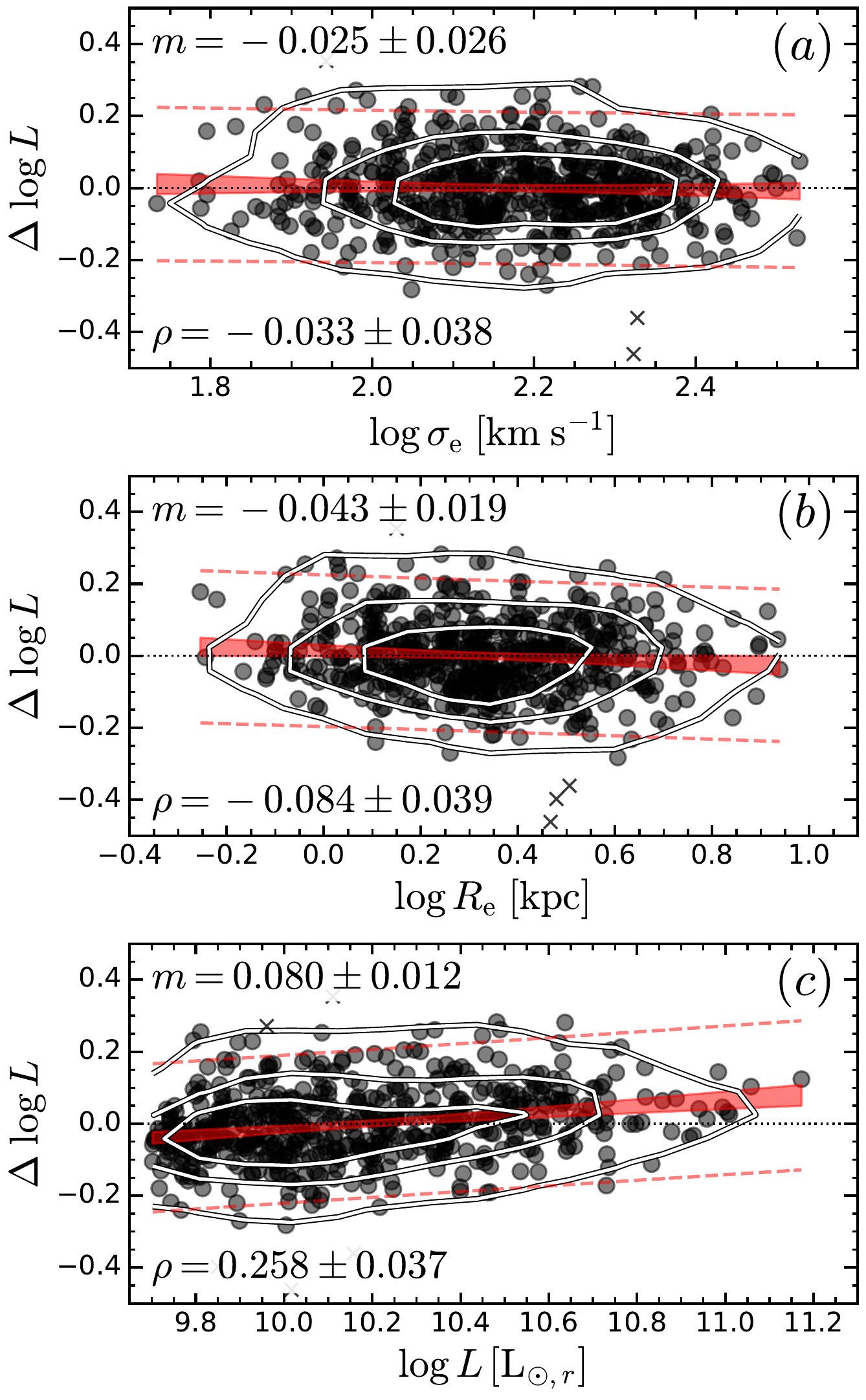}
  {\phantomsubcaption\label{f.a.galev.resid.nnl.a}
   \phantomsubcaption\label{f.a.galev.resid.nnl.b}
   \phantomsubcaption\label{f.a.galev.resid.nnl.c}}
  \caption{The residuals of the LTS FP exhibit the expected correlations
  with the FP observables: \logsige (top), \logre (middle) and \logl
  (bottom). Each circle represents a SAMI galaxy; the white contours
  enclose the 90\textsuperscript{th}, 67\textsuperscript{th} and
  30\textsuperscript{th} percentiles of the distribution. The red line
  traces the best-fit linear relation; the red regions are the 95\% confidence
  intervals and the dashed red lines are the 95\% prediction intervals. The
  best-fit linear slope $m$ and the Spearman rank correlation coefficient are
  reported at the top left and bottom left of each panel.
  }\label{f.a.galev.resid.nnl}
\end{figure}

\begin{figure}
  \includegraphics[type=pdf,ext=.pdf,read=.pdf,width=1.0\columnwidth]{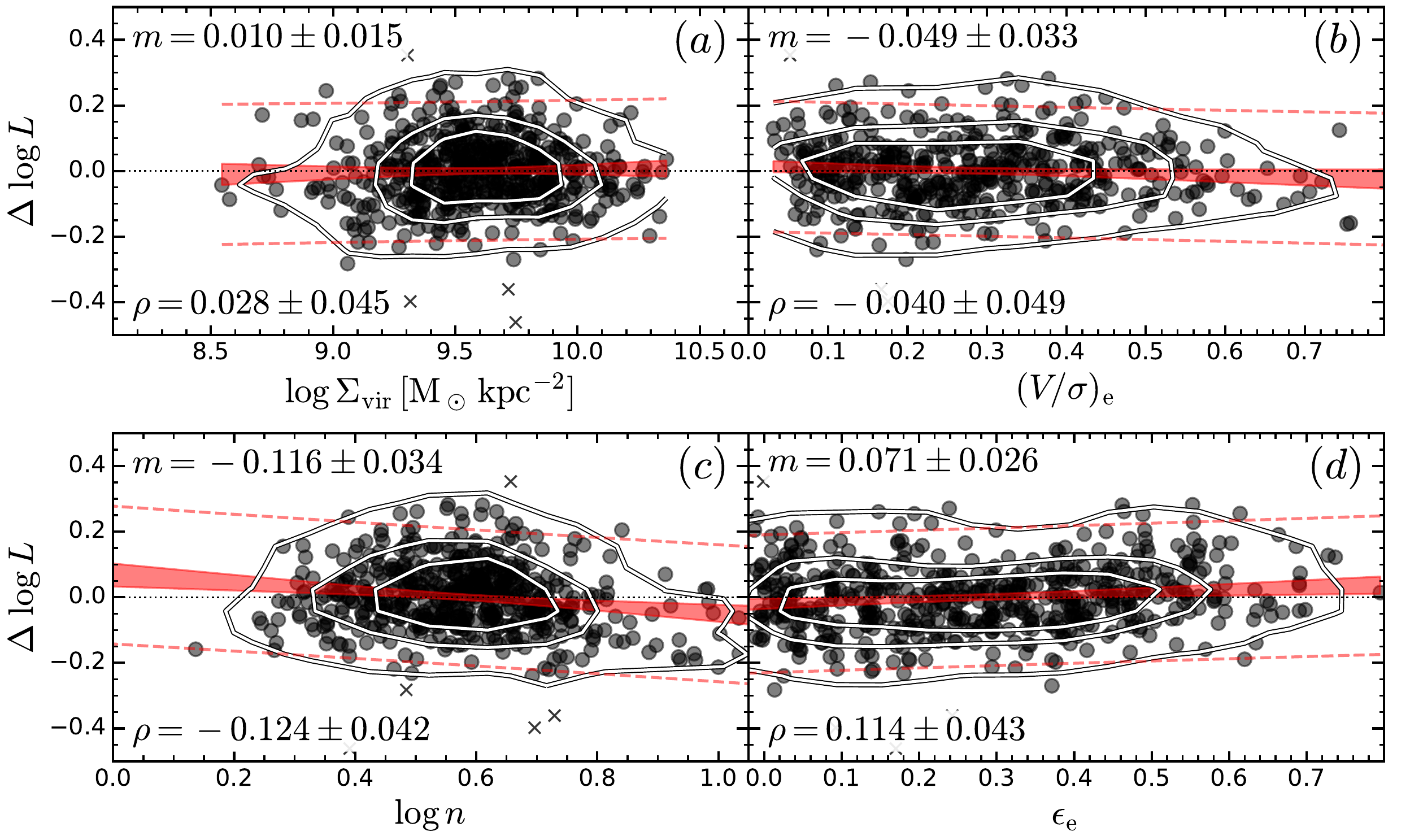}
  {\phantomsubcaption\label{f.a.galev.resid.str.a}
   \phantomsubcaption\label{f.a.galev.resid.str.b}
   \phantomsubcaption\label{f.a.galev.resid.str.c}
   \phantomsubcaption\label{f.a.galev.resid.str.d}}
  \caption{The residuals of the LTS FP have weak or no correlation with
  the structural observables considered here: dynamical surface mass density
  \Sigvir (panel~\subref{f.a.galev.resid.str.a}), $(V/\sigma)_\mathrm{e}$
  (panel~\subref{f.a.galev.resid.str.b}), S{\'e}rsic index
  (panel~\subref{f.a.galev.resid.str.c}) and projected ellipticity
  (\subref{f.a.galev.resid.str.d}). The symbols are the same as in
  \reffig{f.a.galev.resid.nnl}. We find evidence of a correlation only for
  $n$ ($\approx 3 \, \sigma$ significance).
  }\label{f.a.galev.resid.str}
\end{figure}

\begin{figure}
  \includegraphics[type=pdf,ext=.pdf,read=.pdf,width=1.0\columnwidth]{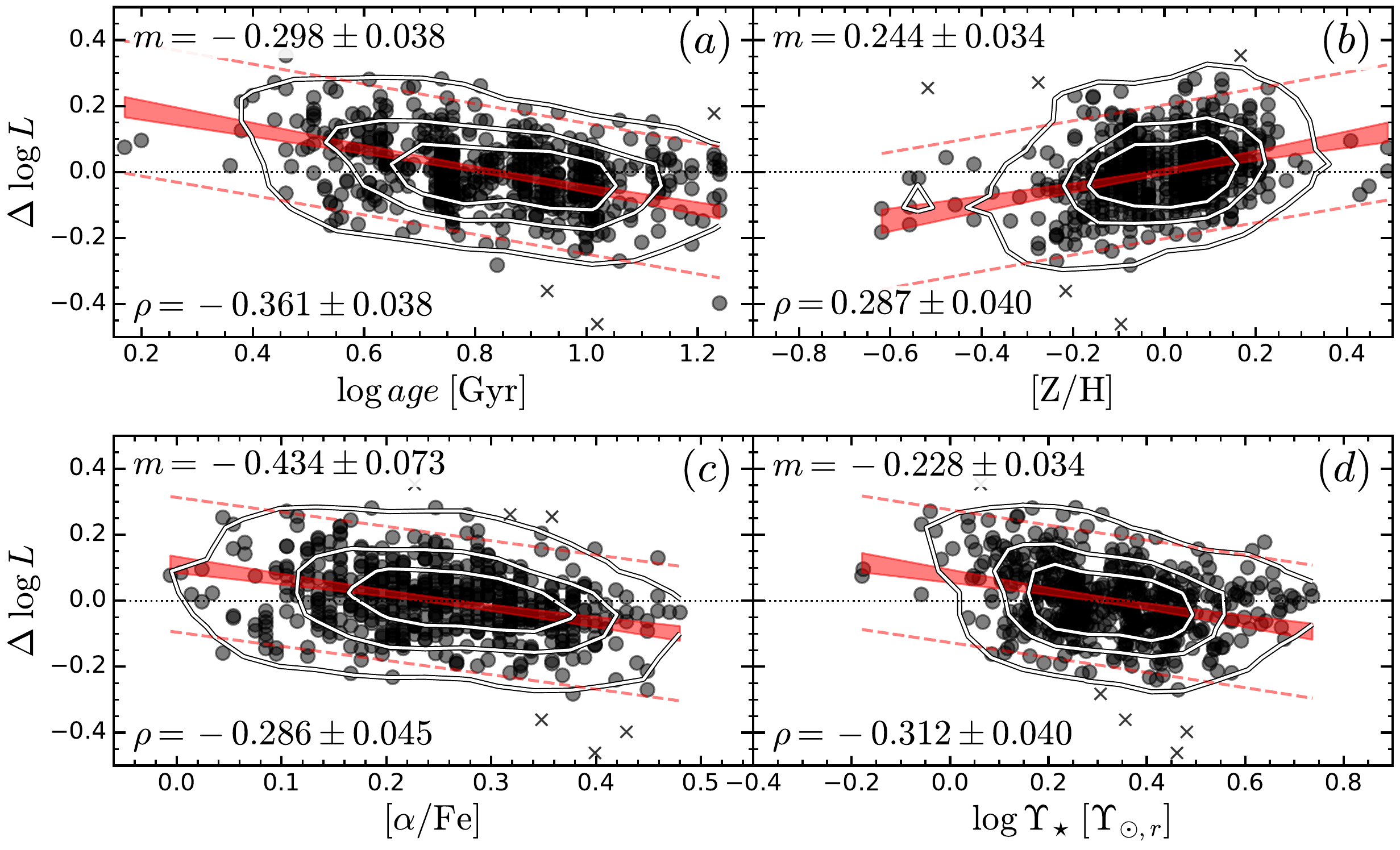}
  {\phantomsubcaption\label{f.a.galev.resid.ssp.a}
   \phantomsubcaption\label{f.a.galev.resid.ssp.b}
   \phantomsubcaption\label{f.a.galev.resid.ssp.c}
   \phantomsubcaption\label{f.a.galev.resid.ssp.d}}
  \caption{Like the residuals of the fiducial FP (Fig.~\ref{f.galev.resid.ssp}),
  also the residuals of the LTS FP correlate most strongly with
  stellar population \logage ($\approx 8 \, \sigma$ significance;
  panel~\subref{f.a.galev.resid.ssp.a}). The other stellar population properties
  considered here all show significant correlations, including \ZH
  (panel~\subref{f.a.galev.resid.ssp.b}), \aFe (\subref{f.a.galev.resid.ssp.c})
  and \logups (\subref{f.a.galev.resid.ssp.d}). The symbols are the same as in
  \reffig{f.a.galev.resid.nnl}.
  }\label{f.a.galev.resid.ssp}
\end{figure}

If we repeat our analysis using the {\sc lts\_planefit} algorithm instead of
the 3dG algorithm we find a FP different to the fiducial FP
(cf.\ Figs \ref{f.r.fpbench} and \ref{f.a.fpbench}), as well as different slope and
significance for the residual trends (as expected from the underlying
probabilistic models, \S~\ref{s.da.ss.rescorr}; Figs \ref{f.a.galev.resid.nnl},
\ref{f.a.galev.resid.str} and \ref{f.a.galev.resid.ssp}).
However, the ranking of the most significant residual 
trends is the same between the two methods, so our conclusions are
qualitatively the same and are not an artefact of the particular algorithm
adopted. In \reffig{f.a.fpbench} we show the LTS FP. The best-fit value of the
coefficient $b = 0.896 \pm 0.024$ is statistically consistent with the 3dG
equivalent, but both $a = 1.149 \pm 0.033$ and $c = 7.389 \pm 0.072$ are
statistically different. The origin of this difference is
due to different underlying models, yet the normal to the LTS FP is very
close ($3.2\pm1.0\degree$) to the normal to the fiducial FP (Table~\ref{t.r.eigen}, column~5).

\section{Extinction across the Fundamental Plane}\label{a.s.dust}

The GAMA database provides two different SED fits, therefore two different
values of the $r-$band extinction $A_r$. Given that the extinction measurements
are very noisy, and are not present for the cluster galaxies, we proceed as follows.

We assume that extinction correlates primarily with stellar population age and
inclination, so we start by smoothing the $A_r$ distribution on the
$\log age$--$\epsilon$ plane; we then interpolate the smooth distribution,
and finally for each galaxy in the mock we infer the interpolated value of $A_r$.
\Lumsynth is then corrected down to account for the inferred extinction.
The resulting FP has lower scatter, by 0.02\,dex \citep[when using
$E(B-V)$ from][]{taylor+2011} or by 0.01\,dex \citep[when using
$\tau_V$ derived from {\sc \href{http://astronomy.swinburne.edu.au/~ecunha/ecunha/MAGPHYS.html}{magphys}}, ][]{dacunha+2008}.

We conclude that extinction
could play a role in the FP, by reducing the scatter due to the spread in stellar
population age, but given the available precision of the measurements of $A_r$, the inferred reduction is marginal.


\bsp	
\label{lastpage}
\end{document}